\documentstyle[amssymb,epic,epsfig,multirow,pifont]{mn}

\newif\ifAMStwofonts

\def\fig#1{Figure~\ref{#1}}
\def\Fig#1{Figure~\ref{#1}}

\def\tab#1{Table~\ref{#1}}
\def\sec#1{Section~\ref{#1}}

\def\appen#1{Appendix~\ref{#1}}

\newcommand{\micron}[1]{$\mu\rm{m}{#1}$}
\newcommand{\hubble}{H$_{\circ}$}
\newcommand{\hubbleunits}{kms$^{-1}$Mpc$^{-1}$}
\newcommand{\omegao}{$\Omega_{\circ}$}
\newcommand{\luminsub}{L$_{850\mu m}$}
\newcommand{\radiopower}{P$_{151{\rm MHz}}$}
\newcommand{\alpharadio}{$\alpha_{radio}$}
\newcommand{\aj}{AJ}
\newcommand{\apj}{ApJ}
\newcommand{\apjs}{ApJS}
\newcommand{\apjl}{ApJ}
\newcommand{\mnras}{MNRAS}
\newcommand{\memras}{MmRAS}
\newcommand{\nat}{Nature}
\newcommand{\aap}{A\&A}
\newcommand{\aaps}{A\&AS}
\newcommand{\qjras}{QJRAS}
\newcommand{\procspie}{Proc. SPIE}
\newcommand{\pasp}{PASP}

\title[The star-formation history of radio galaxies]{A submillimetre survey of the star-formation history of radio galaxies}
\author[E. N. Archibald et al.]
       {E. N. Archibald$^{1,2}$\thanks{email: e.archibald@jach.hawaii.edu}, J. S. Dunlop$^{2}$, D. H. Hughes$^{2,3}$, S. Rawlings$^{4}$, S. A. Eales$^{5}$, \and and R. J. Ivison$^{6}$\\
$^{1}$Joint Astronomy Centre, 660 N. A`oh\={o}k\={u} Place, University
Park, Hilo, Hawaii, 9620, USA \\
$^{2}$Institute for Astronomy, University of Edinburgh, Blackford Hill,
Edinburgh, EH9 3HJ, Scotland\\
$^{3}$Instituto Nacional de Astrofisica, Optica y Electronica (INAOE), 
Apartado Postal 51 y 216, 72000 Puebla, Pue., Mexico\\
$^{4}$Astrophysics, Department of Physics, Keble Road, Oxford, 
OX1 3RH, England\\
$^{5}$Department of Physics and Astronomy, University of Wales Cardiff,
P.O. Box 913, Cardiff, CF2 3YB, Wales\\
$^{6}$Department of Physics \& Astronomy, University College London, Gower 
Street, London, WC1E 6BT, England\\}
\date{Accepted ;
      Received ;
      in original form }

\pagerange{\pageref{firstpage}--\pageref{lastpage}}
\pubyear{2000}

\begin{document}

\maketitle

\label{firstpage}

\begin{abstract}
We present the results of the first major systematic submillimetre
survey of radio galaxies spanning the redshift range $1 < z < 5$. The
primary aim of this work is to elucidate the star-formation history of
this sub-class of elliptical galaxies by tracing the cosmological
evolution of dust mass. Using SCUBA on the JCMT we have obtained
850-\micron{} photometry of 47 radio galaxies to a consistent rms
depth of $1\;$mJy, and have detected dust emission in 14 cases.  The
radio galaxy targets have been selected from a series of low-frequency
radio surveys of increasing depth (3CRR, 6CE, etc), in order to allow
us to separate the effects of increasing redshift and increasing radio
power on submillimetre luminosity.  Although the dynamic range of our
study is inevitably small, we find clear evidence that the typical
submillimetre luminosity (and hence dust mass) of a powerful radio
galaxy is a strongly increasing function of redshift; the detection
rate rises from $\simeq$15 per cent at $z < 2.5$ to $\gtrsim$75 per
cent at $z > 2.5$, and the average submillimetre luminosity rises at a
rate $\propto (1+z)^3$ out to $z \simeq 4$.  Moreover our extensive
sample allows us to argue that this behaviour is not driven by
underlying correlations with other radio galaxy properties such as
radio power, radio spectral index, or radio source size/age.  Although
radio selection may introduce other more subtle biases, the redshift
distribution of our detected objects is in fact consistent with the
most recent estimates of the redshift distribution of comparably
bright submillimetre sources discovered in blank field surveys. The
evolution of submillimetre luminosity found here for radio galaxies
may thus be representative of massive ellipticals in general.
\end{abstract}


\section{Introduction}

Although large numbers of `normal' galaxies have now been discovered
out to $z \simeq 5$ (e.g. Steidel et al. 1999\nocite{sag99}), radio
galaxies continue to offer the best opportunity to study examples of
{\em massive elliptical galaxies} (or the progenitors thereof) back to
comparably early cosmic epochs. The reason for this is that a powerful
radio source requires a massive black hole, and it is relatively
certain that nowadays all such massive black holes reside in massive
ellipticals \cite{magorrian98,mkd99}.  Thus radio selection offers a
relatively efficient way of studying the properties of a sub-set of
massive ellipticals as a function of cosmic epoch. Moreover this
subset may well be representative of massive ellipticals in general,
especially since it can be argued that a substantial fraction of all
present-day ellipticals brighter than $2L^{\star}$ must have been
active at $z \simeq 2.5$ (Dunlop et al. 2000\nocite{dunlopetal2000}).

At low redshift, radio galaxies are dominated by well-evolved stellar
populations \cite{nolan2000}, have rather low
dust/gas masses \cite{kp91}, and lie in the same region of the
fundamental plane as normal inactive ellipticals (McLure et al. 1999;
Dunlop et al. 2000\nocite{mkd99,dunlopetal2000}).  Moreover current
evidence suggests that they have evolved only passively since at least
$z \simeq 1$ \cite{ll84,mcd00}.  This, coupled with the relatively old
ages derived for a few radio galaxies at $z \simeq 1.5$
\cite{dunlophy}, points towards a high redshift of formation, $z > 3$,
for the bulk of their stellar populations.  This means that both the
star-formation rate and gas mass in these galaxies should be a
strongly increasing function of redshift as one approaches their
primary formation epoch(s).

As has been argued by many authors, a massive starburst at high
redshift is expected to produce rapid chemical enrichment and to be
largely enshrouded in dust.  Consequently , submillimetre luminosity
should be a good indicator of the evolutionary state of an elliptical
galaxy, being expected to peak roughly half-way through the production
of a galaxy from primordial material \cite{ee96,frayerbrown97}, or
even earlier (i.e. as soon as the gas receives sufficient heating)
given pre-existing enrichment (Hughes, Dunlop \& Rawlings 1997\nocite{hdr97}).
Indeed submillimetre luminosity, viewed as a tracer of gas mass, is
arguably the best way to assess the evolutionary status of a massive
galaxy.  There are two reasons for this.  Firstly, the sensitivity and
bandwidth limitations of present-day instruments makes detecting
molecular line emission from high-redshift objects extremely
difficult.  Secondly, in the case of radio galaxies, optical-UV
measures may be confused by the direct or indirect effects of AGN
activity.

The detectability of high-redshift radio galaxies at submillimetre
wavelengths was first demonstrated by Dunlop et al. \shortcite{dhr94},
when they detected 4C41.17, at the time the most distant known galaxy
at $z=3.8$.  However the observation, made with the single element
bolometer detector UKT14 on the James Clerk Maxwell Telescope (JCMT),
required 4 hours of integration in exceptional weather conditions.  In
fact, the sensitivity of UKT14 permitted only the most extreme objects
to be detected at high redshift, and it was often a struggle to detect
even those.

The advent of the Submillimetre Common-User Bolometer Array (SCUBA) on
the JCMT offered the first real opportunity to rectify this
situation. Although photometric observations do not fully exploit the
multiplex advantage offered by an array camera, the individual
bolometers in the SCUBA array offered almost an order of magnitude
improvement in sensitivity over UKT14.  This made it feasible to
consider undertaking the first major submillimetre study of radio
galaxies spanning a wide range of redshifts, and it is the results of
the first major SCUBA survey of radio galaxies which we report here.

The layout of the paper is as follows: In \sec{samp} we give a more
detailed overview of pre-existing submillimetre observations of radio
galaxies, and explain the motivation for observing a sample of
galaxies compiled from flux-limited radio surveys of increasing
depth. The resulting sample is then described and summarized, before
the results of our new submillimetre observations are presented in
\sec{submmobs}.  \sec{synchcorrect} then gives details of the radio
properties of each galaxy, and explains how the total and, where
possible, core radio spectrum has been extrapolated to submillimetre
wavelengths to estimate (or at least constrain) the potential level of
non-thermal contamination at 850$\;$\micron{}. The coverage of the
radio-luminosity:redshift plane provided by our observed sample is
presented in \sec{pzplanesec}, and then in \sec{seclumin} we calculate
the rest-frame 850-\micron{} luminosities or upper limits for all the
observed galaxies.  In \sec{evolstats} we present a detailed
statistical exploration of the evidence for genuine cosmological
evolution of \luminsub{} in our sample. Finally in \sec{concsec} we
conclude by considering our results in the context of the recently
published blank-field submillimetre surveys.

We have deliberately confined this paper to a determination and
discussion of the relative behaviour of submillimetre luminosity in
our sample, and have postponed our analysis and interpretation of more
model-dependent properties, such as inferred gas mass and galaxy age,
to a subsequent paper.

\subsection{Conventions}

For clarity, we summarize here a number of conventions which have been 
adopted throughout this paper.
\begin{enumerate}
\item For each galaxy in the sample, information has been collated
from several references.  Each reference has been given a code; for
example, ER93\nocite{er93} corresponds to Eales S.A., Rawlings S.,
1993, ApJ, 411, 67.  These codes are included in the bibliography.
\item Throughout the paper, upper limits are calculated using the
following prescription: Consider an observation with signal S and
standard error $\varepsilon$.  If the signal is positive, the
$n$-$\sigma$ upper limit is S$+(n\times\varepsilon)$.  If the signal
is negative, the $n$-$\sigma$ upper limit is $n\times\varepsilon$.  It
cannot be guaranteed that the upper limits taken from other papers
were calculated in this manner.
\item For a power-law spectrum, the spectral index, $\alpha$, is
defined as ${\rm S}_{\nu}\propto \nu^{-\alpha}$, where S$_{\nu}$ is
the flux density at frequency $\nu$.  Thus, a radio synchrotron
spectrum has a positive spectral index, and the Rayleigh-Jeans tail of
a thermal spectrum has a negative spectral index.
\item The cosmological constant, $\Lambda$, is assumed to be zero
throughout the paper.
\end{enumerate}

\section{Project background and sample selection}
\label{samp}

\subsection{Pre-existing millimetre/submillimetre detections of high-redshift radio galaxies}

This work was carried out using SCUBA \cite{scubapaper}, the
submillimetre bolometer array mounted on the JCMT.  When the project
was planned, SCUBA had not yet been commissioned, and only three
high-redshift radio galaxies had been unambiguously detected at
millimetre/submillimetre wavelengths: 6C0902+34 (z=3.395), 4C41.17
(z=3.8), and 8C1435+635 (z=4.25).  While the millimetre observations
of 6C0902+34 appeared to be dominated by non-thermal emission from the
radio core \cite{dss96}, the observations of 4C41.17 and 8C1435+635
were striking in their similarity, with high emission levels
attributed to the presence of large reservoirs of dust.  The observed
flux densities are detailed in \tab{previoussubmmobs}.

\begin{figure*}
\epsfig{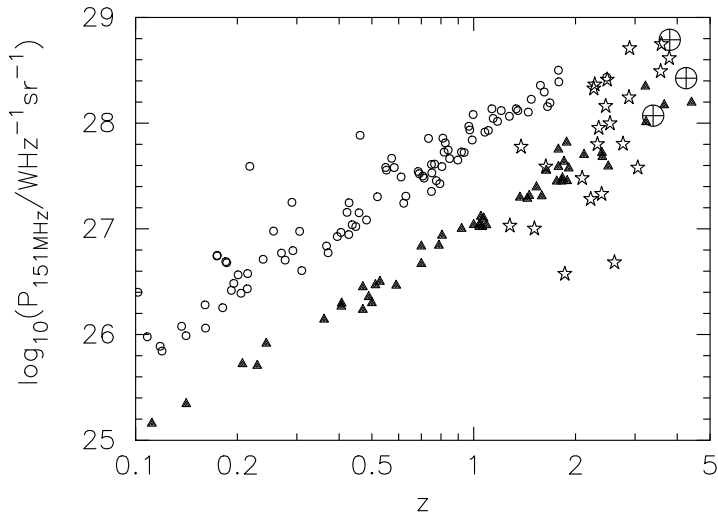}
\caption{Radio luminosity-redshift plane as defined by radio surveys
with successively deeper flux-density limits.  Open circles denote
galaxies from the 3C/3CRR survey and triangles denote galaxies from the 6CE
survey.  Stars indicate additional galaxies targeted as part of our
SCUBA survey from the 4C*, 7CRS, 8C*, MIT-Green Bank MG*, and Texas
TX* surveys.  4C41.17, 8C1435+635, and 6C0902+34 are each indicated by
a large cross surrounded by a circle.  As can be seen from this
figure, they are some of the most distant, radio-luminous objects
known.  Note, 53W069 is too faint to appear on this version (and all
subsequent versions) of the P-$z$ plane.  H$_{\circ}=50$
kms$^{-1}$Mpc$^{-1}$, $\Omega_{\circ} = 1.0$.}
\label{prescubapz}
\end{figure*}

4C14.17 and 8C1435+635 are both ultra-steep spectrum ($\alpha>1.0$)
radio galaxies, and both held the title, at the time of their
respective discovery, of being the most distant radio galaxy known.
In addition, as can be seen in \fig{prescubapz}, they are two of the
most-luminous (at 151$\;$MHz) radio sources known.

4C41.17 was detected at both 800 \micron{} and 1300 \micron{}
\cite{dhr94,ck94}, with the contribution from the non-thermal radio
spectrum being less than a few percent at these wavelengths.  Further
support for a thermal emission mechanism came from the submillimetre
spectral index: $\alpha^{800}_{1300}\sim -4$ (for self-absorbed
synchrotron emission, $\alpha\geq -2.5$).

8C1435+635 was first detected at 1250 \micron{} \cite{i95}.  As for
4C41.17, the contribution to the flux density from radio synchrotron
emission is $\sim 1$ per cent.  In addition, the flux density is
roughly the same as that detected for 4C41.17 at 1300 \micron{}.

Given its similarity to 4C41.17, and its tantalizing detection at
1250$\;$\micron{}, 8C1435+635 was a prime target for SCUBA.  When this
project was given its initial allocation of time, 8C1435+635 was the
first object attempted, and we were able to study its submillimetre
SED in unprecedented detail.  Firm detections were made at
850$\;$\micron{} and 450$\;$\micron{}, confirming that the
submillimetre spectrum rises well above the declining radio emission.
Observations at 350$\;$\micron{}, 750$\;$\micron{}, and
175$\;$\micron{} resulted in upper limits to the continuum flux
density; the last observation was made with the {\em Infrared Space
Observatory (ISO)}.  In addition, a sensitive upper limit,
$3\sigma<5\times10^{10}\;{\rm K\;km\,s^{-1}pc^2}$ was obtained for the
CO(4$-3$) line luminosity using the IRAM 30-m telescope.  These data
have been published in Ivison et al. \shortcite{idha98}; the observed flux
densities (or limits) between $\lambda=175$\micron{} and
$\lambda=1250$\micron{} are listed in \tab{previoussubmmobs}.

Isothermal fits to the submillimetre spectrum suggest the majority of
the emission comes from $2\times 10^8\;{\rm M_{\odot}}$ of dust with a
temperature $40\pm 5\;$K and emissivity index $\beta=2$ (assuming 
${\rm H_{\circ}=50\;km\,s^{-1}\,Mpc}$ and $\Omega_{\circ}=1$).  However,
higher temperatures, and correspondingly lower dust masses, are not
ruled out by the data.  The far-infrared luminosity implied by the
preferred fit is ${\rm L_{FIR}\sim 10^{13}L_{\odot}}$.

The amount of gas available for future star formation can be
calculated from the dust mass if a gas-to-dust ratio is assumed.  This
assumption is notoriously uncertain; however, robust upper and lower
limits on the amount of gas present can be calculated if the highest
and lowest reasonable values for the ratio are considered.  For
8C1435+635, such an analysis restricts the gas mass to 
${\rm 4\times10^{10} <M_{gas}< 1.2\times10^{12}\;M_{\odot}}$.  In addition,
if the dust is heated {\em solely} by star formation, the far-infrared
luminosity indicates a star-formation rate of several thousand solar
masses per year.

The dust masses responsible for the detections of 4C41.17 and
8C1435+635 are in excess of $10^8\;{\rm M_{\odot}}$ \cite{dhr94,i95}.
This is at least an order or magnitude higher than what is observed
for radio galaxies in the low-redshift Universe \cite{kp91},
indicating that, at $z=4$, a significant amount of gas is yet to be
converted into stars, suggesting that the host galaxies are not fully
formed \cite{hdr97}.

\subsection{Motivation for a complete sample of radio galaxies}

The distant radio galaxies initially detected at submillimetre
wavelengths are some of the most distant, radio-luminous objects known
(\fig{prescubapz}).  On their own, they cannot be used to determine
the evolutionary status of massive ellipticals at high redshift -
their extreme submillimetre properties could be associated with either
their redshift or the extreme nature of the radio source.  There is a
clear need for a proper sample of radio galaxies spanning a range of
radio luminosities and redshifts.  If the radio luminosity-redshift
plane (the P-$z$ plane) is properly sampled, the effects of radio
luminosity and cosmological evolution can be disentangled.  The goal
of this project was to study such a sample using SCUBA.

The galaxies were chosen based on their radio luminosity and their
redshift.  Before doing so, it was necessary to determine the
frequency at which the radio luminosity should be calculated.

The radio luminosity of a classical double radio source is controlled
by three key physical parameters: the bulk kinetic power in the jets,
the age of the source (measured from the jet-triggering event) as it
is observed on our light cone, and the environmental density
(e.g. Blundell, Rawlings \& Willott 1999\nocite{brw99}).  Provided the
rest-frame frequency is high enough so that it exceeds the synchrotron
self-absorption frequency, and low enough that the radiative timescale
of electrons in the lobes is not so short that the lobes are
completely extinguished, the selection effects inherent in any
flux-limited sample lead to a reasonably close mapping between radio
luminosity and jet power (e.g. Willott et al. 1999\nocite{wrbl99}).  These
considerations render 151 MHz the rest-frame frequency of choice if
one wishes to use radio luminosity to track the underlying jet power.

In any flux-density limited sample, there is a tight correlation
between luminosity and redshift: for large distances, only the
brightest objects will be observable - the faint objects will fall
below the detection limit, and for small distances there are few
bright sources in the small available cosmological volumes.  Suppose
galaxies were selected from a flux-limited radio survey and observed
with SCUBA, with larger submillimetre flux densities measured for the
high-redshift galaxies than for the low-redshift galaxies.  It would
be impossible to determine whether this was genuine cosmological
evolution, or if the higher-redshift galaxies were brighter in the
submillimetre because, for example, they are associated with objects
with more powerful jets, perhaps because they are associated with more
massive black holes and/or dark matter haloes.  This problem can be
overcome by using several radio surveys with different flux density
limits to select our sample. The details of the redshift surveys of
radio sources from which the sample was selected are described in
\sec{actualsample}.  The region of the 151$\;$MHz radio luminosity
versus redshift plane covered by these surveys is shown in
\fig{prescubapz}.

\subsection{The sample}
\label{actualsample}

Initially, a sample of radio galaxies was compiled from the 3C
(Bennett 1962\nocite{bennett1962}; Laing, Riley \& Longair
1983\nocite{lrl83} describe the doubly-revised version of the 3C
sample, 3CRR) and 6CE (Eales 1985; Rawlings, Eales \& Lacy
2000\nocite{eales85,rel00}) radio surveys, for which complete redshift
information exists.  The galaxies were chosen to span a range of
redshifts, $0.2<z<4.5$, but to lie in a narrow band of radio
luminosity, ${\rm
10^{27}\;WHz^{-1}sr^{-1}<P_{151MHz}<10^{29}\;WHz^{-1}sr^{-1}}$.  This
selection was made to minimise any radio luminosity bias and to
determine if the submillimetre emission from radio galaxies truly
evolves with redshift.  Note that the top two decades in radio
luminosity were chosen, allowing galaxies to be studied out to very
high redshift.  If there is a positive correlation between
submillimetre emission and radio luminosity, perhaps via galaxy mass,
this choice should also maximise the chance of making detections.

In order to avoid strong radio synchrotron emission at submillimetre
wavelengths, a further restriction was imposed on the selected
sources: they were required to have steep radio spectra, with a
spectral index $\alpha>0.5$ for $\nu \sim 1\;$GHz.  This selection
criterion was aided by the fact that the radio galaxies were selected
to be very luminous at 151$\;$MHz - the most intrinsically luminous
objects in low-frequency surveys tend to be steep-spectrum sources,
whereas high-frequency surveys are biased towards flat-spectrum
sources.

The 6CE galaxies have a declination limit $\delta < 40^{\circ}$.  In
order to minimise airmass, the criterion $\delta < 40^{\circ}$ was
also imposed on the 3CRR catalogue sources.

As the project progressed, it became apparent that sources selected
from the 3CRR and 6CE samples alone could not produce an even
spread in radio luminosity within the chosen luminosity band.
Additional sources, often with declinations $\delta > 40^{\circ}$,
were added to the sample from:
\begin{enumerate}
\item two flux-density limited radio surveys with complete, or
near-complete, redshift information: the 7C Redshift Survey (7CRS) -
Blundell et al. (2000); Willott et
al. (2000)\nocite{brr00,willott7crs}; and the Leiden-Berkeley Deep
Survey (LBDS) - Waddington et al. (2000) and references
therein\nocite{wadlbds}.
\item several `filtered surveys' for which a filter, spectral index
and/or angular size for example, has been applied to identify high
redshift galaxies in a radio survey.  The names and references for
these filtered surveys are as follows: 4C* - Chambers et
al. (1996)\nocite{cmvb96}; 6C* - Blundell et al. (1998)\nocite{bre98},
Jarvis et al. in preparation; 8C* - Lacy (1992)\nocite{lthesis}; TX*
(Texas) - van Ojik et al. (1996); and MG* (MIT-Green Bank) - Stern et
al. (1999)\nocite{sds99}, Spinrad private communication.
\end{enumerate}

\begin{table*}
\caption{Positions and redshifts of the radio galaxy sample observed with SCUBA.  The parent samples, described in \sec{actualsample}, are listed in Column 5.  Column 6 (Pos. ID) indicates how the position was measured:  `c' - a radio core ID exists, `mc' - a radio ID exists for a marginal radio core, `h' - the radio ID is the mid-point of the hotspot peaks, `CSS' - same as `h' but for an extremely compact radio source, `O' - an optical ID exits, `IR' - an infrared ID exists, `HST' - a Hubble Space Telescope ID exists.  The position and redshift references are given in column 8, with references containing redshift information in bold.  Notes: $^{\S}$ 6C0930+38 is also referred to as 6C0929+38 in the literature.  In the pre-release version of the 6C survey, astrometry indicated an RA of 09h 30m.  However, in the final release of the survey, the position of the radio source appears as 09 29 59.7 +38 55 2.  $^{\#}$ The redshifts of 6C0919+38 and 6C1159+36 are not yet confirmed.  For 6C0919+38, the estimate may not be too far off; for 6C1159+36, the redshift is based on a highly tentative Ly$_{\alpha}$ emission line which is no longer considered secure (Rawlings et al. 2000).  $^{\dag}$ For 3C356 the telescope was pointed at the midpoint of the two possible identifications (BLR97).  $^{\ddag}$ The redshift of MG1744+18 is unpublished.  A reference to a paper which quotes the unpublished value is included.}
\label{posreds}
\begin{tabular}{llllllll}
\hline
Common Name&IAU Name&\multicolumn{1}{c}{RA (B1950.0)}&\multicolumn{1}{c}{Dec (B1950.0)}&Parent&Pos.&\multicolumn{1}{c}{z}&References.\\
                &(B1950.0)      &\multicolumn{1}{c}{(h m s)}&\multicolumn{1}{c}{($^{\circ}$ $'$ $''$)}&Sample&ID&\\
\hline
6C0032+412      &0032+412       &00 32 10.73    &+41 15 00.2    &6C*    &c      &3.66           &BRE98, {\bf Jprep}\\
6C0140+326      &0140+326       &01 40 51.53    &+32 38 45.8    &6C*    &h      &4.41           &BRE98, {\bf RLB96}\\
4C60.07         &0508+604       &05 08 26.12    &+60 27 17.0    &4C*    &mc, O  &3.788          &CMvB96, {\bf RvO97}\\
4C41.17         &0647+415       &06 47 20.57    &+41 34 03.9    &4C*    &c, O   &3.792          &COH94, {\bf RvO97}\\
6C0820+36       &0820+367       &08 20 33.96    &+36 42 28.9    &6CE    &c, IR  &1.86           &ERLG97, LLA95, {\bf REL00}\\
5C7.269         &0825+256       &08 25 39.48    &+25 38 26.5    &7CRS   &IR     &2.218          &{\bf ER96},{\bf W7CRS}, BRR00\\
6C0901+35       &0901+358       &09 01 25.02    &+35 51 01.8    &6CE    &IR     &1.904          &ERLG97, {\bf REW90}, {\bf REL00}\\
6C0902+34       &0902+343       &09 02 24.77    &+34 19 57.8    &6CE    &c, O, IR       &3.395          &COH94, {\bf L88}, {\bf REL00}\\
6C0905+39       &0905+399       &09 05 04.95    &+39 55 34.9    &6CE    &c, IR  &1.882          &{\bf LGE95}, {\bf REL00}\\
3C217           &0905+380       &09 05 41.34    &+38 00 29.9    &3CRR   &mc, HST        &0.8975         &BLR97, Lpc, RS97, {\bf SD84b}\\
6C0919+38       &0919+381       &09 19 07.99    &+38 06 52.5    &6CE    &IR     &1.65$^{\#}$            &ER96, {\bf REL00}\\
6C0930+38$^{\S}$&0930+389       &09 30 00.77    &+38 55 09.1    &6CE    &h      &2.395          &NAR92, {\bf ER96}, {\bf REL00}\\
3C239           &1008+467       &10 08 38.98    &+46 43 08.8    &3CRR   &mc, HST        &1.781          &Lpc, BLR97, {\bf MSvB95}\\
MG1016+058      &1016+058       &10 16 56.82    &+05 49 39.3    &MG*    &c, O, IR       &2.765          &{\bf DSD95}\\
3C241           &1019+222       &10 19 09.38    &+22 14 39.7    &3CRR   &CSS, HST&1.617         &BLR97, {\bf SD84b}\\
8C1039+68       &1039+681       &10 39 07.75    &+68 06 06.6    &8C*    &O      &2.53           &{\bf Lthesis}\\
6C1113+34       &1113+349       &11 13 47.64    &+34 58 46.6    &6CE    &h, IR  &2.406          &LLA95, {\bf REL00}\\
3C257           &1120+057       &11 20 34.55    &+05 46 46.0    &3C     &O      &2.474          &Spriv, {\bf vBS98}\\
3C265           &1142+318       &11 42 52.35    &+31 50 26.6    &3CRR   &c, HST &0.8108         &FBP97, BLR97, {\bf SJ79}\\
3C266           &1143+500       &11 43 04.22    &+50 02 47.4    &3CRR   &mc, HST        &1.272          &BLR97, LPR92, {\bf SD84b}\\
3C267           &1147+130       &11 47 22.07    &+13 03 60.0    &3CRR   &c, HST &1.144          &BLR97, Lpc, {\bf SD84b}\\
6C1159+36       &1159+368       &11 59 20.94    &+36 51 36.2    &6CE    &CSS    &3.2$^{\#}$     &NAR92, {\bf REL00}\\
6C1204+37       &1204+371       &12 04 21.75    &+37 08 20.0    &6CE    &IR     &1.779          &ERLG97, {\bf REL00}\\
6C1232+39       &1232+397       &12 32 39.12    &+39 42 09.4    &6CE    &c, IR  &3.221          &ERD93, NAR92, {\bf REW90}, {\bf REL00}\\
TX1243+036      &1243+036       &12 43 05.40    &+03 39 44.5    &TX*    &c, O   &3.57           &{\bf vOR96}, RMC95, {\bf RvO97}\\
MG1248+11       &1248+113       &12 48 29.11    &+11 20 40.3    &MG*    &O      &2.322          &{\bf SDS99}\\
3C277.2         &1251+159       &12 51 03.85    &+15 58 47.1    &3CRR   &c, HST &0.766          &Lpc, McC97, {\bf SDM85}\\
4C24.28         &1345+245       &13 45 54.66    &+24 30 44.7    &4C*    &O      &2.879          &CMvB96, {\bf RvO97}\\
3C294           &1404+344       &14 04 34.06    &+34 25 40.0    &3CRR   &c      &1.786          &{\bf MS90}\\
8C1435+635      &1435+635       &14 35 27.50    &+63 32 12.8    &8C*    &O      &4.25           &{\bf LMR94, SDG95}\\
3C322           &1533+557       &15 33 46.27    &+55 46 47.4    &3CRR   &mc     &1.681          &LRL, LLA95, {\bf SRS90}\\
3C324           &1547+215       &15 47 37.14    &+21 34 41.0    &3CRR   &c, HST, IR&1.2063      &BCG98, BLR97, {\bf SD84a}\\
3C340           &1627+234       &16 27 29.42    &+23 26 42.2    &3CRR   &c      &0.7754         &Lpc, LRL, {\bf SD84b}\\
53W002          &1712+503       &17 12 59.83    &+50 18 51.3    &LBDS   &O      &2.39           &{\bf WBM91}\\
53W069          &1718+499       &17 18 46.50    &+49 47 47.7    &LBDS   &c, O   &1.432          &Wthesis, {\bf D99}\\
3C356$^{\dag}$\ \ (a)&1732+510  &17 23 06.77    &+51 00 17.8    &3CRR   &c, IR  &1.079          &BLR97, {\bf S82}\\
\hspace{1.152cm}(b)&            &17 23 06.95    &+51 00 14.2    &3CRR   &c, IR  &               &LR94\\
MG1744+18       &1744+183       &17 44 55.30    &+18 22 11.3    &MG*    &IR     &2.28$^{\ddag}$ &Bthesis, {\bf S-unpub}, {\bf ER93}\\
4C13.66         &1759+138       &17 59 21.64    &+13 51 22.8    &3CRR   &IR     &1.45           &{\bf RLL96}\\
3C368           &1802+110       &18 02 45.63    &+11 01 15.8    &3CRR   &c, HST &1.132          &BLR97, DSP87, {\bf S82}\\
4C40.36         &1809+407       &18 09 19.42    &+40 44 38.9    &4C*    &O      &2.265          &CMvB88, {\bf RvO97}\\
4C48.48         &1931+480       &19 31 40.03    &+48 05 07.1    &4C*    &c, O   &2.343          &CMvB96, {\bf RvO97}\\
4C23.56         &2105+233       &21 05 00.96    &+23 19 37.7    &4C*    &c, O   &2.483          &CMvB96, {\bf RvO97}\\
MG2141+192      &2141+192       &21 41 46.95    &+19 15 26.7    &MG*    &h      &3.592          &CRvO97, {\bf Mprep}, \\
3C437           &2145+151       &21 45 01.58    &+15 06 36.1    &3CRR   &HST    &1.48           &BLR97, {\bf MSvB95}\\
MG2305+03       &2305+033       &23 05 52.31    &+03 20 47.9    &MG*    &O      &2.457          &{\bf SDS99}\\
4C28.58         &2349+288       &23 49 26.94    &+28 53 47.2    &4C*    &c, O   &2.891          &CMvB96, {\bf RvO97}\\
3C470           &2356+438       &23 56 02.90    &+43 48 03.6    &3CRR   &c, HST &1.653          &Lpc, BLR97, {\bf MSvB95}\\
\hline
\end{tabular}
\end{table*}

Names, positions, and redshifts for the full sample observed with
SCUBA are given in \tab{posreds}.  In total, 47 radio galaxies were
observed (27 from our original 3C/3CRR/6CE list).  Note, for 3C356 two
unresolved radio `cores' have been detected within 5$''$ of each
other; they each have a galaxy at z=1.079 associated with them.  It is
not clear which galaxy is the host of the radio source; for further
discussion of the merits of the two positions, refer to Best, Longair,
\& R\"{o}ttgering \shortcite{blr97}.  SCUBA was pointed midway between
the two identifications, and will have been offset 2$''$ from the true
position of the host galaxy; the flux lost should be negligible, $<5$
per cent at 850 \micron{}.

\section{Submillimetre observations}
\label{submmobs}

In this section, the results of all the submillimetre observations of
the 47 galaxies in the sample are presented.

\subsection{The SCUBA survey}
\label{radiocontaminatedfluxes}

SCUBA contains two bolometer arrays cooled to $\sim75\;$mK to achieve
sky background-limited performance.  Each array is arranged in a
close-packed hexagon with a $2.3'$ instantaneous field of view.  The
diffraction-limited beamsizes delivered by the JCMT to SCUBA are
$14.7''$ at 850$\;$\micron{} and $7.5''$ at 450$\;$\micron{}; the
bolometer feedhorns are sized for optimal coupling to the beams.
Consequently, the long-wavelength (LW) array, optimised for an
observing wavelength of 850$\;$\micron{}, contains 37 pixels.  The
short-wavelength (SW) array, optimised for operation at
450$\;$\micron{}, contains 91 pixels.  A dichroic beamsplitter allows
observations to be made with both arrays simultaneously.

For point sources, SCUBA's photometry mode is recommended, where the
target is observed with a single pixel.  During commissioning, it was
discovered that the best photometric accuracy could be obtained by
averaging the source signal over a slightly larger area than the beam.
This is achieved by `jiggling' the secondary mirror such that the
chosen bolometer samples a $3\times3$ grid with 2$''$ spacing centered
on the source.  The on-source integration time is 1 second per jiggle
point; excluding overheads, the jiggle pattern takes 9$\;$s to complete.
While making the mini jiggle map, the telescope was chopped at 45$''$ in
azimuth at a frequency of 7Hz.  After the first 9-point jiggle, the
telescope was nodded to the reference position, an event which
occurred every 18 seconds thereafter.

Between June 1997 and March 1999, a total of 192 hours was used to
carry out the observations, all in excellent weather conditions.  The
atmospheric zenith opacities were at the very most 0.3 at
850$\;$micron and 2.5 at 450$\;$micron, although they were frequently
much better than this.  Skydips and observations of calibrators were
obtained regularly.  For each source, the data have been flatfielded,
despiked, and averaged over 18 seconds.  The residual sky background
that chopping and nodding are unable to remove has been subtracted by
using the data taken by the off-source bolometers.  The
Kolmogorov-Smirnov (KS) test has been applied to the data to check for
consistency, and the data have been calibrated using observations of
Mars, Uranus and several secondary calibrators.

The 850$\;$\micron{} and 450$\;$\micron{} flux densities measured for
the radio galaxies in the sample are listed in
\tab{introducingmydata}.  This table includes both $3\sigma$ upper
limits for sources whose signal-to-noise ratio (S/N) does not exceed
3.0 and $2\sigma$ upper limits for sources whose S/N does not exceed
2.0.  The reason for this is that while traditionally the detection of
a given source might only be regarded as robust if it is greater than
$3\sigma$, genuine $2\sigma$ detections should be taken seriously,
especially given that the telescope was pointed at known galaxies
instead of the submillimetre source being identified in a blank-field
survey.  Furthermore, the adoption of a $2\sigma$ detection level is
necessary to implement the statistical analysis presented in
\sec{survsub}.  It should be noted, however, that the results
presented in this paper hold whether a $3\sigma$ or a $2\sigma$
detection threshold is adopted.

\begin{table*}
\caption{Observed submillimetre flux densities (S$_{\nu}$) and standard errors for the sample.  $3\sigma$ upper limits are shown for sources whose S/N does not exceed 3.0; $2\sigma$ upper limits are shown for sources whose S/N does not exceed 2.0}
\label{introducingmydata}
\begin{tabular}{lllrrllrrrrr}
\hline
&&&\multicolumn{4}{c}{850$\;$\micron{}}&&\multicolumn{4}{c}{450$\;$\micron{}}\\
Source&\multicolumn{1}{c}{$z$}&&\multicolumn{1}{c}{S$_{\nu}$}&\multicolumn{1}{c}{S/N}&\multicolumn{1}{c}{$3\sigma$ limit}&\multicolumn{1}{c}{$2\sigma$ limit}&&\multicolumn{1}{c}{S$_{\nu}$}&\multicolumn{1}{c}{S/N}&\multicolumn{1}{c}{$3\sigma$ limit}&\multicolumn{1}{c}{$2\sigma$ limit}\\
&&&\multicolumn{1}{c}{(mJy)}&&\multicolumn{1}{c}{(mJy)}&\multicolumn{1}{c}{(mJy)}&&\multicolumn{1}{c}{(mJy)}&&\multicolumn{1}{c}{(mJy)}&\multicolumn{1}{c}{(mJy)}\\
\hline
3C277.2      &0.766     &&$ 1.01\pm 1.04$ &$  1.0 $&$<  4.13 $&$<  3.09$      &&$ -10.6\pm  10.7$ &$  -1.0 $&$<    32$&$<    21$\\
3C340        &0.7754    &&$ 0.85\pm 0.87$ &$  1.0 $&$<  3.46 $&$<  2.59$      &&$   6.8\pm   9.0$ &$   0.8 $&$<    34$&$<    25$\\
3C265        &0.8108    &&$-1.41\pm 0.98$ &$ -1.4 $&$<  2.94 $&$<  1.96$      &&$  -2.0\pm  11.2$ &$  -0.2 $&$<    34$&$<    22$\\
3C217        &0.8975    &&$ 1.03\pm 0.83$ &$  1.2 $&$<  3.52 $&$<  2.69$      &&$  -0.7\pm   9.8$ &$  -0.1 $&$<    30$&$<    20$\\
3C356        &1.079     &&$ 1.66\pm 1.04$ &$  1.6 $&$<  4.78 $&$<  3.74$      &&$  17.3\pm  19.8$ &$   0.9 $&$<    77$&$<    57$\\
3C368        &1.132     &&$ 4.08\pm 1.08$ &$  3.8 $&$        $&$       $      &&$  39.6\pm  15.1$ &$   2.6 $&$<    85$&$       $\\
3C267        &1.144     &&$ 1.93\pm 0.96$ &$  2.0 $&$<  4.81 $&$       $      &&$  27.0\pm  14.6$ &$   1.8 $&$<    71$&$<    56$\\
3C324        &1.2063    &&$ 1.75\pm 0.87$ &$  2.0 $&$<  4.36 $&$       $      &&$   4.1\pm  11.8$ &$   0.3 $&$<    39$&$<    28$\\
3C266        &1.272     &&$ 0.46\pm 1.30$ &$  0.4 $&$<  4.36 $&$<  3.06$      &&$  -2.1\pm  30.5$ &$  -0.1 $&$<    91$&$<    61$\\
53W069       &1.432     &&$-2.70\pm 1.04$ &$ -2.6 $&$<  3.12 $&$<  2.08$      &&$  14.7\pm  11.5$ &$   1.3 $&$<    49$&$<    38$\\
4C13.66      &1.45      &&$ 3.53\pm 0.96$ &$  3.7 $&$        $&$       $      &&$ -16.2\pm  18.2$ &$  -0.9 $&$<    55$&$<    36$\\
3C437        &1.48      &&$-1.18\pm 0.98$ &$ -1.2 $&$<  2.94 $&$<  1.96$      &&$   2.9\pm  17.3$ &$   0.2 $&$<    55$&$<    37$\\
3C241        &1.617     &&$ 1.81\pm 0.94$ &$  1.9 $&$<  4.63 $&$<  3.69$      &&$  14.8\pm  12.6$ &$   1.2 $&$<    52$&$<    40$\\
6C0919+38    &1.65      &&$-0.88\pm 1.05$ &$ -0.8 $&$<  3.15 $&$<  2.10$      &&$  10.5\pm  10.1$ &$   1.0 $&$<    41$&$<    31$\\
3C470        &1.653     &&$ 5.64\pm 1.08$ &$  5.2 $&$        $&$       $      &&$  57.8\pm  32.9$ &$   1.8 $&$<   156$&$<   124$\\
3C322        &1.681     &&$-0.05\pm 1.06$ &$  0.0 $&$<  3.18 $&$<  2.12$      &&$ -37.5\pm  16.0$ &$  -2.3 $&$<    48$&$<    32$\\
6C1204+37    &1.779      &&$ 0.16\pm 1.25$ &$  0.1 $&$<  3.91 $&$<  2.66$      &&$  45.1\pm  26.6$ &$   1.7 $&$<   125$&$<    98$\\
3C239        &1.781     &&$ 0.83\pm 1.00$ &$  0.8 $&$<  3.83 $&$<  2.83$      &&$  -2.1\pm  18.3$ &$  -0.1 $&$<    55$&$<    37$\\
3C294        &1.786     &&$ 0.19\pm 0.78$ &$  0.2 $&$<  2.53 $&$<  1.75$      &&$   5.4\pm  13.4$ &$   0.4 $&$<    45$&$<    32$\\
6C0820+36    &1.86      &&$ 2.07\pm 0.96$ &$  2.2 $&$<  4.95 $&$       $      &&$  13.6\pm  18.0$ &$   0.8 $&$<    68$&$<    50$\\
6C0905+39    &1.882     &&$ 3.62\pm 0.89$ &$  4.1 $&$        $&$       $      &&$  31.2\pm  16.2$ &$   1.9 $&$<    80$&$<    64$\\
6C0901+35    &1.904      &&$-1.83\pm 1.15$ &$ -1.6 $&$<  3.45 $&$<  2.30$      &&$ -19.3\pm   8.2$ &$  -2.4 $&$<    25$&$<    16$\\
5C7.269      &2.218     &&$ 1.68\pm 1.00$ &$  1.7 $&$<  4.68 $&$<  3.68$      &&$   5.1\pm   9.2$ &$   0.6 $&$<    33$&$<    23$\\
4C40.36      &2.265     &&$ 0.67\pm 1.06$ &$  0.6 $&$<  3.85 $&$<  2.79$      &&$   7.3\pm  23.2$ &$   0.3 $&$<    77$&$<    54$\\
MG1744+18    &2.28      &&$ 0.83\pm 1.02$ &$  0.8 $&$<  3.89 $&$<  2.87$      &&$  13.2\pm  17.7$ &$   0.7 $&$<    66$&$<    49$\\
MG1248+11    &2.322     &&$ 1.10\pm 1.06$ &$  1.0 $&$<  4.28 $&$<  3.22$      &&$ -20.9\pm  12.9$ &$  -1.6 $&$<    39$&$<    26$\\
4C48.48      &2.343     &&$ 5.05\pm 1.05$ &$  4.8 $&$        $&$       $      &&$  17.9\pm  24.9$ &$   0.7 $&$<    93$&$<    68$\\
53W002       &2.39      &&$ 1.03\pm 1.10$ &$  0.9 $&$<  4.33 $&$<  3.23$      &&$   2.9\pm  14.3$ &$   0.2 $&$<    46$&$<    32$\\
6C0930+38    &2.395     &&$ 0.36\pm 1.00$ &$  0.4 $&$<  3.36 $&$<  2.36$      &&$   8.3\pm  10.8$ &$   0.8 $&$<    41$&$<    30$\\
6C1113+34    &2.406       &&$ 0.33\pm 1.14$ &$  0.3 $&$<  3.75 $&$<  2.61$      &&$ -11.4\pm  16.8$ &$  -0.7 $&$<    50$&$<    34$\\
MG2305+03    &2.457     &&$ 2.31\pm 0.99$ &$  2.3 $&$<  5.28 $&$       $      &&$  -9.5\pm  41.8$ &$  -0.2 $&$<   126$&$<    84$\\
3C257        &2.474     &&$ 5.40\pm 0.95$ &$  5.7 $&$        $&$       $      &&$  17.8\pm  15.4$ &$   1.2 $&$<    64$&$<    49$\\
4C23.56      &2.483     &&$ 1.72\pm 0.98$ &$  1.8 $&$<  4.66 $&$<  3.68$      &&$  -3.3\pm  17.0$ &$  -0.2 $&$<    51$&$<    34$\\
8C1039+68    &2.53      &&$ 0.38\pm 0.98$ &$  0.4 $&$<  3.32 $&$<  2.34$      &&$  31.8\pm  17.0$ &$   1.9 $&$<    83$&$<    66$\\
MG1016+058   &2.765     &&$ 2.40\pm 0.92$ &$  2.6 $&$<  5.16 $&$       $      &&$  28.7\pm  10.1$ &$   2.8 $&$<    59$&$       $\\
4C24.28      &2.879     &&$ 2.59\pm 1.16$ &$  2.2 $&$<  6.07 $&$       $      &&$  11.2\pm  18.7$ &$   0.6 $&$<    67$&$<    48$\\
4C28.58      &2.891     &&$ 3.93\pm 0.95$ &$  4.1 $&$        $&$       $      &&$  23.0\pm  15.5$ &$   1.5 $&$<    69$&$<    54$\\
6C1232+39    &3.221      &&$ 3.86\pm 0.72$ &$  5.4 $&$        $&$       $      &&$  -2.4\pm   6.1$ &$  -0.4 $&$<    18$&$<    12$\\
6C1159+36    &3.2     &&$ 1.20\pm 1.08$ &$  1.1 $&$<  4.44 $&$<  3.36$      &&$  28.3\pm  16.0$ &$   1.8 $&$<    76$&$<    60$\\
6C0902+34    &3.395     &&$ 2.83\pm 1.00$ &$  2.8 $&$<  5.83 $&$       $      &&$  11.6\pm  11.5$ &$   1.0 $&$<    46$&$<    35$\\
TX1243+036   &3.57      &&$ 2.28\pm 1.11$ &$  2.1 $&$<  5.61 $&$       $      &&$   1.5\pm  19.0$ &$   0.1 $&$<    59$&$<    40$\\
MG2141+192   &3.592     &&$ 4.61\pm 0.96$ &$  4.8 $&$        $&$       $      &&$  21.4\pm  18.9$ &$   1.1 $&$<    78$&$<    59$\\
6C0032+412   &3.66      &&$ 2.64\pm 1.20$ &$  2.2 $&$<  6.24 $&$       $      &&$  -5.5\pm  13.3$ &$  -0.4 $&$<    40$&$<    27$\\
4C60.07      &3.788     &&$17.11\pm 1.33$ &$ 12.9 $&$        $&$       $      &&$  69.0\pm  23.0$ &$   3.0 $&$       $&$        $\\
4C41.17      &3.792       &&$12.10\pm 0.88$ &$ 13.8 $&$        $&$       $      &&$  22.5\pm   8.5$ &$   2.7 $&$<    48$&$        $\\
8C1435+635   &4.25      &&$ 7.77\pm 0.76$ &$ 10.2 $&$        $&$       $      &&$  23.6\pm   6.4$ &$   3.7 $&$       $&$        $\\
6C0140+326   &4.41      &&$ 3.33\pm 1.49$ &$  2.2 $&$<  7.80 $&$       $      &&$ -20.5\pm  17.1$ &$  -1.2 $&$<    51$&$<    34$\\
\hline
\end{tabular}
\end{table*}

\subsection{Radio galaxies with other millimetre/submillimetre observations}

Several attempts were made to observe the millimetre/submillimetre
spectrum of radio galaxies before SCUBA existed.  The attempts, made
with the IRAM 30-m telescope and UKT14 (SCUBA's predecessor), were
largely unsuccessful.  More recently, Best et al. \shortcite{brb98} also looked
at 3C324 with SCUBA and Benford et al. \shortcite{bco99} observed 4C41.17
at 350$\;$\micron{} with the CSO.  All of these observations,
including the full submillimetre spectrum of 8C1435+635, are listed in
\tab{previoussubmmobs}.

\begin{table*}
\caption{All other millimetre and submillimetre observations of
galaxies in the sample.  All flux densities are quoted in mJy.
The references are: 3C324 - BRB98; 53W002 - HDR97; 3C257 - HDR97; MG1016+058 - CFR98; 6C1232+39 - C\&K94; 6C0902+34 - C\&K94, DSS96, HDR97; TX1243+036 - CFR98; MG2141+192 -  HDR97; 6C0032+412 -  HDR97; 4C41.17 - BCO99, C\&K94, DHR94; 8C1435+635 - HDR97, I95, IDHA98}
\label{previoussubmmobs}
\begin{tabular}{lllllllll}
\hline
Source          &S$_{3000\mu m}$&S$_{1300\mu m}$  &S$_{850\mu m}$  &S$_{800\mu m}$   &S$_{750\mu m}$   &S$_{450\mu m}$  &S$_{350\mu m}$ &S$_{175\mu m}$\\
\hline
3C324           &               &                 &$3.65\pm1.17$   &                 &                 &$3\sigma<21$    &               &\\
53W002          &               &                 &                &$6.9\pm2.3$      &                 &                &               &\\
3C257           &               &                 &                &$3\sigma<11$     &                 &                &               &\\
MG1016+058      &               &$2.13\pm0.47$    &                &$14.7\pm4.6$     &                 &                &               &\\
6C1232+39       &               &$3\sigma<3.0$    &                &                 &                 &                &               &\\
6C0902+34       &$4.2\pm0.6$    &$3.1\pm0.6$      &                &$3\sigma<14$     &                 &                &               &\\
TX1243+036      &               &$3\sigma<2.6$    &                &$3\sigma<9.3$    &                 &                &               &\\
MG2141+192      &               &                 &                &$3\sigma<11$     &                 &                &               &\\
6C0032+412      &               &                 &                &$3\sigma<14$     &                 &                &               &\\
4C41.17         &               &$2.5\pm0.4$      &                &$17.4\pm3.1$     &                 &$3\sigma<56$    &$37\pm9$       &\\
8C1435+643      &               &$2.57\pm0.42$    &$7.77\pm0.76$   &$3\sigma<13$     &$8.74\pm3.31$    &$23.6 \pm 6.4$  &$3\sigma<87.0$ &$3\sigma<40.1$\\
\hline
\end{tabular}
\end{table*}

In most cases, these data are consistent with what we observed with
SCUBA (for 3C324, although we find a lower flux density than Best et
al. \shortcite{brb98}, the two results are equivalent within errors).
There are two exceptions:
\begin{enumerate}
\item 53W002 - Even though 53W002 was detected at 800-\micron{} with
UKT14 \cite{hdr97}, we have failed to detect it
with SCUBA.  Assuming a typical submillimetre spectral index of $-4.0$,
the 800-\micron{} detection of 53W002 implies a flux density of at
least 3.6$\;$mJy at 850-\micron{}.  Note the spectral index is
expected to lie between the values of $-3.0$ and $-4.0$ (e.g. Hildebrand
1983\nocite{hildebrand}), and adopting a value of $-4.0$ will give the
lowest estimate of 850-\micron{} flux.  Thus, if the UKT14 data are
trustworthy, we should have detected 53W002 with SCUBA, even if the
large error bar on the 800-\micron{} measurement is taken into
account.
\item MG1016+058 - For MG1016+058, previous detections existed at both
1300$\;$\micron{} and 800$\;$\micron{} \cite{cfr98}.  The minimum
submillimetre spectral index consistent with these data points and
their error bars is $\sim-2.8$.  This can be used to estimate a lower
limit on the flux density expected at 850$\;$\micron{}.  The
850-\micron{} flux density should be well in excess of 5$\;$mJy,
easily detectable with SCUBA.  An obvious objection to this analysis
is that the 1300-\micron{} flux density could be solely due to radio
synchrotron emission.  This is unlikely; the radio spectrum is highly
curved and dives down well below the 1300-\micron{} flux density
(\appen{sedfig1}).  However, assume for a moment that the
1300-\micron{} flux density was contaminated by radio emission.
Again, adopting a typical submillimetre spectral index of -4.0 and
applying it to the 800$\;$\micron{} detection still predicts an
850-\micron{} flux density in excess of 5$\;$mJy.
\end{enumerate}

These two discrepancies are perhaps not all that surprising.  UKT14
was a single-element bolometer, relying solely on chopping and nodding
for sky removal.  However, a chop/nod set-up is unable to account for
quick variations in the background sky, and there will be residual sky
noise present in the data (Jenness, Lightfoot \& Holland
1998\nocite{remskypaper}).  SCUBA is an array, and the off-source
bolometers can be used to remove this residual sky noise
\cite{remskypaper}.  Omont et al. \shortcite{omc96} investigated the
difference between single- and multi-bolometer detectors, and found
that the less-reliable sky-cancellation offered by single-bolometer
systems often resulted in fake detections of weak sources.
Furthermore, they failed to detect three high-redshift quasars with a
multi-bolometer array that Andreani, La Franca \& Cristiani
\shortcite{alc93} had apparently detected with a single bolometer.

\section{Correcting radio galaxy dust emission for potential contamination}
\label{synchcorrect}

For the majority of galaxies in the sample which we have actually
detected, a detection has only been obtained at one submillimetre
wavelength (i.e. 850$\;$\micron{}).  Without a reliable estimate of
the submillimetre spectral index, care needs to be taken to ensure the
detected submillimetre emission is clearly in excess of the radio
synchrotron spectrum, and can thus be safely attributed to dust.
However, as will become clear, due to the fact that the sample was
confined to steep-spectrum lobe-dominated radio galaxies at $z>1$, for
almost all objects, detectability with SCUBA at 850$\;$\micron{}
corresponds to a successful detection of dust.  Radio spectra have
been compiled for each source with the aim of extrapolating the radio
emission to submillimetre wavelengths and subtracting it off.  This
correction should give a good estimate of the flux density produced by
thermal dust emission.

It might appear that such a correction would be excessive, or even
completely unnecessary where it is known that the radio lobes lie
outside the SCUBA beam.  However, at high frequencies, $\geq8\;$GHz,
experience with individual well-studied objects (e.g. 4C41.17)
indicates that much of the emission can arise from structures
coincident or close to the radio core (e.g. Carilli, Owen \& Harris
1994\nocite{coh94}).  Therefore we have adopted the conservative but
consistent approach of estimating submillimetre synchrotron
contamination from the extrapolation of the total high-frequency radio
emission in every case.  In practice, for radio sources larger than
the JCMT beam, this correction makes a significant impact on the
estimated dust mass in only a very small number of sources.  These are
discussed at the end of \sec{contam}.

The radio spectra are shown in \appen{sedfig1}, and include both integrated
and core radio flux densities.  Core flux densities are generally much
fainter than the integrated radio emission at low frequencies, but
they can also be much flatter.  If the core has a flat spectral index,
it may contribute significantly to the submillimetre flux density even
if the integrated radio emission seems to fall well below this.

Fits to the radio spectra are also shown in \appen{sedfig1}.  These
fits are described in \sec{contam} (with full details in
\appen{sedfig1}) and are subsequently used to estimate the non-thermal
contamination at submillimetre wavelengths.  Details on how the
individual radio spectra were compiled are given in Archibald
\shortcite{mythesis}.

Note the radio spectrum of 53W069 has not been presented here.  53W069
is only $3.73\;$mJy at 1.4$\;$GHz, and synchrotron contamination
will not be a problem \cite{wvhk84}.

\subsection{Synchrotron contamination}
\label{contam}

A simple synchrotron spectrum is a power law of the form
$S_{\nu}\propto \nu^{-\alpha}$.  However, real radio spectra turn over
at low frequencies owing to synchrotron self-absorption, or through
running out of electrons at low energies.  At high frequencies the
spectra steepen as they age - the highest-energy electrons radiate
away their energy the fastest.  In log-space, such spectra appear
curved.

Consider, as an example, the radio spectrum of the hotspots of
Cygnus-A, described in detail by Muxlow, Pelletier \& Roland
\shortcite{mpr88}.  The spectrum is a power law at low frequencies
$0.4\;{\rm GHz} < \nu < 1.5\; {\rm GHz}$, with a spectral index
$\alpha\sim 0.5$, and a low-frequency turn-over at $\nu \sim 0.2
\;{\rm GHz}$.  At $2 {\rm GHz}$ the spectrum steepens; for $\nu > 2
\;{\rm GHz}$, the spectrum is consistent with a power law of index
$\alpha\sim 1.0$.  Robson et al. \shortcite{rlsh98} and Eales,
Alexander \& Duncan \shortcite{ead89} have found that this steep power
law is an accurate description of the spectrum right out to
submillimetre wavelengths.  Thus in log-space, at high frequencies a
straight-line fit would be accurate.  However, when fitting to the
whole spectrum, a model with some curvature would be required.

Another good example is the study of the dominant hotspot of 3C273, by
Meisenheimer \& Heavens \shortcite{mh86}.  For low frequencies, $\nu <
1.5\; {\rm GHz}$, the spectral index is $\alpha\sim 0.7$.  For $1.5\;
{\rm GHz} < \nu < 5\; {\rm GHz}$, the spectrum steepens to an index of
$\alpha\sim 0.9$.  At high frequencies $5\; {\rm GHz} < \nu < 15\;
{\rm GHz}$, the spectrum steepens even further to an index of
$\alpha\sim 1.0$.

More recently, Murgia et al. \shortcite{mff99} published a study of
compact steep-spectrum radio sources.  The majority of objects in
their sample exhibit a clear spectral steepening, occurring anywhere
from a few hundred MHz to tens of GHz.

In order to successfully correct for synchrotron contamination at
submillimetre wavelengths, the high-frequency radio spectrum needs to
be well determined.  Unfortunately, for most galaxies in the sample,
the highest-frequency observation is $\nu \sim 20 \; {\rm GHz}$.  A
straight-line extrapolation (in log-space) from the highest frequency
data point could be used to decontaminate the submillimetre flux densities.
However, in the interval between $20 \; {\rm GHz}$ and the SCUBA
wavebands, the spectrum may steepen, and the straight-line
extrapolation would over-correct the SCUBA flux density.

Evidence to support this steepening comes from 3C324, MG1744+18, and
MG1016+058 - the only sources in the sample with measured radio flux
densities at $\nu > 20 \; {\rm GHz}$.  In each case the high-frequency
observation indicates a significant steepening of the radio spectrum.
4C23.56 and 8C1435+635, two of the higher-redshift galaxies in the
sample, even appear significantly curved for $\nu < 20 \; {\rm GHz}$.

Note that if a straight-line extrapolation of the high-frequency radio
emission is correct, three of 850-\micron{} detections made with SCUBA
appear to be seriously contaminated with non-thermal synchrotron -
4C13.66, 3C470, and 3C257.

More high-frequency radio observations are needed to be certain what
the radio spectrum of each galaxy in the sample does as it approaches
the submillimetre waveband.  For now, both possibilities will be
considered: a {\em linear fit}, a straight-line extrapolation of the
high-frequency data, as could be accurately applied to Cygnus-A; and a
{\em parabolic fit} which reflects a degree of steepening, or
curvature, at higher frequencies.  The details of the two fits,
together with plots of the SEDs, are given in \appen{sedfig1}.

It would be surprising if the radio spectra flattened off at high
frequencies, as this only occurs if the source has an extremely
bright, flat radio core (except for 6C0902+34, the radio galaxies
discussed here all have relatively faint cores - \sec{cores}).  Thus,
the linear fit is a strong upper limit on the synchrotron emission at
submillimetre wavelengths, especially since a significant fraction of
the emission may lie outside the JCMT beam.  Deviations from this
standard power-law take the form of a steepening at high frequencies -
the parabolic fit mimics this curvature and is a good lower limit on
the amount of synchrotron emission at submillimetre wavelengths.
Given the lack of high-frequency radio observations, it is often
difficult to determine which is the better fit.  The best estimate of
the synchrotron contamination has been taken to be the midpoint of
these two models, with the error in the estimate being the difference
between the midpoint and either the upper or lower limit.

These estimates of synchrotron emission were thus subtracted from the
measured SCUBA flux densities, with the errors being added in
quadrature (as prescribed by the propagation of errors).  However,
there were four special cases for which particular care had to be
taken:
\begin{enumerate}
\item 3C324, MG1744+18, MG1016+058, and 4C23.56 all have
high-frequency data that the linear fit has trouble fitting to.  For
these galaxies, the parabolic fit looks to be the more realistic
extrapolation of the radio spectra, and was used to decontaminate the
SCUBA flux densities.
\item For the 14 radio sources whose lobes lie outside the JCMT beam,
we may have over-corrected for synchrotron contamination.  However,
for all of these but one, the source was either undetected by SCUBA,
or the applied correction was insignificantly small.  For 3C470, the
applied correction turned a detection into a non-detection, reducing
the 850-\micron{} flux from 5.6$\;$mJy to 3.1$\;$mJy.  However, as
mentioned previously, our method of correction may be reasonable for
sources that lie outside the JCMT beam.  Furthermore, if the
correction is not applied to 3C470, the results presented in the paper
are unaffected.
\item For some galaxies SCUBA measured a negative signal.  They
have not been corrected for synchrotron contamination; performing the
correction would have unnecessarily increased the noise of the
measurement and inappropriately made the signal even more negative.
\item For some of the undetected galaxies, the estimated synchrotron
contamination is larger than the signal measured by SCUBA, resulting
in a negative submillimetre flux density if it is corrected for.
Claiming a `negative emission' from dust is not physically
meaningful.  It is better to set the SCUBA flux densities equal to zero: it
appears that there is no thermal emission from dust at these
wavelengths.  This leaves the question of how to handle the errors.
It seems best to leave them as they are - as they have been measured
by SCUBA.  The statement then being made is that, as far as it can be
ascertained, all of the measured signal may be radio synchrotron
emission, and any dust emission from the source is definitely less
than $2\times$ the measured noise (for a $2\sigma$ upper limit).  This
seems a good, conservative upper limit - in each case the limit is
larger than the original measured flux density.
\end{enumerate}

\subsection{Radio cores}
\label{cores}

When searching for thermal dust emission, the telescope is pointed at
the optical/infrared ID of the galaxy.  This position is coincident
with the radio core (if one has been detected).  A flat radio core
could contribute at submillimetre wavelengths.  A useful exercise is
to estimate the upper limit on this contribution for each source -
if it tends to be negligible, the SCUBA flux densities will not need
to be corrected.

Only 8 of the galaxies in the sample have their radio core detected at
more than one frequency: 3C267, 3C324, 3C241, 4C23.56, 6C1232+39,
6C0902+34, TX1243+036, and 4C41.17.  6C0902+34 is a special case that
is excluded from the following discussion; it will be treated
separately at the end of this section.  For the remaining 7 galaxies,
the radio core spectral index has been estimated using the two
highest-frequency core detections.  3C324 and 3C241 have relatively
flat core spectra, with $\alpha \sim 0.35$.  The majority, however,
have very steep spectra, ranging from $\alpha \sim 0.7$ to $\alpha
\sim 1.8$.  Note that no detected core appears either completely flat
or inverted (i.e. flux density {\em increasing} with frequency)
between 1$\;$GHz and 15$\;$GHz.  If these cores are extrapolated to
submillimetre wavelengths, in all cases, save one, the core
contribution is negligible: $< 0.05{\rm mJy}$ at 850$\;$\micron{}.
For 3C241, on the other hand, the core strength at 850$\;$\micron{} is
likely to be $\sim 1{\rm mJy}$.

For several galaxies in the sample, the radio core has been detected
at a single frequency.  Adopting the flattest measured spectral index
of $\alpha \sim 0.35$, the corresponding core flux densities at
850$\;$\micron{} have been estimated.  This is a pessimistic estimate,
many of these galaxies may have steeper core spectra, but it is a good
indication of the worst-case scenario.  In all cases but three, the
predicted core flux densities are negligible, $< 0.2{\rm mJy}$.
Higher 850$\;$\micron{} core flux densities are expected for 3C322 and
3C265.  However, for these galaxies a negative signal was measured
with SCUBA, and the core brightness is irrelevant.  For MG1016+058, if
the spectrum is flat, a core flux density of $\sim 0.8{\rm mJy}$ is
expected at 850$\;$\micron{}.  If, on the other hand, the spectral index
is steeper, $\alpha \sim 0.75$ for example, the core contamination is
negligible.

A lack of sources with core detections at more than one frequency
potentially hinders this analysis.  However, given the available
information it seems unlikely that the cores are significant except
for 3C241 and MG1016+058.  The core contribution in the case of 3C241
has been corrected for, as the core spectrum is definitely flat.  For
MG1016+058, on the other hand, it is impossible to know whether the
core is flat or steep.  Given this uncertainty, the SCUBA observations
have not been corrected for core contamination.  Instead, it is simply
noted that MG1016+058 may have a large core contribution at
submillimetre wavelengths.

It is mentioned in the literature that three galaxies have completely
flat (or possibly inverted) cores at $\nu < 5\;{\rm GHz}$: 3C294
(MS90\nocite{ms90}), 6C0905+39 (LGE95\nocite{lge95}), and 6C0032+412
(BRE98\nocite{bre98}; Jarvis et al., in preparation).  These cores
could contribute significantly to the observed SCUBA flux densities.
However, Rudnick, Jones \& Fiedler \shortcite{rjf86} have studied flat
cores in powerful radio galaxies.  They found the cores often turned
over or steepened at $\nu > 5\;{\rm GHz}$ - a completely flat core at
low frequencies does not mean that it will remain flat out to the
submillimetre wavebands, indeed all evidence suggests it is very
unlikely to do so.

In addition, note that for 3C356, two radio core identifications
exist: one faint and steep, the other brighter and flat.  Neither
should significantly contribute at submillimetre wavelengths.

6C0919+38 could have a very bright radio core, $\sim6\;$mJy at
5$\;$GHz.  However, it is not clear whether the bright feature
detected by Naundorf et al. \shortcite{nar92} is the radio core or a
knot in the radio jet.  Given the uncertainty, the possible radio core
contamination is merely noted here and is not corrected for.

\subsubsection{6C0902+34}
\label{6c0902}

6C0902+34 is a special case because of the brightness of its core,
$\sim 10 {\rm mJy}$, between 1.5 and 15$\;$GHz.  In addition, two
slightly conflicting observations at 8$\;$GHz make it difficult to be
certain how steep the core spectrum is.  If the flatter spectral index
is used ($\alpha\sim 0.3$), the 3$\;$mm and 1.3$\;$mm detections of
Downes et al. \shortcite{dss96} and Chini \& Kr\"ugel \shortcite{ck94}
look to be completely dominated by the radio core.  Downes et
al. \shortcite{dss96} came to the same conclusion, and predict the
contribution of dust emission at 1.3$\;$mm to be $<0.6\;$mJy.  Even if
the steeper spectral index is used ($\alpha\sim 1.0$), the detections
could easily be dominated by the integrated radio spectrum (although
higher-frequency radio detections are needed to confirm this).

When observed with SCUBA at 850$\;$\micron{}, a $3\sigma$ upper limit of
5.83$\;$mJy was measured.  Combining this with Chini \& Kr\"ugel's
detection of $3.1\pm0.6\;{\rm mJy}$ at 1300$\;$\micron{}, the
submillimetre spectral index is $|\alpha| \leq 2.0$.  This is
inconsistent with thermal dust emission being present, for which
$|\alpha| > 2.0$.  (Note that the submillimetre spectral indices are
negative).

6C0902+34 appears completely dominated by radio synchrotron emission.
It has thus been left out of the sample.  Hereafter, we discuss a
sample of 46 radio galaxies which excludes 6C0902+34.

\subsection{Submillimetre confusion}

\subsubsection{Galactic cirrus confusion}

Extrapolating the galactic cirrus confusion that {\em IRAS} measured
at 100$\;$\micron{} to 850$\;$\micron{} at SCUBA resolution yields a
cirrus confusion noise of $<<1\;{\rm mJy\,beam^{-1}}$ in the direction of
our targets.  Thus, contamination by galactic cirrus should not be a problem for this study (see Hughes et al. 1997 \nocite{hdr97} and
Helou \& Beichman 1990 \nocite{hb90} for detailed discussions).

\subsubsection{Confusion by extragalactic sources}

Hughes et al. \shortcite{hdfnature} performed the deepest submillimetre survey
when they mapped the Hubble Deep Field with SCUBA.  For 850$\;$\micron{},
confusion was found to be a problem for sources weaker than 2$\;$mJy.  At the
2$\;$mJy level, the source density of their map is $\sim 1$ source per 30
beams.  In a large {\em blank-field} survey reaching 2$\;$mJy, some detections
will inevitably result from confusion, particularly if sources are strongly
clustered.  However, the chances that such false confusion-induced detections
should coincide in position with known massive objects at high redshift is
clearly extremely small.  Of course, given sufficient resolution our detected
sources may break up into sub-components, as for example has the optical image
of 4C41.17 under HST resolution \cite{mcvb92}.  However, since such
sub-components are still extremely likely to be associated with the radio
galaxy itself, this cannot be regarded as a problem.  See also Blain et al.
\shortcite{blainconfusion} for an in-depth discussion of source confusion at
SCUBA wavelengths.

\subsubsection{Confusion by sources in the off-beam}

The measured flux densities could be affected by chopping onto a nearby
source.  This is unlikely to be an important problem for this study, even
though SCUBA maps of the fields of radio galaxies have revealed an
over-density of sources in some cases (Ivison et al. 2000a\nocite{4cmap};
Ivison et al., in preparation\nocite{rgmapping}).  For the galaxies in our
sample which have been mapped and shown to have companion sources (4C41.17 and
8C1435+635), our chop did not land on the companions.  Furthermore, our
observations were made with an azimuthal chop, and the off-beam will have
moved across a substantial angle (several degrees) of the sky as we tracked a
source, diluting the effect of off-beam contamination.  There is also no {\em
  a priori} reason why this possible source of confusion would result in a
redshift bias and affect the analysis presented in \sec{evolstats}.

\subsection{A final note on corrections: 6C0140+326}

As a final note on correcting the SCUBA flux densities, it appears
that 6C0140+326 is gravitationally lensed \cite{rlb96}.  However, the
predicted amplification factor is small ($<2$).  Given the
uncertainties in estimating this factor and the synchrotron
contamination, a correction has not been applied.

\subsection{Corrected flux densities}

The final 850-\micron{} flux densities, corrected for synchrotron
contamination, and in the case of 3C241 for radio core contamination,
are presented in \tab{decontaminateddata}.  Note, we do not include
the corrected 450-\micron{} flux densities; at 450$\;$\micron{} the
corrections are tiny compared with the errors.  As mentioned
earlier, we adopt a detection threshold of S/N$\ge2.0$, with upper
limits taken at the $2\sigma$ level.  Out of a sample of 46 radio
galaxies (excluding 6C0902+34), thermal emission from dust has been
detected in 14 galaxies at 850$\;$\micron{}, and in 5 of these
galaxies at 450$\;$\micron{}.

\begin{table}
\centering
\caption{850-\micron{} flux densities (S$_{\nu}$) and errors corrected
for radio synchrotron contamination.  In the case of 3C241, radio core
contamination has also been taken into account.  A `\ding{51}' in column 2
indicates that applying the correction has changed the flux density by
less than 1$\sigma$.  $2\sigma$ upper limits are shown for sources
whose S/N does not exceed 2.0.}
\label{decontaminateddata}
\begin{tabular}{lcrrl}
\hline
&&\multicolumn{3}{c}{850$\;$\micron{}}\\
Source&$<1\sigma$&\multicolumn{1}{c}{S$_{\nu}$}&\multicolumn{1}{c}{S/N}&\multicolumn{1}{c}{$2\sigma$ limit}\\
&&\multicolumn{1}{c}{(mJy)}&&\multicolumn{1}{c}{(mJy)}\\
\hline
3C277.2      &\ding{51} &$ 0.00\pm 1.04$ & 0.00 &$<  2.08$\\
3C340        &\ding{51} &$ 0.00\pm 0.87$ & 0.00 &$<  1.74$\\
3C265        &\ding{51} &$-1.41\pm 0.98$ &-1.44 &$<  1.96$\\
3C217        &          &$ 0.00\pm 0.83$ & 0.00 &$<  1.66$\\
3C356        &          &$ 0.08\pm 1.25$ & 0.06 &$<  2.58$\\
3C368        &\ding{51} &$ 3.70\pm 1.11$ & 3.32 &$       $\\
3C267        &          &$ 0.00\pm 0.96$ & 0.00 &$<  1.92$\\
3C324        &          &$ 0.80\pm 0.89$ & 0.90 &$<  2.57$\\
3C266        &\ding{51} &$ 0.00\pm 1.30$ & 0.00 &$<  2.60$\\
53W069       &\ding{51} &$-2.70\pm 1.04$ &-2.60 &$<  2.08$\\
4C13.66      &\ding{51} &$ 2.62\pm 1.24$ & 2.11 &$       $\\
3C437        &\ding{51} &$-1.18\pm 0.98$ &-1.20 &$<  1.96$\\
3C241        &          &$ 0.01\pm 1.15$ & 0.00 &$<  2.30$\\
6C0919+38    &\ding{51} &$-0.88\pm 1.05$ &-0.84 &$<  2.10$\\
3C470        &          &$ 3.07\pm 1.56$ & 1.96 &$<  6.20$\\
3C322        &\ding{51} &$-0.05\pm 1.06$ &-0.05 &$<  2.12$\\
6C1204+37    &\ding{51} &$ 0.00\pm 1.25$ & 0.00 &$<  2.50$\\
3C239        &\ding{51} &$ 0.00\pm 1.00$ & 0.00 &$<  2.00$\\
3C294        &\ding{51} &$ 0.00\pm 0.78$ & 0.00 &$<  1.56$\\
6C0820+36    &\ding{51} &$ 1.80\pm 0.98$ & 1.83 &$<  3.76$\\
6C0905+39    &\ding{51} &$ 3.52\pm 0.89$ & 3.95 &$       $\\
6C0901+35    &\ding{51} &$-1.83\pm 1.15$ &-1.59 &$<  2.30$\\
5C7.269      &\ding{51} &$ 1.41\pm 1.00$ & 1.41 &$<  3.41$\\
4C40.36      &\ding{51} &$ 0.63\pm 1.06$ & 0.60 &$<  2.75$\\
MG1744+18    &\ding{51} &$ 0.02\pm 1.02$ & 0.02 &$<  2.06$\\
MG1248+11    &\ding{51} &$ 0.96\pm 1.07$ & 0.90 &$<  3.10$\\
4C48.48      &\ding{51} &$ 4.72\pm 1.06$ & 4.44 &$       $\\
53W002       &\ding{51} &$ 0.99\pm 1.10$ & 0.90 &$<  3.19$\\
6C0930+38    &\ding{51} &$ 0.00\pm 1.00$ & 0.00 &$<  2.00$\\
6C1113+34    &\ding{51} &$ 0.00\pm 1.14$ & 0.00 &$<  2.28$\\
MG2305+03    &          &$ 0.08\pm 1.02$ & 0.08 &$<  2.13$\\
3C257        &          &$ 1.62\pm 2.25$ & 0.72 &$<  6.12$\\
4C23.56      &\ding{51} &$ 1.68\pm 0.98$ & 1.72 &$<  3.64$\\
8C1039+68    &\ding{51} &$ 0.02\pm 0.99$ & 0.02 &$<  2.01$\\
MG1016+058   &\ding{51} &$ 2.31\pm 0.92$ & 2.51 &$       $\\
4C24.28      &\ding{51} &$ 2.35\pm 1.17$ & 2.01 &$       $\\
4C28.58      &\ding{51} &$ 3.89\pm 0.95$ & 4.10 &$       $\\
6C1232+39    &\ding{51} &$ 3.84\pm 0.72$ & 5.34 &$       $\\
6C1159+36    &\ding{51} &$ 0.79\pm 1.15$ & 0.68 &$<  3.09$\\
TX1243+036   &\ding{51} &$ 2.14\pm 1.11$ & 1.92 &$<  4.36$\\
MG2141+192   &\ding{51} &$ 4.57\pm 0.96$ & 4.75 &$       $\\
6C0032+412   &\ding{51} &$ 2.40\pm 1.20$ & 2.00 &$       $\\
4C60.07      &\ding{51} &$17.08\pm 1.33$ &12.84 &$       $\\
4C41.17      &\ding{51} &$12.06\pm 0.88$ &13.70 &$       $\\
8C1435+635   &\ding{51} &$ 7.76\pm 0.76$ &10.21 &$       $\\
6C0140+326   &\ding{51} &$ 3.31\pm 1.49$ & 2.22 &$       $\\
\hline
\end{tabular}
\end{table}

For 3C368, MG1016+058, 4C60.07, 4C41.17, and 8C1435+635, an estimate
of the 850-450$\;$\micron{} spectral index can be made.  For thermal
dust emission, the slope along the Rayleigh-Jeans tail is expected to
be around 3.0-4.0.  The spectral indices of the lower-$z$ detected
sources, 3C368 and MG1016+058, are consistent with this:
$|\alpha^{850}_{450}| \sim 4$.  The indices of 4C60.07, 4C41.17 and
8C1435+635 are on the low side, $|\alpha^{850}_{450}| \leq 2$.
However, this is to be expected.  They have large redshifts, and for
thermal greybody emission at 30 K, the spectral turnover should be
redshifted into or near the 450$\;$\micron{} waveband.  Using the
450$\;$\micron{} observation would underestimate the slope along the
Rayleigh-Jeans tail in this situation.

\section{The radio luminosity-redshift plane for radio galaxies observed with SCUBA}
\label{pzplanesec}

It is useful to plot radio luminosity against redshift (the P-$z$
plane) highlighting the galaxies we have observed with SCUBA.  This
ensures that even luminosity coverage of the P-$z$ plane has been
achieved, and identifies the redshifts at which submillimetre emission
from radio galaxies is detected.

\begin{figure}
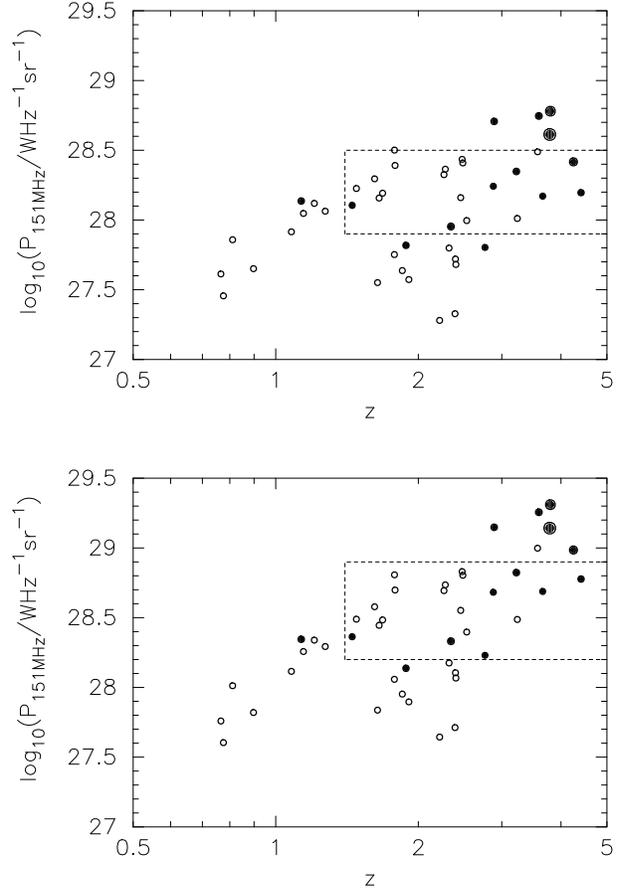

\centering
\epsfig{file=fig2a.eps,height=90mm,angle=270}
\epsfig{file=fig2b.eps,height=90mm,angle=270}
\caption{Radio luminosity-redshift plane for $\Omega_{\circ} = 1.0$ (upper plot) and $\Omega_{\circ} = 0.1$ (lower plot), assuming H$_{\circ}=50$ kms$^{-1}$Mpc$^{-1}$.  The solid circles indicate radio galaxies for which thermal emission from dust was detected at 850$\;$\micron{} with S/N$>2$, and the open circles indicate galaxies for which no dust emission was detected.  For the detections, the size of the solid circle represents the brightness of the galaxy at 850 \micron{}.  The 151-MHz radio luminosities, P$_{151{\rm MHz}}$, were estimated using the radio SEDs presented and referenced in \appen{sedfig1}.  Note, 53W069 is too faint to appear on this version of the radio luminosity-redshift plane.  The dashed-line boxes depict Subset A (upper plot) and Subset B (lower plot) which are defined in \sec{survsub}.}
\label{pzom_1_01}
\end{figure}

As previously mentioned, the radio luminosities have been calculated
at rest-frame 151$\;$MHz to ensure that the radio luminosity tracks
the power of the radio jets as accurately as possible. Simple physical
models for the time development of the 151$\;$MHz luminosity and the
linear size (D$_{linear}$) of a double radio source suggest that over
a large range of D$_{linear}$, the 151$\;$MHz radio luminosity gives a
reasonably accurate guide to the jet power Q of the vast majority of
objects detected in complete radio flux-limited samples \cite{wrbl99}.

Both the 151$\;$MHz luminosities and the linear sizes for the sample
have been calculated for two cosmologies: $\Omega_{\circ} = 1.0$, and
$\Omega_{\circ} = 0.1$, where H$_{\circ}=50$ kms$^{-1}$Mpc$^{-1}$.
The values are presented in \tab{sourcesizes}.  \fig{pzom_1_01} shows
the corresponding P-$z$ plane for each cosmology.

A proper statistical analysis of these data will be presented in
\sec{evolstats}, but at first glance, \fig{pzom_1_01} seems to
indicate that submillimetre emission is more predominant in
high-redshift radio galaxies than in low-redshift radio galaxies.

\begin{table*}
\caption{Radio sizes and luminosities for the sample.  Column 3 gives the largest angular size (LAS) of the radio source, measured in arcseconds.  The corresponding linear size (D$_{linear}$) and 151-MHz radio luminosity (P$_{151{\rm MHz}}$) have been calculated for both $\Omega_{\circ} = 1.0$ and $\Omega_{\circ} = 0.1$ (H$_{\circ}=50$ kms$^{-1}$Mpc$^{-1}$).  The radio luminosities were estimated using the radio spectral energy distributions presented and referenced in \appen{sedfig1}.  The references for the LAS values are given in column 8.  Where necessary, the LAS was measured off published radio maps.  For the majority of the 3CRR sources, the LAS was taken from Blundell et al. (2000), which summarises the radio properties of these sources.}
\label{sourcesizes}
\begin{tabular}{llrrcrcl}
\hline
&&&\multicolumn{2}{c}{$\Omega_{\circ}=1.0$}&\multicolumn{2}{c}{$\Omega_{\circ}=0.1$}&\\
Source&\multicolumn{1}{c}{z}&\multicolumn{1}{c}{LAS}&\multicolumn{1}{c}{D$_{linear}$}&\multicolumn{1}{c}{log P$_{151{\rm MHz}}$}&\multicolumn{1}{c}{D$_{linear}$}&\multicolumn{1}{c}{log P$_{151{\rm MHz}}$}&Refs.\\
&&\multicolumn{1}{c}{($''$)}&\multicolumn{1}{c}{(kpc)}&\multicolumn{1}{c}{(WHz$^{-1}$sr$^{-1}$)}&\multicolumn{1}{c}{(kpc)}&\multicolumn{1}{c}{(WHz$^{-1}$sr$^{-1}$)}&\\
\hline
3C277.2      &0.766    &  58.0     & 473  &27.61     &  559   &27.76      &BRR00\\
3C340        &0.7754   &  46.7     & 382  &27.46     &  452   &27.60      &BRR00\\
3C265        &0.8108   &  78.0     & 643  &27.86     &  768   &28.01      &BRR00\\
3C217        &0.8975   &  12.0     & 101  &27.65     &  122   &27.82      &BRR00\\
3C356        &1.079    &  75.0     & 643  &27.92     &  809   &28.12      &BRR00\\
3C368        &1.132    &  7.9      &  68  &28.14     &   86   &28.35      &BRR00\\
3C267        &1.144    &  38.0     & 327  &28.05     &  416   &28.26      &BRR00\\
3C324        &1.2063   &  10.0     &  86  &28.12     &  111   &28.34      &BRR00\\
3C266        &1.272    &   4.5     &  39  &28.06     &   51   &28.29      &BRR00\\
53W069       &1.432    & $<5.1 $   &$<44$ &24.54     &$<59$   &24.80      &WvHK84\\
4C13.66      &1.45     &   6.0     &  51  &28.11     &   69   &28.36      &BRR00\\
3C437        &1.48     &  34.4     & 294  &28.23     &  398   &28.49      &BRR00\\
3C241        &1.617    &   0.9     &   8  &28.30     &   11   &28.58      &BRR00\\
6C0919+38    &1.65     &  10.4     &  88  &27.55     &  122   &27.84      &NAR92\\
3C470        &1.653    &  24.0     & 203  &28.16     &  283   &28.45      &BRR00\\
3C322        &1.681    &  33.0     & 279  &28.19     &  390   &28.48      &BRR00\\
6C1204+37    &1.779     &  51.7     & 433  &27.75     &  615   &28.06      &LLA95\\
3C239        &1.781    &  11.2     &  94  &28.50     &  133   &28.81      &BRR00\\
3C294        &1.786    &  15.0     & 125  &28.39     &  179   &28.70      &BRR00\\
6C0820+36    &1.86     &  23.0     & 191  &27.64     &  275   &27.95      &LLA95\\
6C0905+39    &1.882    & 111.0     & 920  &27.82     & 1330   &28.14      &LGE95\\
6C0901+35    &1.904     &   2.7     &  22  &27.57     &   32   &27.90      &NAR92\\
5C7.269      &2.218    &   7.7     &  62  &27.28     &   94   &27.64      &BRR00\\
4C40.36      &2.265    &   4.0     &  32  &28.33     &   49   &28.69      &CRvO97\\
MG1744+18    &2.28     &   7.5     &  60  &28.36     &   91   &28.74      &CRvO97\\
MG1248+11    &2.322    & $<1.2$    &$<9$  &27.80     &$<15$   &28.18      &LBH86\\
4C48.48      &2.343    &  14.0     & 110  &27.95     &  171   &28.33      &CRvO97\\
53W002       &2.39     &   0.7     &   5  &27.33     &    9   &27.71      &WBM91\\
6C0930+38    &2.395    &   3.7     &  29  &27.72     &   45   &28.11      &NAR92\\
6C1113+34    &2.406      &  16.5     & 129  &27.68     &  201   &28.07      &LLA95\\
MG2305+03    &2.457    &   3.0     &  23  &28.16     &   37   &28.55      &SDS99\\
3C257        &2.474    &  12.0     &  93  &28.44     &  147   &28.83      &vBS98\\
4C23.56      &2.483    &  53.0     & 411  &28.41     &  647   &28.80      &CRvO97\\
8C1039+68    &2.53     &  15.0     & 116  &28.00     &  183   &28.40      &Lthesis\\
MG1016+058   &2.765    &   1.3     &  10  &27.80     &   16   &28.23      &DSD95\\
4C24.28      &2.879    &   2.3     &  17  &28.24     &   28   &28.68      &CRvO97\\
4C28.58      &2.891    &  16.2     & 119  &28.71     &  198   &29.15      &CMvB96\\
6C1232+39    &3.221     &   7.7     &  54  &28.35     &   94   &28.82      &NAR92\\
6C1159+36    &3.2    &   1.2     &   8  &28.01     &   15   &28.49      &LLA95\\
TX1243+036   &3.57     &   8.8     &  60  &28.49     &  107   &29.00      &vOR96\\
MG2141+192   &3.592    &   8.9     &  60  &28.75     &  108   &29.26      &CRvO97\\
6C0032+412   &3.66     &   2.3     &  15  &28.17     &   28   &28.69      &BRE98\\
4C60.07      &3.788    &   9.0     &  59  &28.61     &  109   &29.14      &CRvO97\\
4C41.17      &3.792      &  13.1     &  86  &28.78     &  159   &29.31      &COH94\\
8C1435+635   &4.25     &   3.9     &  24  &28.42     &   47   &28.99      &LMR94\\
6C0140+326   &4.41     &   2.5     &  15  &28.20     &   30   &28.78      &RLB96, BRE98\\
\hline
\end{tabular}
\end{table*}

\section{Submillimetre luminosities and dust masses}
\label{seclumin}

To calculate the rest-frame 850-\micron{} luminosity (\luminsub{}) and
dust mass (${\rm M_{dust}}$) of each source, we adopt an optically
thin isothermal dust emission template with $\beta=1.5$ and ${\rm
T_{dust}}=40\;$K.  This template is consistent with the studies
attempting to constrain $\beta$ and ${\rm T_{dust}}$ that have been
published in the literature.  For example, Benford et
al. \shortcite{bco99} have observed a sample of high-redshift
radio-quiet quasars and radio galaxies at 350-\micron{}.  Assuming a
critical wavelength of ${\rm \lambda_{\circ}=125\mu m}$, isothermal
fits indicate $\beta \sim 1.5$ and ${\rm T_{dust} \sim 50 K}$.  The
study of {\em IRAS} galaxies by Dunne et al. \shortcite{dunne2000}
indicates $\beta \sim 1.2$ and ${\rm T_{dust} \sim 36 K}$ (although
the value $\beta$ is thought to have been artificially lowered by the
presence of a cold dust component).  Finally, for the handful of
galaxies identified in the recent SCUBA surveys that have known
redshifts (refer to Smail et al. 2000\nocite{sibk2000} for a review),
the favoured value of ${\rm T_{dust}}$ is $\sim 40\;$K (e.g. Ivison et
al. 1998b, 2000b\nocite{islb98,isbk00}; Barger, Cowie \& Sanders
1999\nocite{bargercs99}).

It is currently impossible to determine the precise values of $\beta$
and ${\rm T_{dust}}$, and even though we have assumed the most likely
template, it could very well be wrong.  This has important
consequences for the analysis presented here.  For greybody emission,
the values of $\beta$ and ${\rm T_{dust}}$ determine the shape and
position of the spectral turnover, but have less of an effect on the
Rayleigh-Jeans tail.  As the redshift of the source increases, the
spectral peak is brought into view.  Thus the K-correction used to
calculate the 850-\micron{} luminosity is more sensitive to changing
the values of $\beta$ and ${\rm T_{dust}}$ at high redshift than it is
at low redshift.  If the template is wrong, the error in the
estimations of \luminsub{} will be larger if the source has a high
redshift.  To explore this, the analysis of evolutionary trends has
been conducted using several $\beta$-${\rm T_{dust}}$ combinations, in
addition to the template itself, to ensure the trends are real.

Furthermore, applying a template to the entire sample assumes that the
individual radio galaxies all have similar dust parameters.  This is a
reasonable assumption if the same mechanisms for creating and heating
dust are expected in each source.  If, however, the dust
characteristics do vary from source to source, they would have to do
so in a redshift- (or radio power-) dependent manner to seriously
affect the evolutionary trends presented here.

The choice of cosmology is also an important issue.  As the relative
properties of the sample are of most interest in the analysis
presented here, the value of the Hubble constant is irrelevant.
Changing the Hubble constant simply scales all the luminosities up or
down by the same amount.  However, changing the value of the density
parameter affects the luminosities in a redshift-dependent manner.  In
order to encompass all possibilities, we thus consider two extreme
values: \omegao{}$\;$=$\;$1.0 and \omegao{}$\;$=$\;$0.1, with
\hubble{}$\;$=$\;$50$\;$\hubbleunits{}.  In \sec{binningsec} a
low-density Universe (\omegao{}$\;$=$\;$0.1) with \hubble{}$\;$=$\;$67
is also considered.  The reason for this last choice is that the age
of a Universe with \omegao{}$\;$=$\;$1.0, \hubble{}$\;$=$\;$50, is
$\sim 13\;$Gyr.  A Universe with \omegao{}$\;$=$\;$0.1,
\hubble{}$\;$=$\;$67 has the same age - hence the `absolute values' of
the luminosities in these high and low density cases can be
realistically compared.  Recall the cosmological constant, $\Lambda$,
is assumed to be zero throughout.  It is worth noting that the
evidence from the Supernova Cosmology Project is currently in favour
of a cosmology with $\Omega_{\rm M} = 0.3$ and $\Omega_{\Lambda}=0.7$
(e.g. Perlmutter et al. 1999\nocite{perl99}).  This cosmology produces
intermediate results to the two cosmologies considered here.

\subsection{Implication of the adopted template}

\begin{figure}
\centering
\epsfig{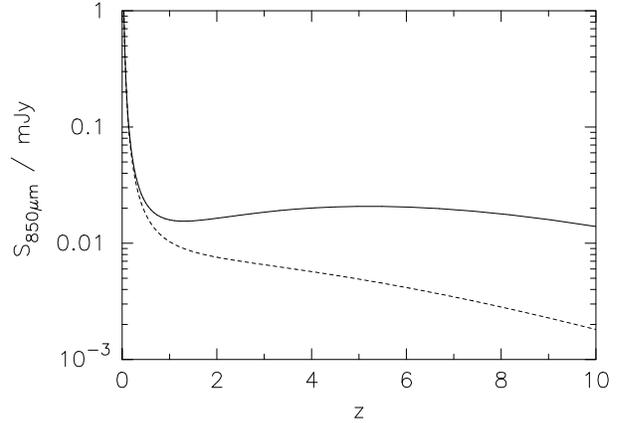}
\caption{The 850 \micron{} flux density that would be observed for the starburst galaxy M82 (Hughes, Gear \& Robson 1994) if it were placed at progressively higher redshifts, assuming the dust emission template adopted here ($\beta=1.5$, T$=40$K).  The solid line represents an \omegao{}=$\;$1.0 Universe; the dashed line represents an \omegao{}=$\;$0.1 Universe (\hubble{}=$\;$50$\;$km$^{-1}$Mpc$^{-1}$).}
\label{m82beta15temp40}
\end{figure}

It is worth investigating what happens to the 850-\micron{} flux
density of M82 as it is placed at progressively higher redshifts,
assuming the chosen dust template of $\beta=1.5$ and ${\rm T_{dust} =
40\;K}$.

\fig{m82beta15temp40} demonstrates that in an \omegao{}$\;$=$\;$1.0
Universe, the 850-\micron{} flux density is almost flat between $z=1$
and $z=4$; the 850-\micron{} flux density should effectively trace the
850-\micron{} luminosity.

In an \omegao{}$\;$=$\;$0.1 Universe, on the other hand, SCUBA is less
sensitive to $z=4$ galaxies than it is to those at $z=1$.  Thus, the
higher redshift objects in the sample should be more luminous than
those at lower redshifts if they are detected at the same level.

The act of increasing $\beta$ and ${\rm T_{dust}}$ will effectively
make SCUBA more sensitive to submillimetre emission at $z=4$ than
$z=1$.  For a given set of fluxes, the luminosities at $z=4$ will be
reduced relative to those at $z=1$ if $\beta$ and ${\rm T_{dust}}$ are
increased.  Likewise, decreasing $\beta$ and ${\rm T_{dust}}$
increases the inferred 850-\micron{} luminosities at $z=4$ relative to
those at $z=1$.

\subsection{\boldmath \luminsub{} for the radio galaxy sample}

\tab{lumin850values} gives the 850-\micron{} luminosities and dust
masses for the radio galaxy sample.  The optically thin, isothermal
template spectrum with $\beta=1.5$ and ${\rm T_{dust}=40K}$ has been
assumed.  The errors on the values of \luminsub{} and ${\rm M_{dust}}$
have been calculated using the errors in the flux density measurements
only; they do not account for the uncertainties in cosmology, $\beta$,
or ${\rm T_{dust}}$.

The dust mass of each galaxy in the sample was calculated from the
850-\micron{} observation using:

\begin{equation}
M_{dust} = \frac{S_{obs}\;D_L^2}{(1+z)\;\kappa_{\nu_{rest}}\;B_{\nu_{rest}}(T_{dust})}
\label{dustmasseq}
\end{equation}

where $S_{obs}$ is the observed flux-density, $\nu_{rest}$ is the
corresponding rest-frame frequency, $D_L$ is the luminosity distance
to the source, $\kappa_{\nu_{rest}}$ is the mass absorption
coefficient of the dust at $\nu_{rest}$, and
$B_{\nu_{rest}}(T_{dust})$ is the intensity of a blackbody at
$\nu_{rest}$ assuming isothermal emission from dust grains at a
temperature $T_{dust}$.  The mass absorption coefficient $\kappa$ is
poorly constrained, with published values of $\kappa$(800\micron{}):
0.04$\;$m$^2$kg$^{-1}$ \cite{draineandlee}, 0.12$\;$m$^2$kg$^{-1}$
(Chini, Kr\"ugel, \& Kreysa 1986\nocite{ckk86}),
0.15$\;$m$^2$kg$^{-1}$ \cite{hildebrand}, and 0.3$\;$m$^2$kg$^{-1}$
\cite{mw89}.  These estimates can be extrapolated to
other wavelengths using $\kappa\propto\lambda^{-\beta}$ \cite{ckk86}.
The intermediate value of
$\kappa$(800\micron{})=0.15$\;$m$^2$kg$^{-1}$ will be assumed here.  A
different choice will not affect the relative dust masses of the
sample, and can only alter the absolute dust masses by a factor of
$\sim 4$ at most \cite{hdr97}.

\begin{table*}
\caption{Rest-frame 850-\micron{} luminosities (\luminsub{}) and dust masses (${\rm M_{dust}}$) for the radio galaxy sample.  Optically thin, isothermal greybody emission has been assumed, with $\beta=1.5$ and ${\rm T_{dust}=40K}$.  Errors have been estimated using only the standard errors on the flux density measurements; 2$\sigma$ upper limits are given for undetected (S/N$\;<\;$2) sources.  \luminsub{} and ${\rm M_{dust}}$ have been calculated for three different cosmologies: \omegao{}=1.0, \hubble{}=50$\;$\hubbleunits{}; \omegao{}=0.1, \hubble{}=50$\;$\hubbleunits{}; \omegao{}=0.1, \hubble{}=67$\;$\hubbleunits{}.  Note, column 3 indicates whether the source belongs to either or the subsets defined in \sec{survsub}.}
\label{lumin850values}
\begin{tabular}{llclrrrrrrrr}
\hline
&&&&\multicolumn{2}{c}{\omegao{}=1.0, \hubble{}=50}&&\multicolumn{2}{c}{\omegao{}=0.1, \hubble{}=50}&&\multicolumn{2}{c}{\omegao{}=0.1, \hubble{}=67}\\
Source&\multicolumn{1}{c}{z}&Subset&&\multicolumn{1}{c}{log$\;$L$_{850\mu m}$}&\multicolumn{1}{c}{log$\;{\rm M_{dust}}$}&&\multicolumn{1}{c}{log$\;$L$_{850\mu m}$}&\multicolumn{1}{c}{log$\;{\rm M_{dust}}$}&&\multicolumn{1}{c}{log$\;$L$_{850\mu m}$}&\multicolumn{
1}{c}{log$\;{\rm M_{dust}}$}\\
&&&&\multicolumn{1}{c}{(WHz$^{-1}$sr$^{-1}$)}&\multicolumn{1}{c}{(${\rm M_{\odot}}$)}&&\multicolumn{1}{c}{(WHz$^{-1}$sr$^{-1}$)}&\multicolumn{1}{c}{(${\rm M_{\odot}}$)}&&\multicolumn{1}{c}{(WHz$^{-1}$sr$^{-1}$)}&\multicolumn{1}{c}{(${\rm M_{\odot}}$)}\\
\hline
3C277.2         &  0.766    &           && $<$  22.70  & $<$   8.18    && $<$  22.85  & $<$   8.32    && $<$  22.59  & $<$   8.07\\
3C340           &  0.775    &           && $<$  22.63  & $<$   8.10    && $<$  22.77  & $<$   8.25    && $<$  22.52  & $<$   8.00\\
3C265           &  0.811    &           && $<$  22.69  & $<$   8.16    && $<$  22.84  & $<$   8.32    && $<$  22.59  & $<$   8.06\\
3C217           &  0.897    &           && $<$  22.63  & $<$   8.10    && $<$  22.80  & $<$   8.27    && $<$  22.54  & $<$   8.02\\
3C356           &  1.079    &           && $<$  22.84  & $<$   8.31    && $<$  23.04  & $<$   8.51    && $<$  22.78  & $<$   8.26\\
3C368           &  1.132    &           &&      23.00  &       8.47    &&      23.20  &       8.68    &&      22.95  &       8.43\\
3C267           &  1.144    &           && $<$  22.71  & $<$   8.19    && $<$  22.92  & $<$   8.40    && $<$  22.67  & $<$   8.14\\
3C324           &  1.206    &           && $<$  22.84  & $<$   8.32    && $<$  23.06  & $<$   8.54    && $<$  22.81  & $<$   8.28\\
3C266           &  1.272    &           && $<$  22.85  & $<$   8.32    && $<$  23.08  & $<$   8.55    && $<$  22.82  & $<$   8.30\\
53W069          &  1.432    &           && $<$  22.75  & $<$   8.22    && $<$  23.00  & $<$   8.48    && $<$  22.75  & $<$   8.22\\
4C13.66         &  1.45     &A,B        &&      22.85  &       8.32    &&      23.11  &       8.58    &&      22.85  &       8.33\\
3C437           &  1.48     &A,B        && $<$  22.72  & $<$   8.20    && $<$  22.98  & $<$   8.46    && $<$  22.73  & $<$   8.20\\
3C241           &  1.617    &A,B        && $<$  22.79  & $<$   8.26    && $<$  23.07  & $<$   8.55    && $<$  22.82  & $<$   8.29\\
6C0919+38       &  1.65     &           && $<$  22.75  & $<$   8.22    && $<$  23.03  & $<$   8.51    && $<$  22.78  & $<$   8.25\\
3C470           &  1.653    &A,B        && $<$  23.21  & $<$   8.69    && $<$  23.50  & $<$   8.98    && $<$  23.25  & $<$   8.72\\
3C322           &  1.681    &A,B        && $<$  22.75  & $<$   8.22    && $<$  23.04  & $<$   8.52    && $<$  22.79  & $<$   8.26\\
6C1204+37       &  1.779     &          && $<$  22.81  & $<$   8.29    && $<$  23.12  & $<$   8.60    && $<$  22.87  & $<$   8.34\\
3C239           &  1.781    &A,B        && $<$  22.72  & $<$   8.19    && $<$  23.02  & $<$   8.50    && $<$  22.77  & $<$   8.25\\
3C294           &  1.786    &A,B        && $<$  22.61  & $<$   8.09    && $<$  22.92  & $<$   8.39    && $<$  22.66  & $<$   8.14\\
6C0820+36       &  1.86     &           && $<$  22.99  & $<$   8.46    && $<$  23.30  & $<$   8.78    && $<$  23.05  & $<$   8.53\\
6C0905+39       &  1.882    &           &&      22.96  &       8.43    &&      23.28  &       8.75    &&      23.02  &       8.50\\
6C0901+35       &  1.904     &          && $<$  22.77  & $<$   8.25    && $<$  23.10  & $<$   8.57    && $<$  22.84  & $<$   8.32\\
5C7.269         &  2.218    &           && $<$  22.93  & $<$   8.40    && $<$  23.29  & $<$   8.77    && $<$  23.04  & $<$   8.51\\
4C40.36         &  2.265    &A,B        && $<$  22.83  & $<$   8.31    && $<$  23.20  & $<$   8.68    && $<$  22.95  & $<$   8.42\\
MG1744+18       &  2.28     &A,B        && $<$  22.70  & $<$   8.18    && $<$  23.08  & $<$   8.55    && $<$  22.82  & $<$   8.30\\
MG1248+11       &  2.322    &           && $<$  22.88  & $<$   8.36    && $<$  23.26  & $<$   8.73    && $<$  23.00  & $<$   8.48\\
4C48.48         &  2.343    &A,B        &&      23.06  &       8.54    &&      23.44  &       8.92    &&      23.19  &       8.66\\
53W002          &  2.39     &           && $<$  22.89  & $<$   8.36    && $<$  23.27  & $<$   8.75    && $<$  23.02  & $<$   8.49\\
6C0930+38       &  2.395    &           && $<$  22.69  & $<$   8.16    && $<$  23.07  & $<$   8.55    && $<$  22.82  & $<$   8.29\\
6C1113+34       &  2.406      &         && $<$  22.74  & $<$   8.22    && $<$  23.13  & $<$   8.60    && $<$  22.87  & $<$   8.35\\
MG2305+03       &  2.457    &A,B        && $<$  22.71  & $<$   8.18    && $<$  23.10  & $<$   8.57    && $<$  22.84  & $<$   8.32\\
3C257           &  2.474    &A,B        && $<$  23.17  & $<$   8.64    && $<$  23.56  & $<$   9.04    && $<$  23.31  & $<$   8.78\\
4C23.56         &  2.483    &A,B        && $<$  22.94  & $<$   8.42    && $<$  23.34  & $<$   8.81    && $<$  23.08  & $<$   8.56\\
8C1039+68       &  2.53     &A,B        && $<$  22.68  & $<$   8.15    && $<$  23.08  & $<$   8.55    && $<$  22.82  & $<$   8.30\\
MG1016+058      &  2.765    &B          &&      22.73  &       8.20    &&      23.16  &       8.63    &&      22.90  &       8.38\\
4C24.28         &  2.879    &A,B        &&      22.73  &       8.21    &&      23.17  &       8.65    &&      22.92  &       8.39\\
4C28.58         &  2.891    &           &&      22.95  &       8.42    &&      23.39  &       8.86    &&      23.13  &       8.61\\
6C1232+39       &  3.221     &A,B       &&      22.93  &       8.40    &&      23.40  &       8.88    &&      23.15  &       8.62\\
6C1159+36       &  3.2    &A,B  && $<$  22.83  & $<$   8.31    && $<$  23.31  & $<$   8.78    && $<$  23.06  & $<$   8.53\\
TX1243+036      &  3.57     &A          && $<$  22.97  & $<$   8.45    && $<$  23.48  & $<$   8.95    && $<$  23.22  & $<$   8.70\\
MG2141+192      &  3.592    &           &&      22.99  &       8.47    &&      23.50  &       8.98    &&      23.25  &       8.72\\
6C0032+412      &  3.66     &A,B        &&      22.71  &       8.18    &&      23.22  &       8.70    &&      22.97  &       8.45\\
4C60.07         &  3.788    &           &&      23.56  &       9.03    &&      24.08  &       9.56    &&      23.83  &       9.31\\
4C41.17         &  3.792      &         &&      23.40  &       8.88    &&      23.93  &       9.41    &&      23.68  &       9.16\\
8C1435+635      &  4.25     &A          &&      23.20  &       8.68    &&      23.77  &       9.25    &&      23.52  &       8.99\\
6C0140+326      &  4.41     &A,B        &&      22.83  &       8.30    &&      23.41  &       8.89    &&      23.16  &       8.63\\
\hline
\end{tabular}
\end{table*}

\begin{figure}
\centering
\epsfig{file=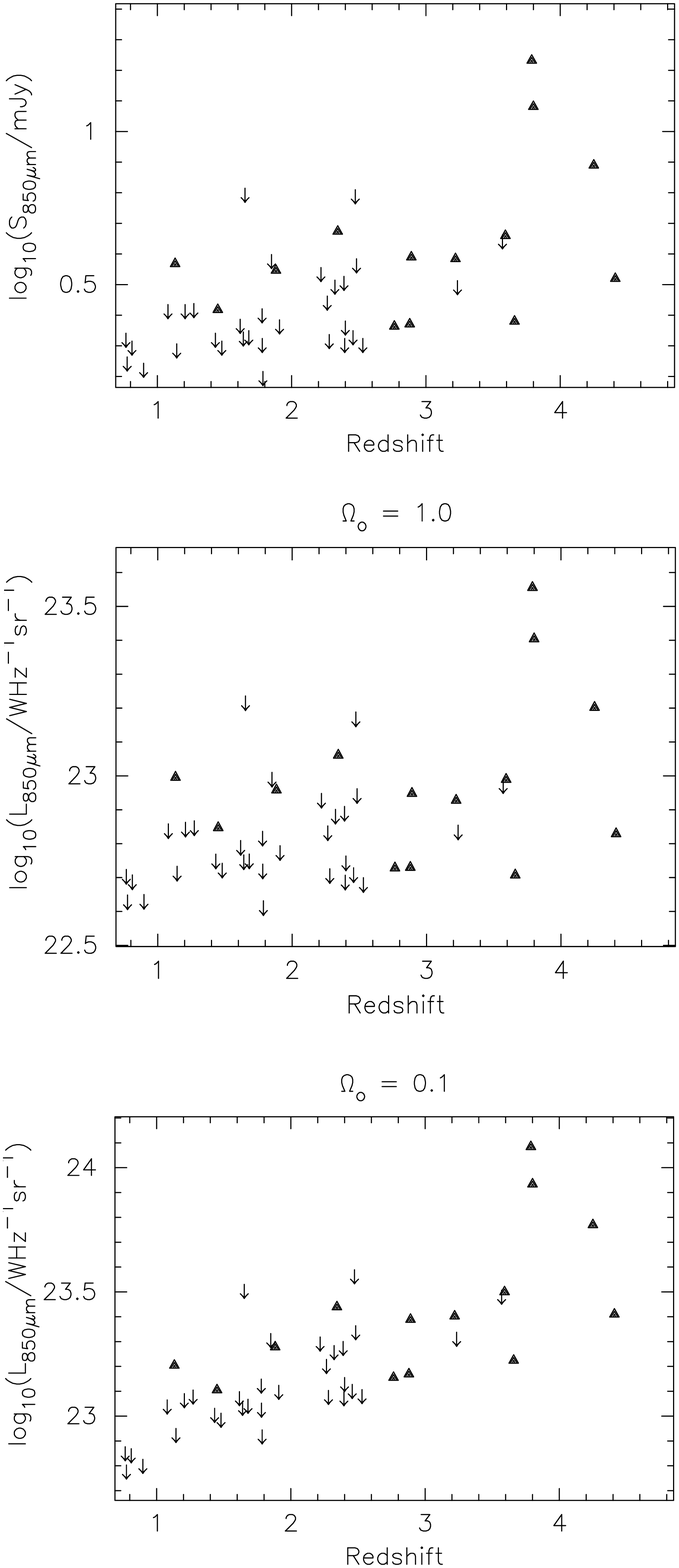,height=15cm}
\caption{Scatter plots of 850-\micron{} flux density and (rest-frame)
luminosity against redshift for the radio galaxy sample.  \luminsub{}
has been calculated assuming $\beta=1.5$ and ${\rm T_{dust}=40K}$, and
is shown for both $\Omega_{\circ}=1.0$ and $\Omega_{\circ}=0.1$
(\hubble{} is assumed to be 50$\;$\hubbleunits{}).}
\label{sampledim}
\end{figure}

In \Fig{sampledim}, the 850-\micron{} flux density is plotted against
redshift for the radio galaxy sample.  Plots of \luminsub{} vs. $z$ in
\omegao$\;$=$\;$1.0 and \omegao$\;$=$\;$0.1 Universes are also shown.
As expected from \fig{m82beta15temp40}, in the \omegao$\;$=$\;$1.0
case the flux density traces the luminosity between $z=1$ and $z=4$ ,
and the two plots look the same.  For \omegao$\;$=$\;$0.1,
the luminosity of the $z \sim 4$ galaxies is higher relative to those
at $z \sim 1$ , as SCUBA is less sensitive to the high-redshift
objects.

The small dynamic range in the measured values of \luminsub{} for this
study are limited by the sensitivity of SCUBA and the characteristic
dust mass in high-redshift objects.  Note the detections of
\luminsub{} appear at roughly the same level as the \luminsub{} upper
limits; most of the galaxies in the sample have been observed to the
same depth in 850-\micron{} {\em luminosity}.  Coupled with the fact
that the upper limits are clustered at the low-redshift end of the
sample, this indicates that the high-redshift radio galaxies are
intrinsically brighter in the submillimetre than the lower-redshift
radio galaxies.  This evidence for evolution of the radio galaxy
population with redshift will now be analysed in detail.

\section{Evidence For evolution}
\label{evolstats}

\subsection{Survival analysis: correlations between radio properties, submillimetre properties, and redshift}
\label{survsub}

A logical approach to finding evidence which either supports or
rejects the evolution scenario is to run correlation tests between the
radio properties, submillimetre properties, and redshifts of the
sample.  For example, a correlation between \luminsub{} and redshift,
in the absence of a correlation between \luminsub{} and \radiopower{},
will imply that radio galaxies undergo cosmological
evolution of their submillimetre emission.  Conversely, a significant
correlation between \luminsub{} and \radiopower{} in the absence of a
correlation between \luminsub{} and redshift, would indicate that the
submillimetre properties of radio galaxies are more closely tied to
the radio properties of the galaxy (perhaps, for example, via host mass) 
than to cosmological epoch.

The obvious correlations to consider are between \luminsub{},
\radiopower{}, and $z$.  It is also instructive to investigate whether
\radiopower{} correlates with D$_{linear}$ (the linear size of the
radio source), and whether the radio spectral index \alpharadio{}
correlates with either \luminsub{}, \radiopower{}, or $z$.

As mentioned previously, for radio sources selected from flux-limited
samples, the 151-MHz radio luminosity should be a good indicator of
the intrinsic power $Q$ of the radio jet. The size of a radio source
depends on its age, the medium it is expanding into, and the jet
power.  Again, considering sources from flux-limited samples, Willott
et al. \shortcite{wrbl99} showed that the dependence of D$_{linear}$
on jet power is very weak; D$_{linear}$ correlates most strongly with
the age of the radio source. Although radio luminosities are expected
to decline throughout the lifetime of radio sources (Kaiser,
Dennett-Thorpe \& Alexander 1997\nocite{kda97}; Blundell et
al. 1999\nocite{brw99}; Blundell \& Rawlings 1999\nocite{bs99}),
selection effects conspire to ensure a very weak anti-correlation of
radio luminosity with D$_{linear}$ for sources selected from a number
of separate flux-limited samples \cite{brw99}.  Although our sample of
SCUBA targets is not purely a combination of flux-limited samples, it
is dominated by 3C/3CRR and 6CE targets so no strong \radiopower{}-D
links are expected.

The criterion that \alpharadio{} is large is often used to identify
high redshift sources in large radio surveys.  This selection is based
on the empirical correlation observed between \alpharadio{} and radio
luminosity (e.g. Veron, Veron \& Witzel 1972\nocite{vvw72}).  In a flux-limited
survey, high values of \alpharadio{} will often correspond to a large
intrinsic luminosity and hence to a high redshift.  

However, it is possible to envisage models in which \alpharadio{} is
related closely to star-formation activity (e.g. Lilly
1989\nocite{lilly89}).  This is relevant to the work presented here,
as several of the high-redshift galaxies in the sample were originally
identified in these ultra-steep spectrum searches (e.g. Chambers,
Miley \& van Breugel 1988, 1990; Chambers et
al. 1996\nocite{cmvb88,cmvb90,cmvb96}; Stern et
al. 1999\nocite{sds99}).  Thus, if \luminsub{} appears correlated with
redshift, it could be a manifestation of how the sources were selected
in the first place, a possibility which can be tested by investigating
whether a correlation exists between \luminsub{} and \alpharadio{}.
Note, for the correlation analysis, \alpharadio{} was calculated at
151$\;$MHz.

\begin{table*}
\caption{Results of the survival analysis correlation tests applied to the entire sample.  The significance ($P$) is the probability of the null hypothesis ({\em no correlation}) being true.  If all three tests yield $P\leq 5$ per cent, the variables are taken to be correlated.  If only some of the tests yield $P\leq 5$ per cent, evidence for a correlation is taken to be uncertain.}
\label{correlationsALL}
\begin{tabular}{|c|cc|c|rrr|c|}
\hline
Cosmology&\multicolumn{2}{|c|}{Variable}&Percentage of&\multicolumn{3}{c|}{Significance ($P$)}&Correlation\\
&Dependent&Independent&Data Censored    &\multicolumn{1}{c}{Cox}&\multicolumn{1}{c}{Kendall}&\multicolumn{1}{c|}{Spearman}&Present?\\
\hline
--              &S$_{850\mu m}$         &z                      &70$\%$         &0.00$\%$   &0.02$\%$   &0.01$\%$       &YES\\
\hline
\multirow{7}{15mm}{$\Omega_{\circ}=1.0$}
                &D$_{linear}$           &log(P$_{151MHz}$)      &4$\%$          &22.48$\%$  &82.69$\%$  &71.17$\%$      &NO\\
\dottedline{4}(55,10)(409,10)
                &log(P$_{151MHz}$)      &z                      &0$\%$          &0.01$\%$   &0.03$\%$   &0.07$\%$       &YES\\
                &log(L$_{850\mu m}$)    &log(P$_{151MHz}$)      &70$\%$         &2.65$\%$   &0.21$\%$   &0.64$\%$       &YES\\
                &log(L$_{850\mu m}$)    &z                      &70$\%$         &0.00$\%$   &0.19$\%$   &0.08$\%$       &YES\\
\dottedline{4}(55,10)(409,10)
                &log(L$_{850\mu m}$)    &$\alpha_{radio}$       &70$\%$         &0.31$\%$   &0.36$\%$   &1.73$\%$       &YES\\
                &$\alpha_{radio}$       &log(P$_{151MHz}$)      &0$\%$          &0.64$\%$   &1.00$\%$   &1.36$\%$       &YES\\
                &$\alpha_{radio}$       &z                      &0$\%$          &4.18$\%$   &13.94$\%$  &10.50$\%$      &MAYBE\\
\hline
\multirow{7}{15mm}{$\Omega_{\circ}=0.1$}
                &D$_{linear}$           &log(P$_{151MHz}$)      &4$\%$          &32.29$\%$  &67.59$\%$  &50.22$\%$      &NO\\
\dottedline{4}(55,10)(409,10)
                &log(P$_{151MHz}$)      &z                      &0$\%$          &0.00$\%$   &0.00$\%$   &0.00$\%$       &YES\\
                &log(L$_{850\mu m}$)    &log(P$_{151MHz}$)      &70$\%$         &0.53$\%$   &0.01$\%$   &0.08$\%$       &YES\\
                &log(L$_{850\mu m}$)    &z                      &70$\%$         &0.00$\%$   &0.00$\%$   &0.00$\%$       &YES\\
\dottedline{4}(55,10)(409,10)
                &log(L$_{850\mu m}$)    &$\alpha_{radio}$       &70$\%$         &0.18$\%$   &0.67$\%$   &1.88$\%$       &YES\\
                &$\alpha_{radio}$       &log(P$_{151MHz}$)      &0$\%$          &0.44$\%$   &1.61$\%$   &1.47$\%$       &YES\\
                &$\alpha_{radio}$       &z                      &0$\%$          &4.18$\%$   &13.94$\%$  &10.50$\%$      &MAYBE\\
\hline
\multicolumn{8}{|c|}{Total Number of Galaxies: 46}\\
\hline
\end{tabular}
\end{table*}

\subsubsection{Survival analysis}

The submillimetre dataset for the radio galaxy sample contains a large
number of upper limits, or {\em censored data}.  Running correlation
tests on the dataset is not a trivial matter, as standard statistical
theory is not equipped to handle this kind of information.

Statisticians have extensively studied the problem of censored
datasets, and a means for successfully handling them has been known
for decades.  It is called {\em survival analysis} and it allows
censored data samples to be treated in a meaningful manner with
minimal loss of information - the values of the upper/lower limits are
formally taken into account when calculating correlation coefficients
and other test statistics.

The application of survival analysis to astronomical data is outlined
in two papers: Feigelson \& Nelson 1985\nocite{fn85} (Paper I) and
Isobe, Feigelson, \& Nelson 1986\nocite{ifn86} (Paper II).  Paper I
deals with univariate problems.  Paper II considers bivariate
techniques to characterise the correlation and line-regression between
two variables.  The test statistics described in these papers are
available in the software package ASURV Rev. 1.1 (Isobe \& Feigelson
1990\nocite{if90}; LaValley, Isobe \& Feigelson 1992\nocite{lif92}).

ASURV contains three different correlation tests - Cox's Proportional
Hazard Model, Generalized Kendall's Tau, and Generalized Spearman's
Rho.  All three methods test the null hypothesis that {\em no
correlation is present}.  The ASURV routines compute the probability
of the null hypothesis being true, $P$, also referred to as the
significance of the correlation.  The convention to be adopted here is
that X and Y are correlated if $P\leq 5$ per cent.  A correlation will
be treated as firm if all three tests yield $P\leq 5$ per cent; the
existence of a correlation will be treated as possible, but
unconfirmed, if only some of the tests indicate $P\leq 5$ per cent.

Note, the Cox test only allows censoring in the dependent variable;
the Kendall and Spearman tests allow censoring in both the dependent
and the independent variables.  The Spearman technique has not been
properly tested yet.  It is known to give misleading results if
$N<30$.

\begin{table*}
\caption{Results of the survival analysis correlation tests applied to Subset A.  The significance ($P$) is the probability of the null hypothesis ({\em no correlation}) being true.  The generalized Spearman test has not been applied as it breaks down if the sample size is less than 30.  If both Cox and Kendall yield $P\leq 5$ per cent, the variables are taken to be correlated.  If only one of them yields $P\leq 5$ per cent, evidence for a correlation is taken to be uncertain.}
\label{correlationsluminbandA}
\begin{tabular}{|c|cc|c|rrc|c|}
\hline
Cosmology&\multicolumn{2}{|c|}{Variable}&Percentage of&\multicolumn{3}{c|}{Significance ($P$)}&Correlation\\
&Dependent&Independent&Data Censored&\multicolumn{1}{c}{Cox}&\multicolumn{1}{c}{Kendall}&\multicolumn{1}{c|}{Spearman}&Present?\\
\hline
--              &S$_{850\mu m}$         &z                      &67$\%$         &0.29$\%$   &4.05$\%$   &--             &YES\\
\hline
\multirow{7}{15mm}{$\Omega_{\circ}=1.0$}
                &D$_{linear}$           &log(P$_{151MHz}$)      &0$\%$          &63.20$\%$  &34.86$\%$  &--             &NO\\
\dottedline{4}(55,10)(409,10)
                &log(P$_{151MHz}$)      &z                      &0$\%$          &74.43$\%$  &50.65$\%$  &--             &NO\\
                &log(L$_{850\mu m}$)    &log(P$_{151MHz}$)      &67$\%$         &50.52$\%$  &77.71$\%$  &--             &NO\\
                &log(L$_{850\mu m}$)    &z                      &67$\%$         &1.91$\%$   &15.70$\%$  &--             &MAYBE\\
\dottedline{4}(55,10)(409,10)
                &log(L$_{850\mu m}$)    &$\alpha_{radio}$       &67$\%$         &70.37$\%$  &85.01$\%$  &--             &NO\\
                &$\alpha_{radio}$       &log(P$_{151MHz}$)      &0$\%$          &21.03$\%$  &16.40$\%$  &--             &NO\\
                &$\alpha_{radio}$       &z                      &0$\%$          &21.22$\%$  &18.32$\%$  &--             &NO\\
\hline
\multicolumn{8}{|c|}{Total Number of Galaxies: 21}\\
\hline
\end{tabular}
\end{table*}
\begin{table*}
\caption{Results of the survival analysis correlation tests applied to Subset B.  The significance ($P$) is the probability of the null hypothesis ({\em no correlation}) being true.  The generalized Spearman test has not been applied as it breaks down if the sample size is less than 30.  If both Cox and Kendall yield $P\leq 5$ per cent, the variables are taken to be correlated.  If only one of them yields $P\leq 5$ per cent, evidence for a correlation is taken to be uncertain.}
\label{correlationsluminbandB}
\begin{tabular}{|c|cc|c|rrc|c|}
\hline
Cosmology&\multicolumn{2}{|c|}{Variable}&Percentage of&\multicolumn{3}{c|}{Significance ($P$)}&Correlation\\
&Dependent&Independent&Data Censored&\multicolumn{1}{c}{Cox}&\multicolumn{1}{c}{Kendall}&\multicolumn{1}{c|}{Spearman}&Present?\\
\hline
--              &S$_{850\mu m}$         &z                      &65$\%$         &0.81$\%$   &5.87$\%$   &--             &MAYBE\\
\hline
\multirow{7}{15mm}{$\Omega_{\circ}=0.1$}
                &D$_{linear}$           &log(P$_{151MHz}$)      &0$\%$          &84.35$\%$  &49.54$\%$  &--             &NO\\
\dottedline{4}(55,10)(409,10)
                &log(P$_{151MHz}$)      &z                      &0$\%$          &34.86$\%$  &27.00$\%$  &--             &NO\\
                &log(L$_{850\mu m}$)    &log(P$_{151MHz}$)      &65$\%$         &79.95$\%$  &96.22$\%$  &--             &NO\\
                &log(L$_{850\mu m}$)    &z                      &65$\%$         &0.13$\%$   &2.01$\%$   &--             &YES\\
\dottedline{4}(55,10)(409,10)
                &log(L$_{850\mu m}$)    &$\alpha_{radio}$       &65$\%$         &64.95$\%$  &92.43$\%$  &--             &NO\\
                &$\alpha_{radio}$       &log(P$_{151MHz}$)      &0$\%$          &0.39$\%$   &2.49$\%$   &--             &YES\\
                &$\alpha_{radio}$       &z                      &0$\%$          &20.64$\%$  &41.66$\%$  &--             &NO\\
\hline
\multicolumn{8}{|c|}{Total Number of Galaxies: 20}\\
\hline
\end{tabular}
\end{table*}
\begin{figure*}
\epsfig{file=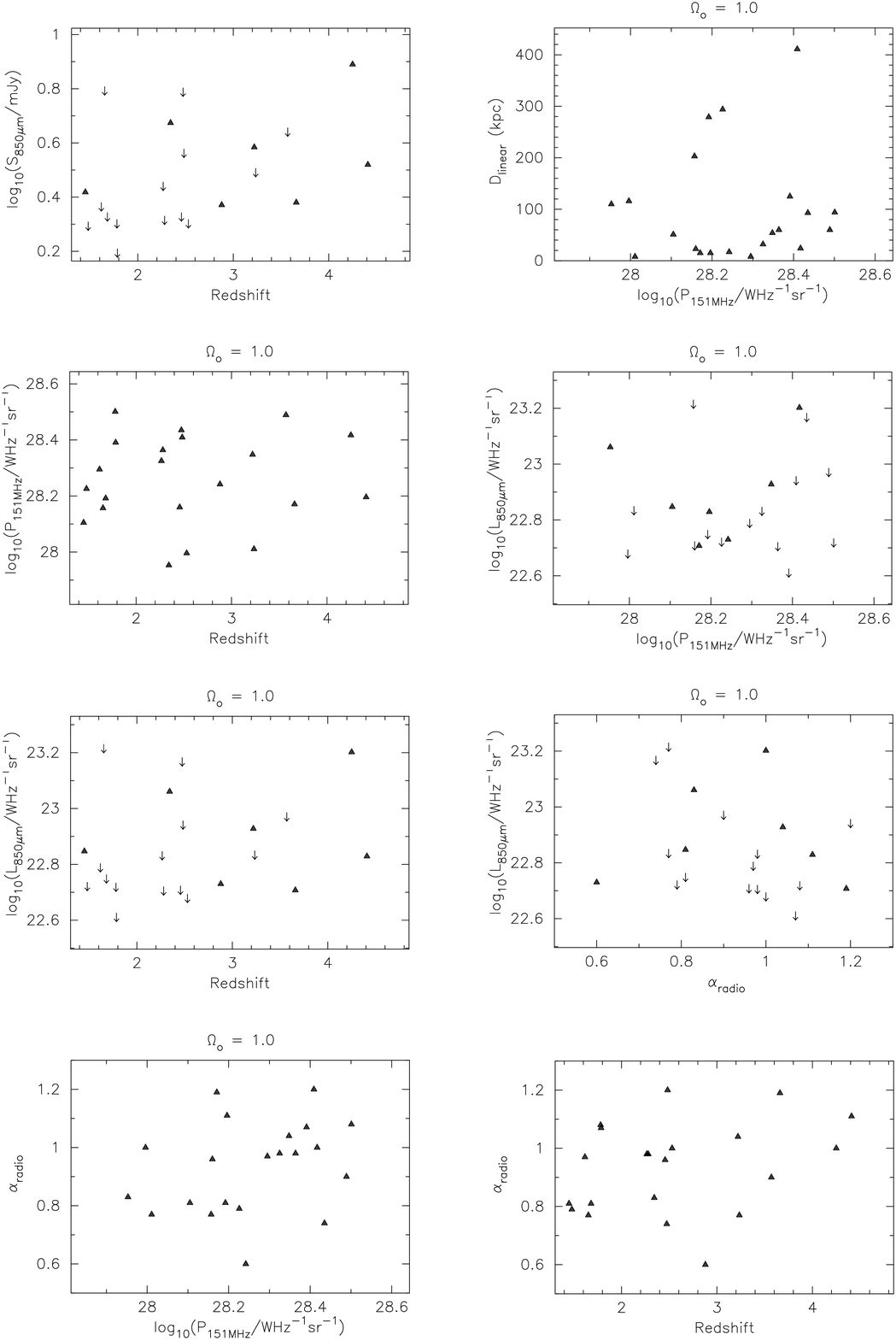,height=19cm}
\caption{Scatter plots corresponding to the correlations investigated for Subset A.}
\label{scatterplotsluminbandA}
\end{figure*}

\subsubsection{Correlations for the entire dataset}

Initially, the entire sample was used to run the correlation tests.
The results of this analysis are given in \tab{correlationsALL}.

\fig{sampledim} predicts that the 850-\micron{} flux density will
mirror the 850-\micron{} luminosity if \omegao{}=1.0.  This is
reflected by the S$_{\rm 850\mu m}$-z and \luminsub{}-z correlations
having similar significances.  As expected, decreasing \omegao{} to
0.1 increases the strength of the \luminsub{}-z correlation.

The absence of a correlation between \radiopower{} and D$_{linear}$ is
indicative of a combined sample in which, as discussed by Willott et
al. \shortcite{wrbl99}, \radiopower{} can be taken to crudely track
jet power, and D$_{linear}$ to crudely track source age.

It is troublesome, but not unexpected, 
that \luminsub{}, \radiopower{}, and $z$ are all
correlated with each other.  The strength of these correlations is
greater in a low-\omegao{} Universe.  The strong correlation between
\radiopower{} and $z$ could be responsible for either the
\luminsub{}-$z$ correlation if \luminsub{} actually correlates with
\radiopower{}, or for the \luminsub{}-\radiopower{} correlation if
\luminsub{} actually correlates with $z$.  Thus, if the effects of
cosmological evolution and radio power are to be disentangled, the
entire dataset cannot be used.

\luminsub{} also appears correlated with \alpharadio{}.  However, this may
be a remnant of the fact that \alpharadio{} and \luminsub{} are both
correlated with \radiopower{} and $z$.  Again, if the entire dataset
is used, discriminating between different effects is difficult.

It was decided to choose a strip in the P-$z$ plane that covered a large
range of redshifts in a single narrow band of radio power.  In the absence
of a correlation between \radiopower{} and $z$, it could then be
determined whether \luminsub{} truly correlates with $z$ or
\radiopower{}.  The strip was chosen to satisfy the following
criteria:
\begin{enumerate}
\item To cover the largest range in redshift possible whilst still maintaining even coverage of radio power.  This automatically imposed the restriction that $z>1.4$ due to the 3C survey having a lack of low-redshift, very high radio-power sources.
\item To contain at least 20 objects to satisfy the requirements of the survival analysis techniques.
\end{enumerate}

Initially a strip was chosen using the P-$z$ plane for \omegao{}=1.0,
hereafter referred to as Subset A.  A different but similar strip was
chosen from the P-$z$ plane for \omegao{}=0.1.  It will be referred to
as Subset B.  The sources included in each subset are listed in
\tab{lumin850values}.

It is important to note that there is nothing special about the
strips.  They have merely been chosen to help disentangle the various
influences affecting \luminsub{}.  The results found for the subsets
should apply to the entire sample and to radio galaxies in general.

\subsubsection{Correlations for Subset A}

Subset A, shown in \fig{pzom_1_01}, contains 21 radio galaxies and is
defined by $z>1.4$ and ${\rm 27.9<log(P_{151MHz})<28.5}$ for
\omegao{}=1.0, \hubble{}$\;$=$\;$50$\;$\hubbleunits{}.

\tab{correlationsluminbandA} contains the results of the survival
analysis applied to Subset A, and \fig{scatterplotsluminbandA}
contains the corresponding scatter plots.

For \omegao{}=1.0, the existence of a correlation between \luminsub{}
and redshift is hinted at, but is not confirmed by all of the
correlation tests.  The important point, however, is that \luminsub{}
definitely fails to correlate with \radiopower{} and \alpharadio{}.
The probability of `no correlation' between \luminsub{} and $z$ is
much lower than for \luminsub{}-\radiopower{} or
\luminsub{}-\alpharadio{}.

Survival analysis could also be applied to Subset A for the
\omegao{}=0.1 case.  However, for Subset A, \radiopower{} and $z$ are
correlated if \omegao{}=0.1.  This is precisely what was trying to be
avoided by selecting the subset in the first place.  Thus, a separate
subsample will need to be chosen to study the statistics for a
low-\omegao{} Universe.

\subsubsection{Correlations for Subset B}

\begin{figure*}
\epsfig{file=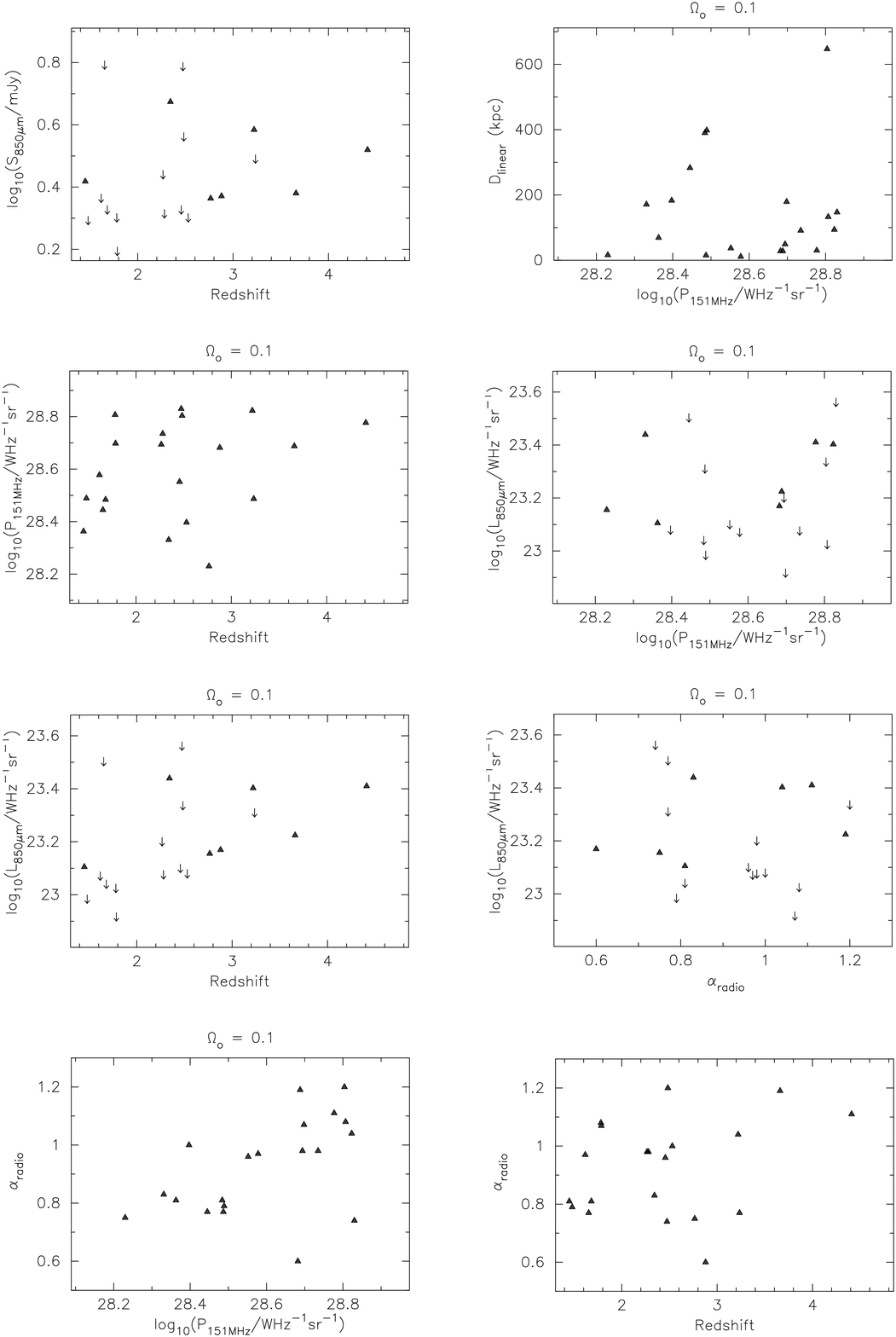,height=19cm}
\caption{Scatter plots corresponding to the correlations investigated for Subset B.}
\label{scatterplotsluminbandB}
\end{figure*}

Subset B, shown in \fig{pzom_1_01}, contains 20 radio galaxies and is
defined by $z>1.4$ and ${\rm 28.2<log(P_{151MHz})<28.9}$ for
\omegao{}=0.1, \hubble{}$\;$=$\;$50$\;$\hubbleunits{}.

\tab{correlationsluminbandB} contains the results of the survival
analysis applied to Subset B, and \fig{scatterplotsluminbandB}
contains the corresponding scatter plots.

The results of the survival analysis are more clear cut in a
low-\omegao{} Universe.  The suspected correlation between \luminsub{}
and $z$ is confirmed as significant by all of the correlation tests.
Neither $z$ nor \luminsub{} correlates with \radiopower{}.  This
indicates that the important correlation for \luminsub{} is most
likely to be with redshift, {\em not} radio luminosity.

For Subset B, \luminsub{} does not correlate with \alpharadio{} or
\radiopower{}.  However, \alpharadio{} and \radiopower{} correlate
with each other.  This implies that the correlation between
\alpharadio{} and \luminsub{} seen for the entire dataset is an
artifact of their mutual correlation with \radiopower{}.

\subsubsection{Considering the dust emission template}

The analysis described so far assumes a template for dust emission
with $\beta=1.5$ and ${\rm T_{dust} = 40K}$.  However, what if this
template was incorrect?  Would the results of the analysis be affected
if the true values of $\beta$ and ${\rm T_{dust}}$ were different?

It has already been observed that increasing $\beta$ and ${\rm
T_{dust}}$ makes SCUBA more sensitive to submillimetre emission at
$z=4$ than it is at $z=1$.  Given the measured fluxes, the
850-\micron{} luminosities at $z=3-4$ will be reduced relative to
those at $z=1-2$ if $\beta$ and ${\rm T_{dust}}$ are increased.  As a
result, correlations between \luminsub{} and $z$ should be weakened.
Likewise, decreasing $\beta$ and ${\rm T_{dust}}$ should strengthen
\luminsub{}-$z$ correlations.

Changing $\beta$ and ${\rm T_{dust}}$ will not alter \luminsub{} in
a radio-power or radio-spectral index dependent manner.  If the dust
template is wrong, the correlations between \luminsub{} and
\radiopower{}/\alpharadio{} should not be affected in a systematic
way.

In order to investigate this fully, the survival analysis was re-run
using several different combinations of $\beta$ and ${\rm T_{dust}}$.
To examine the effect of changing $\beta$, ${\rm T_{dust}}$ was held
at 40$\;$K while $\beta$ was increased to 2.0, and decreased to
1.0.  Likewise, $\beta$ was held at 1.5 while ${\rm T_{dust}}$
was changed to 100$\;$K, 70$\;$K, and 20$\;$K.

The observed effects on the results of the correlation tests can be
summarised as follows:
\begin{enumerate}
\item \luminsub{}-z: Increasing $\beta$ or ${\rm T_{dust}}$ destroys the marginal correlation between \luminsub{} and $z$ for Subset A.  For the entire dataset (both cosmologies) and for Subset B, the correlation is evident regardless of the values of $\beta$ and ${\rm T_{dust}}$.  Decreasing $\beta$ and ${\rm T_{dust}}$ confirms the possible \luminsub{}-z correlation for Subset A.
\item \luminsub{}-\radiopower{}: Changing $\beta$ and ${\rm T_{dust}}$ does not destroy the \luminsub{}-\radiopower{} correlation for the entire dataset, nor does it create a correlation between \luminsub{} and \radiopower{} for Subsets A or B.
\item \luminsub{}-\alpharadio{}: Changing $\beta$ and ${\rm T_{dust}}$ does not destroy the \luminsub{}-\alpharadio{} correlation for the entire dataset, nor does it create a correlation between \luminsub{} and \alpharadio{} for Subsets A or B.
\end{enumerate}

\subsubsection{Summary of correlation analysis}

\begin{table}
\centering
\caption{Summary of the \luminsub{}-$z$ correlation analysis for the two cosmologies considered and for reasonable values of $\beta$.  \ding{51} indicates the survival analysis found a correlation between \luminsub{} and $z$, \ding{55} indicates the survival analysis found no such correlation, and {\bf ?} indicates that some, but not all, of the correlation tests found a correlation to be present.}
\label{correlationsummary}
\begin{tabular}{ccccc}
\hline
&\multicolumn{2}{|c|}{$\Omega_{\circ}=1.0$}&\multicolumn{2}{c|}{$\Omega_{\circ}=0.1$}\\
                &Entire                 &Subset         &Entire                 &Subset\\
                &Dataset                &A              &Dataset                &B\\
\hline  
$\beta=1.0$     &\ding{51}              &\ding{51}      &\ding{51}              &\ding{51}\\
$\beta=1.5$     &\ding{51}              &{\bf ?}        &\ding{51}              &\ding{51}\\
$\beta=2.0$     &\ding{51}              &\ding{55}      &\ding{51}              &\ding{51}\\
\hline
\end{tabular}
\end{table}

For Subsets A and B, survival analysis finds \luminsub{} to be
correlated with redshift unless the highest values of $\beta$ and
\omegao{} are chosen (\tab{correlationsummary}).  Furthermore,
\luminsub{} is not found to correlate with either \radiopower{} or
\alpharadio{} for any of the tested cosmologies or dust emission
templates.  Even for high values of $\beta$ and \omegao{}, there is
evidence that \luminsub{} is dependent on redshift instead of on the
radio properties of the source: the probability of `no correlation' is
substantially smaller for \luminsub{} and $z$ than for \luminsub{} and
\radiopower{} or \alpharadio{}.

This all suggests that the correlation between \luminsub{} and
redshift for the entire dataset is real, and is not an artifact of
their mutual correlation to \radiopower{} (although this may affect
the strength of the observed \luminsub{}-$z$ correlation).

Given the low detection threshold adopted in this paper, it is worth
re-iterating that a detection threshold of S/N$>3$ (with upper limits
taken at the 3$\sigma$ level) yields the same result.

\subsection{Further investigations of the relationship between \boldmath L$_{850\mu m}$ \unboldmath and redshift}
\label{binningsec}

SCUBA has had more success detecting the higher-redshift galaxies in
the sample.  This is shown in \tab{detectionrate}, which display
the 850-\micron{} detection rate for $z\;<2.5$ and $z\;>2.5$.  At
850$\;$\micron{}, SCUBA is almost equally sensitive to all of the
galaxies in the sample.  The striking result that almost none of the
galaxies are detected if $z<2.5$, whereas almost all of them are
detected if $z>2.5$, suggests a trend of increasing submillimetre
luminosity with redshift.  Note, the detection rates observed for
Subset A and Subset B closely mirror those found for the entire
sample, providing further evidence for the cosmological evolution of
the radio galaxy population irrespective of the radio source.

\begin{table}
\centering
\caption{Detection rate of radio galaxies at 850 \micron{} for $z\;<2.5$ and $z\;>2.5$.  For each dataset and redshift interval, the number of galaxies present is given in parenthesis.  Note, 6C1159+36 may be at a lower redshift than previously published, which would increase the detection rate at $z\;>2.5$.}
\label{detectionrate}
\begin{tabular}{llll}
\hline
                                &&\multicolumn{2}{c}{Detection Rate}\\
Dataset                         &&$z\;<2.5$     &$z\;>2.5$\\
\hline
Subset A                        &&15$\%$ (13)   &63$\%$ (8)\\
Subset B                        &&15$\%$ (13)   &71$\%$ (7)\\
All Data                        &&12$\%$ (33)   &77$\%$ (13)\\
\hline
\end{tabular}
\end{table}

There is an alternative to survival analysis for investigating the
relationship between \luminsub{} and $z$.  Instead of running a
correlation test which only takes the strength of the detection or
upper-limit into account, the 850-\micron{} luminosities can be binned
in redshift to calculate the typical submillimetre luminosity as a
function of redshift.  This method considers both the signal and error
for each source and is independent of the precise choice of detection
threshold.  Note, for the following discussion, \hubble{} has been
taken to be 67$\;$\hubbleunits{} for \omegao{}$\;$=$\;$0.1; this
should give a comparable measure of luminosity to the
\omegao{}$\;$=$\;$1.0, \hubble{}$\;$=$\;$50$\;$\hubbleunits{}
cosmology.

\begin{figure*}
\epsfig{file=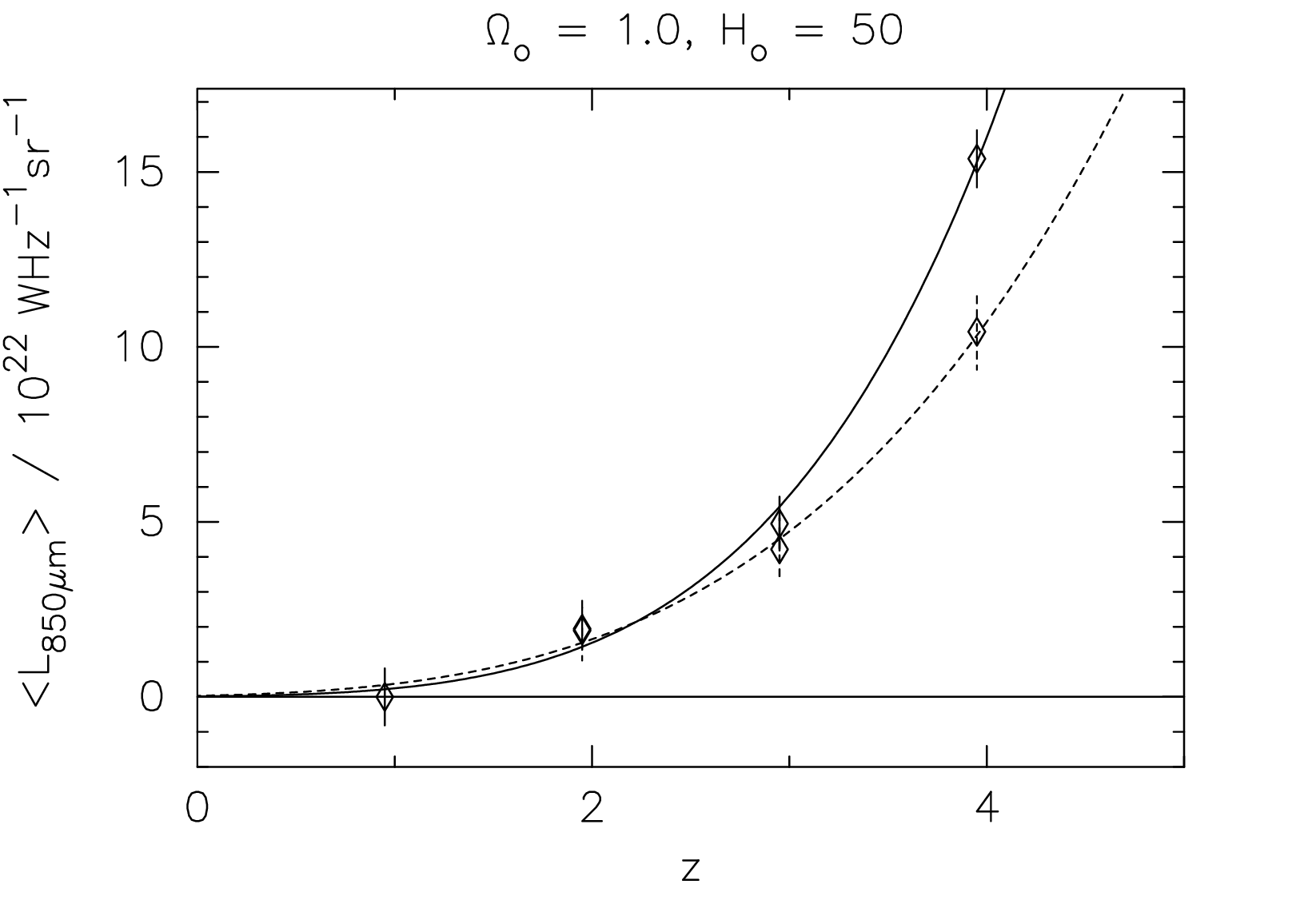,height=9.5cm,angle=270}\vspace*{11pt}
\epsfig{file=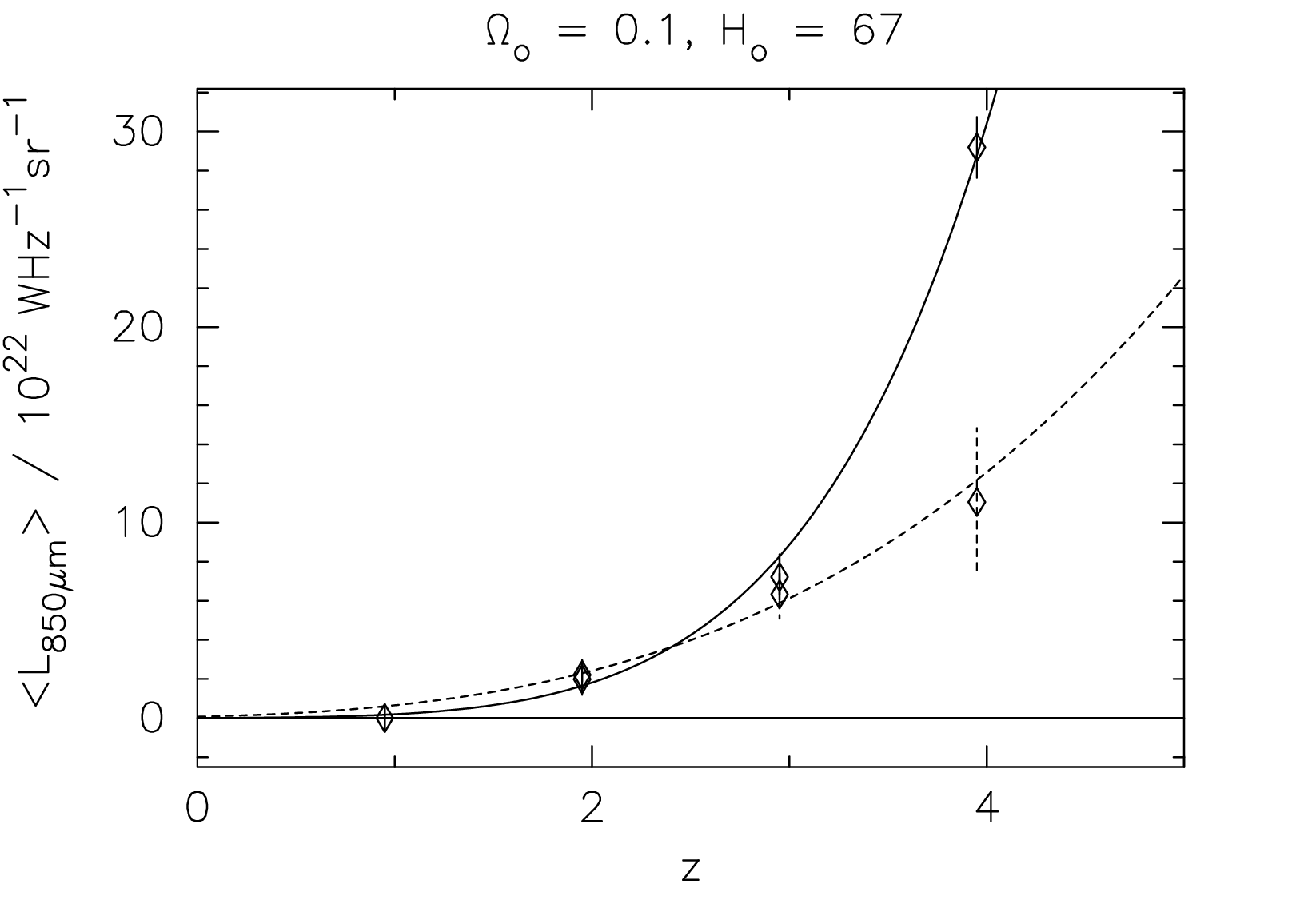,height=9.5cm,angle=270}
\caption{Weighted-mean 850-\micron{} luminosity, $<$\luminsub{}$>$,
binned in redshift.  $<$\luminsub{}$>$ has been plotted at the
midpoint of each bin, which at most varies by 0.2 from the average
redshift of all the objects in the bin.  The error bars only represent
the statistical errors on the observations; they do not account for
uncertainties in either the cosmology or the dust template.  Curves of
the form ${\rm L = a (1+z)^b}$ have been fit to the data using
nonlinear least-squares fitting, and the horizontal line indicates the
origin (or \luminsub{} = 0.0).  The upper plot displays an
\omegao{}$\;$=$\;$1.0, \hubble{}$\;$=$\;$50$\;$\hubbleunits{}
cosmology, and $<$\luminsub{}$>$ is plotted for both the entire
dataset and Subset A (diamond symbols).  The solid curve is the
best-fit for the entire dataset, with ${\rm a = 0.010 \pm 0.008}$,
${\rm b = 4.58 \pm 0.52}$, $\chi^2 = 1.23$.  The dashed curve is the
best-fit for Subset A, with ${\rm a = 0.029 \pm 0.034}$, ${\rm b =
3.67 \pm 0.76}$, $\chi^2 = 0.29$.  Similarly, the lower plot displays
the entire dataset and Subset B (diamond symbols) for an
\omegao{}$\;$=$\;$0.1, \hubble{}$\;$=$\;$67$\;$\hubbleunits{}
cosmology.  The solid curve is the best-fit for the entire dataset,
with ${\rm a = 0.004 \pm 0.003}$, ${\rm b = 5.52 \pm 0.53}$, $\chi^2 =
1.58$.  The dashed curve is the best-fit for Subset B, with ${\rm a =
0.070 \pm 0.099}$, ${\rm b = 3.23 \pm 1.01}$, $\chi^2 = 0.32$.
Subsets A and B do not contain any objects in the $z=1$ bin.}
\label{ombinning}
\end{figure*}

Four redshift bins have been chosen which are adequately sampled by
the dataset and are centered as near as possible on $z =
1,\,2,\,3,\,4$: $0.45\leq z <1.45$, $1.45\leq z <2.45$, $2.45\leq z
<3.45$, and $3.45\leq z <4.45$.  The weighted-mean \cite{bevington}
850-\micron{} luminosity, $<$\luminsub{}$>$, is plotted in
\fig{ombinning} for the entire dataset and Subset A assuming an
\omegao{}$\;$=$\;$1.0 cosmology and for the entire dataset and Subset
B assuming an \omegao{}$\;$=$\;$0.1 cosmology.  Curves of the form
${\rm L = a (1+z)^b}$ have been fit to $<$\luminsub{}$>$ in order to
help quantify the strength of the 850-\micron{} luminosity evolution.

There are several important points to be made from \fig{ombinning}:
\begin{enumerate}
\item For the entire dataset, there are clear steps in $<$\luminsub{}$>$ between each redshift bin.  If this trend is characterised by the function ${\rm L = a (1+z)^b}$, nonlinear least-squares fitting indicates that ${\rm b = 4.58 \pm 0.52}$ if \omegao{}=1.0, and ${\rm b = 5.52 \pm 0.53}$ for \omegao{}=0.1.
\item For Subset A, there are clear steps in $<$\luminsub{}$>$ between the $z=2$ and $z=3$ bins, and between the $z=3$ and $z=4$ bins.  If this trend is characterised by the function ${\rm L = a (1+z)^b}$, nonlinear least-squares fitting indicates that ${\rm b = 3.67 \pm 0.76}$.  Subset A does not contain any objects in the $z=1$ bin.
\item For Subset B, there is a clear step in $<$\luminsub{}$>$ between the $z=2$ and $z=3$ bins.  There may also be an increase in $<$\luminsub{}$>$ from $z=3$ to $z=4$, although the error bars do not preclude $<$\luminsub{}$>$ flattening off between these two redshifts.  If the model ${\rm L = a (1+z)^b}$ is assumed, nonlinear least-squares fitting indicates that ${\rm b = 3.23 \pm 1.01}$.  Subset B does not contain any objects in the $z=1$ bin.
\item For Subsets A and B, the correlation analysis has revealed that \luminsub{} does not depend on the radio properties of the source.  These subsamples are not special, they have been chosen to cover an even spread in radio luminosity over a range of redshifts; their selection has been dictated only by the deficiencies of the original radio surveys.  Thus, the evolution of $<$\luminsub{}$>$ with redshift observed for these two subsamples, which is consistent with ${\rm <L>\sim(1+z)^3}$, should apply to all radio galaxies.  This is supported by the fact that for the $z=2$ and $z=3$ bins, the value of $<$\luminsub{}$>$ for the subset mimics the value of $<$\luminsub{}$>$ for the entire dataset.  At $z\sim4$, however, some of the most extreme objects in the Universe have been observed (e.g. 4C60.07 and 4C41.17), which are simultaneously the most radio-luminous and the most submillimetre-luminous, residing at the highest redshifts.  These galaxies fall outside the subsets which have been constructed to exclude radio-luminosity bias.  Thus at $z\sim4$, $<$\luminsub{}$>$ for the entire sample far exceeds that of the subsets.
\end{enumerate}

As for the correlation analysis, it is important to investigate the
robustness of this result to variations in the dust template.  Thus,
$\beta$ was changed to 1.0 and 2.0, whilst ${\rm T_{dust}}$
was fixed at 40K.  Likewise, holding $\beta$ constant at 1.5, ${\rm
T_{dust}}$ was changed to 20K, 70K, and 100K.  Altering the dust
template has a larger effect on the higher redshift sources.  Thus, if
$\beta$ and ${\rm T_{dust}}$ are increased, the trend of increasing
$<$\luminsub{}$>$ with redshift is weakened; if they are decreased the
trend is strengthened.  More specifically, it was found that:
\begin{enumerate}
\item For the entire dataset, there are always clear steps in $<$\luminsub{}$>$ between all redshift bins, regardless of the values of $\beta$ and ${\rm T_{dust}}$.
\item For Subset A, there is always a clear step in $<$\luminsub{}$>$ between $z=2$ and $z=3$ and between $z=3$ and $z=4$, regardless of the values of $\beta$ and ${\rm T_{dust}}$.
\item For Subset B, there is always a clear step in $<$\luminsub{}$>$ between $z=2$ and $z=3$.  The possible step between $z=3$ and $z=4$ becomes clear (i.e. the error bars do not allow $<$\luminsub{}$>$ to remain flat) if $\beta$ is decreased to 1.0 or if ${\rm T_{dust}}$ is decreased to 20K.
\item For the $z=2$ and $z=3$ bins, $<$\luminsub{}$>$ for the entire dataset always closely mimics $<$\luminsub{}$>$ for the subsets, regardless of the values of $\beta$ and ${\rm T_{dust}}$.
\item Adopting the model ${\rm L = a (1+z)^b}$, the luminosity evolution observed For Subsets A and B upon changing $\beta$ and ${\rm T_{dust}}$ is consistent with `b' lying between 2.5 and 5.0.
\end{enumerate}

Binning the data has been able to confirm the relationship between
\luminsub{} and $z$ that was initially suggested by survival analysis.
It is therefore possible that binning may also reveal a relationship
between \luminsub{} and \radiopower{} or \alpharadio{}.

It is only necessary to check this for Subsets A and B, as \luminsub{}
is known to be strongly correlated with both \radiopower{} and
\alpharadio{} for the entire dataset.  If \luminsub{} remains
independent of the radio source for Subsets A and B, the evidence for
evolution with redshift will be strengthened.

Binning \luminsub{} against radio power and radio spectral index fails
to reveal a dependence of \luminsub{} on the radio source.  Altering
the dust template does not change this result.

\subsection{Addressing potential remaining selection effects}

\subsubsection{Potential contamination by obscured quasars}

Careful scrutiny of the P-$z$ plane defined by combining all the known
high-redshift radio sources shows that there is a lack of
high-redshift, high-radio power quasars, although this is at least
partly due to the selection effects inherent in
steep-spectrum-selected radio samples (Jarvis et al., in
preparation\nocite{jarvisinprep}).  This is also the area of the P-$z$
plane where the 850-\micron{} detections of the radio-galaxy sample
are clustered.

According to Unified Schemes, radio-loud quasars and radio galaxies
are identical phenomena: the former is viewed within the opening angle
of an obscuring dust torus allowing the quasar to be seen, the latter
is viewed outside the opening angle of the dust torus which hides the
quasar from view (e.g. Antonucci \& Barvainis 1990\nocite{ab90} and references
therein).

It can be argued that at higher redshifts, there is an increased
chance of dust hiding the optically-bright nucleus of a radio-loud
quasar, even though the radio-loud quasar is viewed within the opening
angle of the dust torus. Thus, potentially, several high-redshift radio-loud
quasars may have been mis-classified as radio galaxies. Therefore, the
worry for the present study is that our radio galaxy sample could 
become progressively more contaminated by obscured quasars with increasing 
redshift.

Of course, within unified schemes all powerful radio galaxies contain
a buried quasar.  Thus the point of issue for the present study is
whether the submillimetre emission of a powerful radio source is
orientation dependent.  If so, and if our sample was progressively
contaminated by mis-classified radio-loud quasars with redshift, it
could account for the observed increase of \luminsub{} with redshift.
Given the above considerations, this might happen because, for
example, some of our high-redshift objects might be observed in a line
of sight sufficiently close to the jet axis that the contribution from
a more face-on quasar heated torus (which may be optically thick in
the rest-frame mid-far infrared) becomes dominant at 850$\mu m$.

One way to resolve this issue is to check directly whether unification
works at submillimetre wavelengths, and we are currently undertaking a
programme of SCUBA observations of radio-loud quasars to test this
hypothesis. However, even if radio-loud quasars and radio galaxies are
shown to differ at submillimetre wavelengths this need not invalidate 
the conclusions of this study, provided we can demonstrate that
progressive quasar contamination with increasing redshift is not a problem.

\begin{figure}
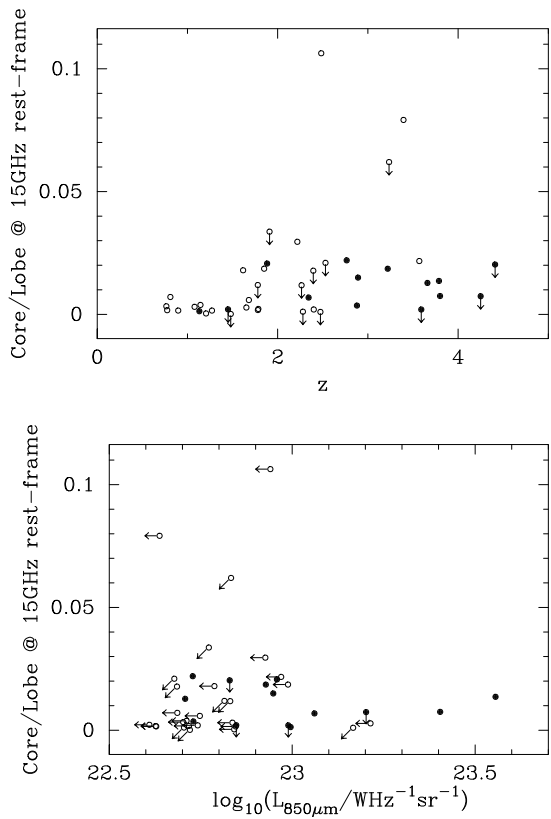

\centering
\epsfig{file=fig8a.eps,height=8cm,angle=270}
\epsfig{file=fig8b.eps,height=8cm,angle=270}
\caption{Investigation of the rest-frame 15$\;$GHz core:lobe ratio for the objects in our sample.  The top plot shows the core:lobe ratio against redshift.  Solid circles represent the sources detected with SCUBA, and open circles represent the non-detections.  Down arrows indicate sources for which no radio core has been detected.  In order to estimate the core flux at 15$\;$GHz rest-frame, the core spectral index is required.  For sources with more than one core detection, we have measured this spectral index.  For sources with only one core detection, we have assumed the core to be flat.  The bottom plot compares the core:lobe ratio with the 850-\micron{} luminosity.  An \omegao{}=1.0 cosmology is assumed; there is no significant difference between this plot and one for which \omegao{}=0.1.}
\label{corelobe}
\end{figure}

We can in fact test the likelihood of this
selection effect directly using the basic radio properties 
of the objects we have observed with SCUBA.  In \Fig{corelobe}, the rest-frame
15$\;$GHz core:lobe ratio is plotted against redshift and \luminsub{}
for our sample.  If the high-redshift objects in the sample are in fact
buried radio-loud quasars masquerading as galaxies, they should be
expected to display systematically higher radio
core:lobe ratios, owing to Doppler boosting of the radio core along
the line of sight.  The typical core:lobe ratio does not appear to
change significantly with redshift, and so this analysis does not
support the idea of a systematic bias towards buried quasars at high
redshift within the SCUBA radio galaxy sample studied here.  Moreover,
there is no evidence that the more core dominated objects are more
detectable with SCUBA.

These mis-classified buried quasars may exist.  Two of the objects in
our sample, 6C0902+34 and 4C23.56, appear significantly more
core-dominated than the rest, and are likely candidates since their
core:lobe radio is comparable ($\simeq 10$ per cent) to that typically
displayed by radio quasars \cite{spencer91}.  Furthermore, 6C0902+34
has a single-sided radio jet characteristic of a radio-loud quasar
\cite{carilli95}.  In addition to the evidence at radio wavelengths,
4C23.56 has a polarized ultraviolet continuum consistent with
scattered quasar light \cite{cdv98}.  It is worth noting that the
core:lobe ratio in these two sources is an order of magnitude higher
than the typical value throughout the rest of the radio galaxy
sample. Moreover, in any case, SCUBA failed to detect dust emission
from either one.

\begin{figure}
\centering
\epsfig{file=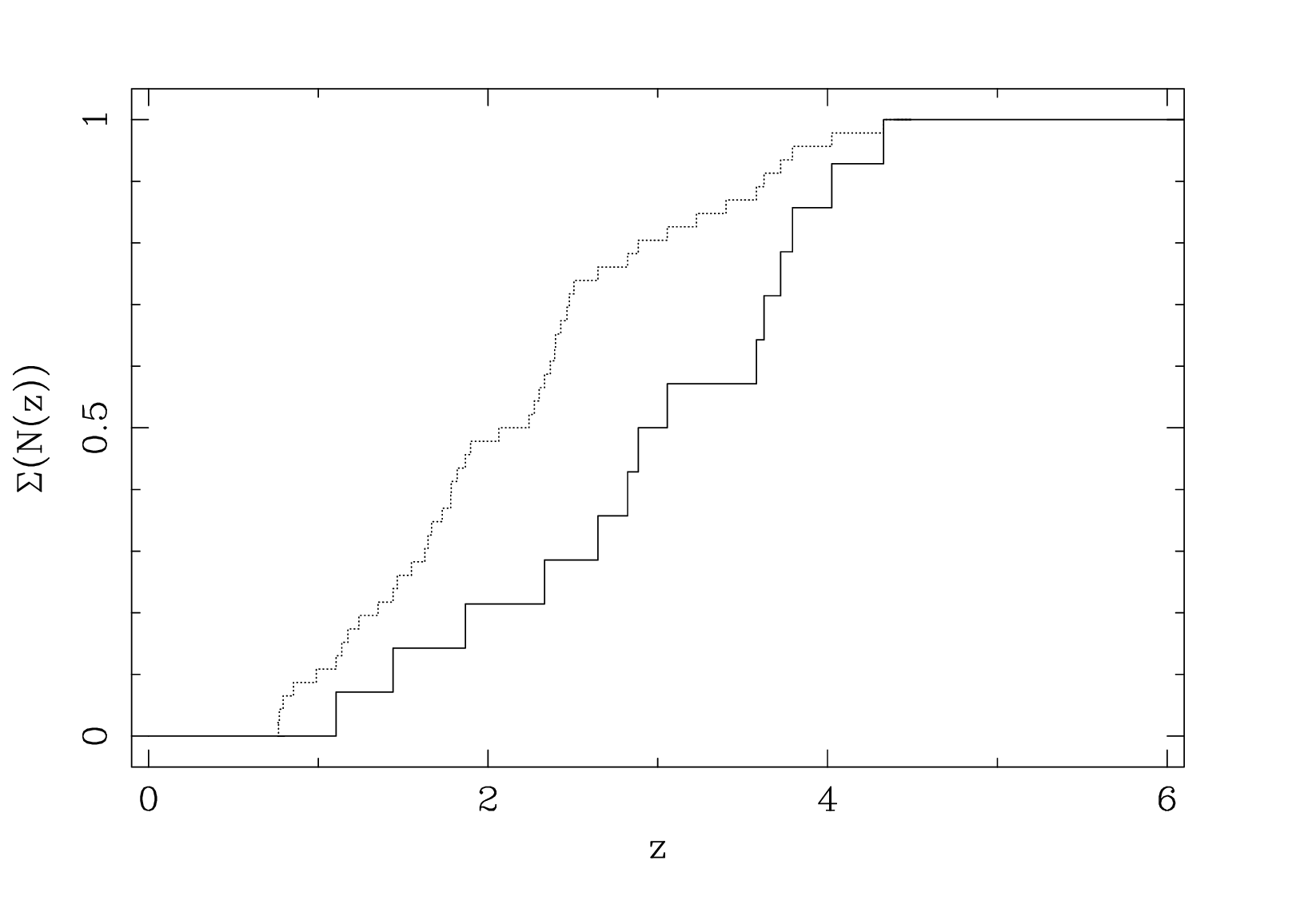,height=9cm,angle=270}
\caption{\footnotesize The cumulative redshift distribution of the radio galaxy sample.  The dotted line is for the entire sample, the solid line only considers the galaxies detected at 850$\;$\micron{}.  For the detections, the median redshift of the distribution is $<z>=3.1$.}
\label{cumulrgzdistrib}
\epsfig{file=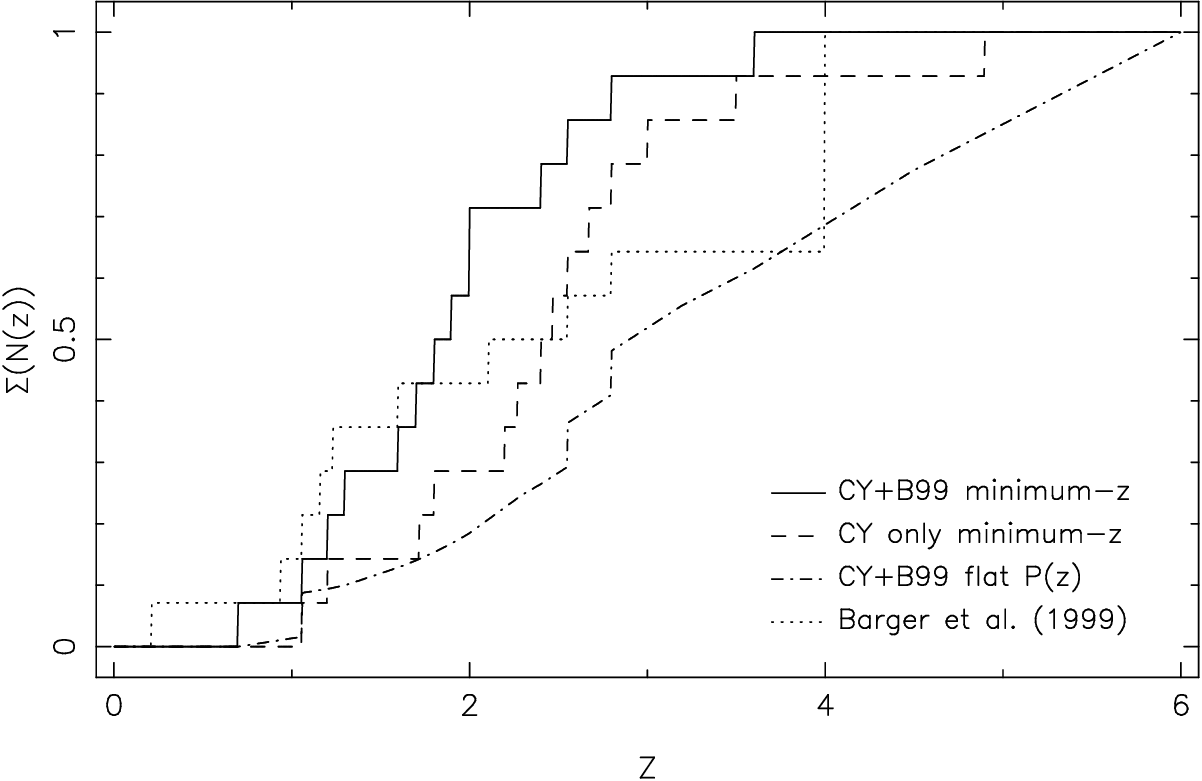,height=8cm,angle=270}
\caption{\footnotesize Figure 2 of Smail et al. 2000b.  The full caption, as it appears in the original paper, is as follows: ``Cumulative redshift distribution for the full submm sample.  We have used the spectroscopic redshifts of those sources thought to be reliable (Table 1) and combined these with the probable redshift ranges of the remaining sources derived from their $\alpha^{850}_{1.4}$ indices or limits.  The solid line shows the cumulative distribution if we assume the minimum redshift distribution that is obtained if all sources lie at their lower $z_{\alpha}$ limit given in Table 1 (dashed line is the equivalent analysis but restricted to the CY models).  The effect of non-thermal radio emission, which drives down the $\alpha^{850}_{1.4}$ indices, means that this is a very conservative assumption if some fraction of the population harbor radio-loud AGNs.  The dash-dotted line assumes a flat probability distribution for the sources within their $z_{\alpha}$ ranges and a maximum redshift of $z=6$ for those sources where we only have a lower limit on $\alpha^{850}_{1.4}$.  Finally, the dotted line is the cumulative redshift distribution from Barger et al. (1999a) with two of the source identifications corrected as in Smail et al. (1999) and all blank-field/extremely red object candidates placed at $z=4$.''} 
\label{smailz}
\end{figure}

\subsubsection{Preferential selection of young radio sources at high redshift}

All realistic models of the time evolution of classical double radio
sources (e.g. Kaiser et al. 1997\nocite{kda97}; Blundell et
al. 1999\nocite{brw99}; Blundell \& Rawlings 1999\nocite{bs99})
predict that their lobe luminosities decline with age once they are
large enough to avoid synchrotron self-absorption.  Older sources are
thus (obviously) more likely to fall below the flux limit of a given
survey if they reside at high redshift. A single flux-limited survey
may thus be biased towards young radio sources at high redshift,
whereas at lower redshifts a large range of radio source ages will be
included in the same survey.

The importance of this selection effect to our SCUBA survey is
probably not a major concern for two reasons.  First, the target
sources are {\em not} selected from a single flux-limited sample, but
rather are selected rather close to the flux-density limit of
progressively deeper surveys - in other words at $z \simeq 4$ the
radio sources must be preferentially young, but the sources we have
studied at, say, $z \simeq 1.5$ must also be fairly young to enter the
brighter 3CRR sample from which our lower-redshift targets are
selected. This fact will undoubtedly weaken any correlation of source
age with redshift (certainly at $ z > 1$), compared with that expected
from a single flux-limited survey as described by Blundell \& Rawlings
(1999).

Second, even if some residual correlation between radio-source
youthfulness and redshift remains in our sample, it is not clear that
this has any implications whatsoever for this submillimetre study. For
it to be responsible for the trend of increasing \luminsub{} with $z$,
radio-source age must be intimately linked to starburst luminosity.
Our data do not provide any evidence in support of this.  For example,
one of the largest radio sources, and hence possibly one of the
oldest, 6C0905+39, was detected by SCUBA.  To examine this more
thoroughly, survival analysis was applied to test for correlations
between D$_{linear}$ (taken as indicative of the age of the radio
source), $z$, and \luminsub{}.  As found for complete samples by
Blundell et al. (1999), we find a negative correlation between
D$_{linear}$ and $z$.  However, no evidence for a correlation between
D$_{linear}$ and \luminsub{} was found (correlation tests yield a
significance of $\sim 40$ per cent). We therefore find no internal
evidence in support of the hypothesis that \luminsub{} is closely
linked to radio-source age, although the mildness of the
D$_{linear}$:$z$ correlation means that caution is advised in
attempting to interpret D$_{linear}$ as a reliable estimator of
radio-source age \cite{brw99,bs99}.

\section{Concluding Remarks}
\label{concsec}

In summary, our attempts to investigate and quantify the various
possible biases which might conceivably afflict this study have simply
served to reaffirm and strengthen our basic result, namely that the
submillimetre luminosity of radio galaxies is primarily a function of
redshift as illustrated in \Fig{ombinning}.

It therefore seems hard to avoid the straightforward conclusion that
the observed increase in submillimetre detection rate and
characteristic luminosity with redshift is due to the increasing
youthfulness of the stellar populations of the radio galaxies in our
sample.

In a separate paper we will explore how the inferred evolution of gas
mass and star-formation rate in these galaxies compares with the
predictions of models of elliptical galaxy formation and evolution.
However, it is interesting to briefly consider whether the apparently
rather extreme evolution is peculiar to radio galaxies, or may in fact
be typical of the cosmological evolution of dust and gas in massive
ellipticals in general.

In \Fig{cumulrgzdistrib} we show the cumulative redshift distribution
of our radio galaxy sample, along with the cumulative redshift
distribution of the subset detected at submillimetre wavelengths. This
figure serves to re-emphasize that the high median redshift of our
detected galaxies ($z = 3$) does not simply reflect the median
redshift of the sample selected for observation ($z = 2$). However,
what is particularly interesting is that the redshift distribution of
our submillimetre detected radio galaxies is statistically
indistinguishable from current best estimates of the redshift
distribution of sources detected in SCUBA surveys, as illustrated in
\Fig{smailz} \cite{smailzdistrib}.  This comparison provides at least
circumstantial evidence that the submillimetre evolution of radio
galaxies found here may indeed be symptomatic of the evolution of
massive elliptical galaxies in general.

At first sight, it may appear contradictory that the average
submillimetre luminosity of the radio galaxy sample continues to rise
beyond $z\sim3$ while the median redshift of the sample is $z\sim3$.
This results from the median redshift of our most luminous detections
being higher at $z\sim3.6$ for sources brighter than 5$\;$mJy.

This raises the interesting possibility that the most massive
dust-enshrouded starbursts are confined to $z>3$.  Therefore the
median redshift of submillimetre sources detected in blank-field
surveys may prove to be a function of flux density.

It will be some time before the redshift information in bright
submillimetre surveys approaches that currently available for
radio-selected samples. However, it will undoubtedly be very
interesting to see how this comparison evolves as
submillimetre-selected samples are studied and refined in the years to
come.

\section*{Acknowledgments}

We wish to thank the staff of the James Clerk Maxwell Telescope,
particularly Wayne Holland, Tim Jenness, Graeme Watt and all the
T.S.S.s, for all their help.  Many thanks to Katherine Blundell for
allowing us access to radio data prior to publication.  Figure 2 of
Smail et al. 2000b, and the corresponding caption text, have been
reproduced by permission of the AAS.  This research has made use of
the NASA/IPAC Extragalactic Database, which is operated by the Jet
Propulsion Laboratory, Caltech, under contract with the National
Aeronautics and Space Administration.  ENA, DHH and RJI acknowledge
support from the UK Particle Physics and Astronomy Research Council
(PPARC).  The JCMT is operated by the Joint Astronomy Centre, on
behalf of PPARC, the Netherlands Organisation for Pure Research, and
the National Research Council of Canada.

\nocite{wthesis}
\nocite{RvO97}
\nocite{mprep}
\nocite{l88}
\nocite{lpcomm}
\nocite{lr94}
\nocite{er96}
\nocite{erlg97}
\nocite{erd93}
\nocite{dsp87}
\nocite{fbp97}
\nocite{msvb95}
\nocite{m82paper}
\nocite{bunkerthesis}
\nocite{mcc97}
\nocite{rmc95}
\nocite{rew90}
\nocite{rll96}
\nocite{rs97}
\nocite{sj79}
\nocite{spriv}
\nocite{sunpub}
\nocite{S82}
\nocite{SD84a}
\nocite{SD84b}
\nocite{SDM85}
\nocite{SDG95}
\nocite{SRS90}


\begin{thebibliography}{}

\bibitem[\protect\citename{{Akujor} et~al.{\ }}{1991}]{asz91}
{Akujor}~C.~E.,  {Spencer}~R.~E.,  {Zhang}~F.~J.,  {Davis}~R.~J.,  {Browne}~I.
  W.~A.,    {Fanti}~C.,  1991, \mnras, 250, 215, (ASZ91)

\bibitem[\protect\citename{{Andreani} et~al.{\ }}{1993}]{alc93}
{Andreani}~P.,  {La Franca}~F.,    {Cristiani}~S.,  1993, \mnras, 261, L35

\bibitem[\protect\citename{{Antonucci} \& {Barvainis}{\ }}{1990}]{ab90}
{Antonucci}~R.,  {Barvainis}~R.,  1990, \apjl, 363, L17

\bibitem[\protect\citename{{Archibald}{\ }}{1999}]{mythesis}
{Archibald}~E.~N.,  1999, The cosmological evolution of dust and gas in radio
  galaxies, Ph.D. thesis, University of Edinburgh

\bibitem[\protect\citename{{Barger} et~al.{\ }}{1999}]{bargercs99}
{Barger}~A.~J.,  {Cowie}~L.~L.,    {Sanders}~D.~B.,  1999, \apjl, 518, L5

\bibitem[\protect\citename{{Becker} et~al.{\ }}{1991}]{bwe91}
{Becker}~R.~H.,  {White}~R.~L.,    {Edwards}~A.~L.,  1991, \apjs, 75, 1,
  (BWE91)

\bibitem[\protect\citename{{Benford} et~al.{\ }}{1999}]{bco99}
{Benford}~D.~J.,  {Cox}~P.,  {Omont}~A.,  {Phillips}~T.~G.,    {McMahon}~R.~G.,
   1999, \apjl, 518, L65, (BCO99)

\bibitem[\protect\citename{{Bennett}{\ }}{1962}]{bennett1962}
{Bennett}~A.~S.,  1962, \memras, 68, 163

\bibitem[\protect\citename{{Best} et~al.{\ }}{1997}]{blr97}
{Best}~P.~N.,  {Longair}~M.~S.,    {R\"ottgering}~H. J.~A.,  1997, \mnras, 292,
  758, (BLR97)

\bibitem[\protect\citename{{Best} et~al.{\ }}{1998a}]{bcg98}
{Best}~P.~N.,  {Carilli}~C.~L.,  {Garrington}~S.~T.,  {Longair}~M.~S.,
  {R\"ottgering}~H. J.~A.,  1998a, \mnras, 299, 357, (BCG98)

\bibitem[\protect\citename{{Best} et~al.{\ }}{1998b}]{brb98}
{Best}~P.~N. et al.,  1998b, \mnras, 301, L15, (BRB98)

\bibitem[\protect\citename{{Bevington}{\ }}{1969}]{bevington}
{Bevington}~P.~R.,  1969, Data Reduction and Error Analysis for the Physical
  Sciences.
McGraw-Hill

\bibitem[\protect\citename{{Blain} et~al.{\ }}{1998}]{blainconfusion}
{Blain}~A.~W.,  {Ivison}~R.~J.,  {Smail}~I.,    {Kneib}~J.-P.,  1998, in
  Colombi~S.,  Mellier~Y.,  eds, Wide Field Surveys in Cosmology, 14th IAP
  meeting held May 26-30, 1998, Paris. Publisher: Editions Frontieres.
p.~364

\bibitem[\protect\citename{{Blundell} \& {Rawlings}{\ }}{1999}]{bs99}
{Blundell}~K.~M.,  {Rawlings}~S.,  1999, \nat, 399, 330

\bibitem[\protect\citename{{Blundell} et~al.{\ }}{1998}]{bre98}
{Blundell}~K.~M.,  {Rawlings}~S.,  {Eales}~S.~A.,  {Taylor}~G.~B.,
  {Bradley}~A.~D.,  1998, \mnras, 295, 265, (BRE98)

\bibitem[\protect\citename{{Blundell} et~al.{\ }}{1999}]{brw99}
{Blundell}~K.~M.,  {Rawlings}~S.,    {Willott}~C.~J.,  1999, \aj, 117, 677

\bibitem[\protect\citename{{Blundell} et~al.{\ }}{2000}]{brr00}
{Blundell}~K.~M.,  {Rawlings}~S.,  {Riley}~J.~M.,  {Willott}~C.~J.,
  {Laing}~R.~A.,  2000, New Astronomy, to be submitted, (BRR00)

\bibitem[\protect\citename{{Blundell} \& et~al.{\ }}{}]{kb17}
{Blundell}~K.~M. et al., in preparation, (KB17)


\bibitem[\protect\citename{{Bunker}{\ }}{1996}]{bunkerthesis}
{Bunker}~A.~J.,  1996, Searches for distant galaxies, Ph.D. thesis, University
  of Oxford, (Bthesis)

\bibitem[\protect\citename{{Carilli}{\ }}{1995a}]{carilli95}
{Carilli}~C.~L.,  1995a, \aap, 298, 77

\bibitem[\protect\citename{{Carilli}{\ }}{1995b}]{c95}
{Carilli}~C.~L.,  1995b, \aap, 298, 77, (C95)

\bibitem[\protect\citename{{Carilli} et~al.{\ }}{1994}]{coh94}
{Carilli}~C.~L.,  {Owen}~F.~N.,    {Harris}~D.~E.,  1994, \aj, 107, 480,
  (COH94)

\bibitem[\protect\citename{{Carilli} et~al.{\ }}{1997}]{crvo97}
{Carilli}~C.~L.,  {R\"ottgering}~H. J.~A.,  {van Ojik}~R.,  {Miley}~G.~K.,
  {van Breugel}~W. J.~M.,  1997, \apjs, 109, 1, (CRvO97)

\bibitem[\protect\citename{{Caswell} \& {Crowther}{\ }}{1969}]{cc69}
{Caswell}~J.~L.,  {Crowther}~J.~H.,  1969, \mnras, 145, 181, (C\&C69)

\bibitem[\protect\citename{{Chambers} et~al.{\ }}{1988}]{cmvb88}
{Chambers}~K.~C.,  {Miley}~G.~K.,    {van Breugel}~W. J.~M.,  1988, \apjl, 327,
  L47, (CMvB88)

\bibitem[\protect\citename{{Chambers} et~al.{\ }}{1990}]{cmvb90}
{Chambers}~K.~C.,  {Miley}~G.~K.,    {van Breugel}~W. J.~M.,  1990, \apj, 363,
  21, (CMvB90)

\bibitem[\protect\citename{{Chambers} et~al.{\ }}{1996}]{cmvb96}
{Chambers}~K.~C.,  {Miley}~G.~K.,  {van Breugel}~W. J.~M.,  {Bremer}~M. A.~R.,
  {Huang}~J.-S.,    {Trentham}~N.~A.,  1996, \apjs, 106, 247, (CMvB96)

\bibitem[\protect\citename{{Chini} \& {Kr\"ugel}{\ }}{1994}]{ck94}
{Chini}~R.,  {Kr\"ugel}~E.,  1994, \aap, 288, L33, (C\&K94)

\bibitem[\protect\citename{{Chini} et~al.{\ }}{1986}]{ckk86}
{Chini}~R.,  {Kr\"ugel}~E.,    {Kreysa}~E.,  1986, \aap, 167, 315

\bibitem[\protect\citename{{Cimatti} et~al.{\ }}{1998a}]{cdv98}
{Cimatti}~A.,  {di Serego Alighieri}~S.,  {Vernet}~J.,  {Cohen}~M.~H.,
  {Fosbury}~R. A.~E.,  1998a, \apjl, 499, L21

\bibitem[\protect\citename{{Cimatti} et~al.{\ }}{1998b}]{cfr98}
{Cimatti}~A.,  {Freudling}~W.,  {R\"ottgering}~H. J.~A.,  {Ivison}~R.~J.,
  {Mazzei}~P.,  1998b, \aap, 329, 399, (CFR98)

\bibitem[\protect\citename{{Condon} et~al.{\ }}{1998}]{cc98}
{Condon}~J.~J.,  {Cotton}~W.~D.,  {Greisen}~E.~W.,  {Yin}~Q.~F.,
  {Perley}~R.~A.,  {Taylor}~G.~B.,    {Broderick}~J.~J.,  1998, \aj, 115, 1693,
  (CC98)

\bibitem[\protect\citename{{Dey} et~al.{\ }}{1995}]{dsd95}
{Dey}~A.,  {Spinrad}~H.,    {Dickinson}~M.,  1995, \apj, 440, 515, (DSD95)

\bibitem[\protect\citename{{Djorgovski} et~al.{\ }}{1987}]{dsp87}
{Djorgovski}~S.,  {Spinrad}~H.,  {Pedelty}~J.,  {Rudnick}~L.,    {Stockton}~A.,
   1987, \aj, 93, 1307, (DSP87)

\bibitem[\protect\citename{{Douglas} et~al.{\ }}{1996}]{dbb96}
{Douglas}~J.~N.,  {Bash}~F.~N.,  {Bozyan}~F.~A.,  {Torrence}~G.~W.,
  {Wolfe}~C.,  1996, \aj, 111, 1945, (DBB96)

\bibitem[\protect\citename{{Downes} et~al.{\ }}{1996}]{dss96}
{Downes}~D.,  {Solomon}~P.~M.,  {Sanders}~D.~B.,    {Evans}~A.~S.,  1996, \aap,
  313, 91, (DSS96)

\bibitem[\protect\citename{{Draine} \& {Lee}{\ }}{1984}]{draineandlee}
{Draine}~B.~T.,  {Lee}~H.~M.,  1984, \apj, 285, 89

\bibitem[\protect\citename{{Dunlop}{\ }}{1999}]{dunlophy}
{Dunlop}~J.~S.,  1999, in Bunker~A.~J.,  van Breugel~W. J.~M.,  eds, ASP Conf.
  Ser. 193: The Hy-Redshift Universe: Galaxy Formation and Evolution at High
  Redshift.
p.~133

\bibitem[\protect\citename{{Dunlop} \& et~al.{\ }}{2000}]{dunlopetal2000}
{Dunlop}~J.~S.,  et~al. 2000, MNRAS, in preparation

\bibitem[\protect\citename{{Dunlop} et~al.{\ }}{1994}]{dhr94}
{Dunlop}~J.~S.,  {Hughes}~D.~H.,  {Rawlings}~S.,  {Eales}~S.~A.,
  {Ward}~M.~J.,  1994, \nat, 370, 347, (DHR94)

\bibitem[\protect\citename{{Dunne} et~al.{\ }}{2000}]{dunne2000}
{Dunne}~L.,  {Eales}~S.,  {Edmunds}~M.,  {Ivison}~R.,  {Alexander}~P.,
  {Clements}~D.~L.,  2000, \mnras, 315, 115



\bibitem[\protect\citename{{Eales}{\ }}{1985}]{eales85}
{Eales}~S.~A.,  1985, \mnras, 217, 149

\bibitem[\protect\citename{{Eales} \& {Edmunds}{\ }}{1996}]{ee96}
{Eales}~S.~A.,  {Edmunds}~M.~G.,  1996, \mnras, 280, 1167

\bibitem[\protect\citename{{Eales} \& {Rawlings}{\ }}{1993}]{er93}
{Eales}~S.~A.,  {Rawlings}~S.,  1993, \apj, 411, 67, (ER93)

\bibitem[\protect\citename{{Eales} \& {Rawlings}{\ }}{1996}]{er96}
{Eales}~S.~A.,  {Rawlings}~S.,  1996, \apj, 460, 68, (ER96)

\bibitem[\protect\citename{{Eales} et~al.{\ }}{1989}]{ead89}
{Eales}~S.~A.,  {Alexander}~P.,    {Duncan}~W.~D.,  1989, \mnras, 240, 817

\bibitem[\protect\citename{{Eales} et~al.{\ }}{1993}]{erd93}
{Eales}~S.~A.,  {Rawlings}~S.,  {Dickinson}~M.,  {Spinrad}~H.,  {Hill}~G.~J.,
   {Lacy}~M.,  1993, \apj, 409, 578, (ERD93)

\bibitem[\protect\citename{{Eales} et~al.{\ }}{1997}]{erlg97}
{Eales}~S.,  {Rawlings}~S.,  {Law-Green}~D.,  {Gotter}~G.,    {Lacy}~M.,  
1997,  \mnras, 291, 593, (ERLG97)

\bibitem[\protect\citename{{Feigelson} \& {Nelson}{\ }}{1985}]{fn85}
{Feigelson}~E.~D.,  {Nelson}~P.~I.,  1985, \apj, 293, 192

\bibitem[\protect\citename{{Fernini} et~al.{\ }}{1993}]{f93}
{Fernini}~I.,  {Burns}~J.~O.,  {Bridle}~A.~H.,    {Perley}~R.~A.,  1993, \aj,
  105, 1690, (F93)

\bibitem[\protect\citename{{Fernini} et~al.{\ }}{1997}]{fbp97}
{Fernini}~I.,  {Burns}~J.~O.,    {Perley}~R.~A.,  1997, \aj, 114, 2292, (FBP97)

\bibitem[\protect\citename{{Ficarra} et~al.{\ }}{1985}]{fgt85}
{Ficarra}~A.,  {Grueff}~G.,    {Tomassetti}~G.,  1985, \aaps, 59, 255, (FGT85)

\bibitem[\protect\citename{{Frayer} \& {Brown}{\ }}{1997}]{frayerbrown97}
{Frayer}~D.~T.,  {Brown}~R.~L.,  1997, \apjs, 113, 221

\bibitem[\protect\citename{{Gower} et~al.{\ }}{1967}]{gsw67}
{Gower}~J. F.~R.,  {Scott}~P.~F.,    {Wills}~D.,  1967, \memras, 71, 49,
  (GSW67)

\bibitem[\protect\citename{{Gregory} \& {Condon}{\ }}{1991}]{gc91}
{Gregory}~P.~C.,  {Condon}~J.~J.,  1991, \apjs, 75, 1011, (GC91)

\bibitem[\protect\citename{{Gregory} et~al.{\ }}{1996}]{gsdc96}
{Gregory}~P.~C.,  {Scott}~W.~K.,  {Douglas}~K.,    {Condon}~J.~J.,  1996,
  \apjs, 103, 427, (GSDC96)

\bibitem[\protect\citename{{Griffith} et~al.{\ }}{1995}]{gwbe95}
{Griffith}~M.~R.,  {Wright}~A.~E.,  {Burke}~B.~F.,    {Ekers}~R.~D.,  1995,
  \apjs, 97, 347, (GWBE95)

\bibitem[\protect\citename{{Hales} et~al.{\ }}{1988}]{6cII}
{Hales}~S. E.~G.,  {Baldwin}~J.~E.,    {Warner}~P.~J.,  1988, \mnras, 234,
919, (6C-II)

\bibitem[\protect\citename{{Hales} et~al.{\ }}{1990}]{6cIII}
{Hales}~S. E.~G.,  {Masson}~C.~R.,  {Warner}~P.~J.,    {Baldwin}~J.~E.,  1990,
  \mnras, 246, 256, (6C-III)

\bibitem[\protect\citename{{Hales} et~al.{\ }}{1993a}]{6cVI}
{Hales}~S. E.~G.,  {Baldwin}~J.~E.,    {Warner}~P.~J.,  1993a, \mnras, 263,
25, (6C-VI)

\bibitem[\protect\citename{{Hales} et~al.{\ }}{1993b}]{6cV}
{Hales}~S. E.~G.,  {Masson}~C.~R.,  {Warner}~P.~J.,  {Baldwin}~J.~E.,
  {Green}~D.~A.,  1993b, \mnras, 262, 1057, (6C-V)

\bibitem[\protect\citename{{Hales} et~al.{\ }}{1995}]{hwr95}
{Hales}~S. E.~G.,  {Waldram}~E.~M.,  {Rees}~N.,    {Warner}~P.~J.,  1995,
  \mnras, 274, 447, (HWR95)

\bibitem[\protect\citename{{Helou} \& {Beichman}{\ }}{1990}]{hb90}
{Helou}~G.,  {Beichman}~C.~A.,  1990, in Kaldeich~B.,  ed., From Ground-Based
  to Space-Borne Sub-mm Astronomy, ESA SP-314.
p.~117

\bibitem[\protect\citename{{Hildebrand}{\ }}{1983}]{hildebrand}
{Hildebrand}~R.~H.,  1983, \qjras, 24, 267

\bibitem[\protect\citename{{Holland} et~al.{\ }}{1999}]{scubapaper}
{Holland}~W.~S. et al.,  1999, \mnras, 303, 659

\bibitem[\protect\citename{{Hughes} et~al.{\ }}{1994}]{m82paper}
{Hughes}~D.~H.,  {Gear}~W.~K.,    {Robson}~E.~I.,  1994, \mnras, 270, 641

\bibitem[\protect\citename{{Hughes} et~al.{\ }}{1997}]{hdr97}
{Hughes}~D.~H.,  {Dunlop}~J.~S.,    {Rawlings}~S.,  1997, \mnras, 289, 766,
  (HDR97)

\bibitem[\protect\citename{{Hughes} et~al.{\ }}{1998}]{hdfnature}
{Hughes}~D.~H. et al.,  1998, \nat, 394, 241

\bibitem[\protect\citename{{Isobe} \& {Feigelson}{\ }}{1990}]{if90}
{Isobe}~T.,  {Feigelson}~E.~D.,  1990, BAAS, 22, 917

\bibitem[\protect\citename{{Isobe} et~al.{\ }}{1986}]{ifn86}
{Isobe}~T.,  {Feigelson}~E.~D.,    {Nelson}~P.~I.,  1986, \apj, 306, 490

\bibitem[\protect\citename{{Ivison}{\ }}{1995}]{i95}
{Ivison}~R.~J.,  1995, \mnras, 275, L33, (I95)

\bibitem[\protect\citename{{Ivison} et~al.{\ }}{1998a}]{idha98}
{Ivison}~R.~J. et al.,  1998a, \apj, 494, 211, (IDHA98)

\bibitem[\protect\citename{{Ivison} et~al.{\ }}{1998b}]{islb98}
{Ivison}~R.~J.,  {Smail}~I.,  {Le Borgne}~J.-F.,  {Blain}~A.~W.,
  {Kneib}~J.-P.,  {B\'ezecourt}~J.,  {Kerr}~T.~H.,    {Davies}~J.~K.,  1998b,
  \mnras, 298, 583

\bibitem[\protect\citename{{Ivison} et~al.{\ }}{2000a}]{4cmap}
{Ivison}~R.~J.,  {Dunlop}~J.~S.,  {Smail}~I.,  {Dey}~A.,  {Liu}~M.~C.,
  {Graham}~J.~R.,  2000a, ApJ, accepted, astro-ph/0005234

\bibitem[\protect\citename{{Ivison} et~al.{\ }}{2000b}]{isbk00}
{Ivison}~R.~J.,  {Smail}~I.,  {Barger}~A.~J.,  {Kneib}~J.-P.,  {Blain}~A.~W.,
  {Owen}~F.~N.,  {Kerr}~T.~H.,    {Cowie}~L.~L.,  2000b, \mnras, 315, 209

\bibitem[\protect\citename{{Ivison} \& et~al.{\ }}{}]{rgmapping}
{Ivison}~R.~J. et al., in preparation

\bibitem[\protect\citename{{Jarvis} \& et~al.{\ }}{}]{jarvisinprep}
{Jarvis}~M.~J. et al., in preparation, (Jprep)

\bibitem[\protect\citename{{Jenness} et~al.{\ }}{1998}]{remskypaper}
{Jenness}~T.,  {Lightfoot}~J.~F.,    {Holland}~W.~S.,  1998, \procspie, 3357,
  548

\bibitem[\protect\citename{{Kaiser} et~al.{\ }}{1997}]{kda97}
{Kaiser}~C.~R.,  {Dennett-Thorpe}~J.,    {Alexander}~P.,  1997, \mnras, 292,
  723

\bibitem[\protect\citename{{Knapp} \& {Patten}{\ }}{1991}]{kp91}
{Knapp}~G.~R.,  {Patten}~B.~M.,  1991, \aj, 101, 1609

\bibitem[\protect\citename{{Lacy}{\ }}{1992}]{lthesis}
{Lacy}~M.~D.,  1992, Samples of radio sources selected at 38 MHz,
Ph.D. thesis, University of Oxford, (Lthesis)

\bibitem[\protect\citename{{Lacy} \& {Rawlings}{\ }}{1994}]{lr94}
{Lacy}~M.,  {Rawlings}~S.,  1994, \mnras, 270, 431, (LR94)

\bibitem[\protect\citename{{Lacy} et~al.{\ }}{1994}]{lmr94}
{Lacy}~M. et al.,  1994, \mnras, 271, 504, (LMR94)

\bibitem[\protect\citename{{Laing}{\ }}{}]{lpcomm}
{Laing}~R.~A., private communication, (Lpc)

\bibitem[\protect\citename{{Laing} \& {Peacock}{\ }}{1980}]{lp80}
{Laing}~R.~A.,  {Peacock}~J.~A.,  1980, \mnras, 190, 903, (LP)

\bibitem[\protect\citename{{Laing} et~al.{\ }}{1983}]{lrl83}
{Laing}~R.~A.,  {Riley}~J.~M.,    {Longair}~M.~S.,  1983, \mnras, 204, 151,
  (LRL)

\bibitem[\protect\citename{{Large} et~al.{\ }}{1981}]{lml81}
{Large}~M.~I.,  {Mills}~B.~Y.,  {Little}~A.~G.,  {Crawford}~D.~F.,
  {Sutton}~J.~M.,  1981, \mnras, 194, 693, (LML81)

\bibitem[\protect\citename{{LaValley} et~al.{\ }}{1992}]{lif92}
{LaValley}~M.,  {Isobe}~T.,    {Feigelson}~E.,  1992, in Worrall~D.~M.,
  Biemesderfer~C.,   Barnes~J.,  eds, ASP Conf. Ser. 25: Astronomical Data
  Analysis Software and Systems I.
Vol.\ ~1, p.~245

\bibitem[\protect\citename{{Law-Green} et~al.{\ }}{1995a}]{lge95}
{Law-Green}~J. D.~B.,  {Eales}~S.~A.,  {Leahy}~J.~P.,  {Rawlings}~S.,
  {Lacy}~M.,  1995a, \mnras, 277, 995, (LGE95)

\bibitem[\protect\citename{{Law-Green} et~al.{\ }}{1995b}]{lla95}
{Law-Green}~J. D.~B.,  {Leahy}~J.~P.,  {Alexander}~P.,
  {Allington-Smith}~J.~R.,  {van Breugel}~W. J.~M.,  {Eales}~S.~A.,
  {Rawlings}~S.~G.,    {Spinrad}~H.,  1995b, \mnras, 274, 939, (LLA95)

\bibitem[\protect\citename{{Lawrence} et~al.{\ }}{1986}]{lbh86}
{Lawrence}~C.~R.,  {Bennett}~C.~L.,  {Hewitt}~J.~N.,  {Langston}~G.~I.,
  {Klotz}~S.~E.,  {Burke}~B.~F.,    {Turner}~K.~C.,  1986, \apjs, 61, 105,
  (LBH86)

\bibitem[\protect\citename{{Lilly}{\ }}{1988}]{l88}
{Lilly}~S.~J.,  1988, \apj, 333, 161, (L88)

\bibitem[\protect\citename{{Lilly}{\ }}{1989}]{lilly89}
{Lilly}~S.~J.,  1989, \apj, 340, 77

\bibitem[\protect\citename{{Lilly} \& {Longair}{\ }}{1984}]{ll84}
{Lilly}~S.~J.,  {Longair}~M.~S.,  1984, \mnras, 211, 833

\bibitem[\protect\citename{{Liu} et~al.{\ }}{1992}]{lpr92}
{Liu}~R.,  {Pooley}~G.,    {Riley}~J.~M.,  1992, \mnras, 257, 545, (LPR92)

\bibitem[\protect\citename{{Magorrian} et~al.{\ }}{1998}]{magorrian98}
{Magorrian}~J. et al.,  1998, \aj, 115, 2285

\bibitem[\protect\citename{{Mathis} \& {Whiffen}{\ }}{1989}]{mw89}
{Mathis}~J.~S.,  {Whiffen}~G.,  1989, \apj, 341, 808

\bibitem[\protect\citename{{Maxfield} \& et~al.{\ }}{}]{mprep}
{Maxfield}~L. et al., in preparation, (Mprep)

\bibitem[\protect\citename{{McCarthy} et~al.{\ }}{1990}]{ms90}
{McCarthy}~P.~J.,  {Spinrad}~H.,  {van Breugel}~W.,  {Liebert}~J.,
  {Dickinson}~M.,  {Djorgovski}~S.,    {Eisenhardt}~P.,  1990, \apj, 365, 487,
  (MS90)

\bibitem[\protect\citename{{McCarthy} et~al.{\ }}{1995}]{msvb95}
{McCarthy}~P.~J.,  {Spinrad}~H.,    {van Breugel}~W.,  1995, \apjs, 99, 27,
  (MSvB95)

\bibitem[\protect\citename{{McCarthy} et~al.{\ }}{1997}]{mcc97}
{McCarthy}~P.~J.,  {Miley}~G.~K.,  {de Koff}~S.,  {Baum}~S.~A.,
  {Sparks}~W.~B.,  {Golombek}~D.,  {Biretta}~J.,    {Macchetto}~F.,  1997,
  \apjs, 112, 415, (McC97)

\bibitem[\protect\citename{{McLure} \& {Dunlop}{\ }}{2000}]{mcd00}
{McLure}~R.~J.,  {Dunlop}~J.~S.,  2000, MNRAS, submitted, astro-ph/9908214

\bibitem[\protect\citename{{McLure} et~al.{\ }}{1999}]{mkd99}
{McLure}~R.~J.,  {Kukula}~M.~J.,  {Dunlop}~J.~S.,  {Baum}~S.~A.,
  {O'Dea}~C.~P.,    {Hughes}~D.~H.,  1999, \mnras, 308, 377

\bibitem[\protect\citename{{Meisenheimer} \& {Heavens}{\ }}{1986}]{mh86}
{Meisenheimer}~K.,  {Heavens}~A.~F.,  1986, \nat, 323, 419

\bibitem[\protect\citename{{Miley} et~al.{\ }}{1992}]{mcvb92}
{Miley}~G.~K.,  {Chambers}~K.~C.,  {van Breugel}~W. J.~M.,    {Macchetto}~F.,
  1992, \apjl, 401, L69

\bibitem[\protect\citename{{Murgia} et~al.{\ }}{1999}]{mff99}
{Murgia}~M.,  {Fanti}~C.,  {Fanti}~R.,  {Gregorini}~L.,  {Klein}~U.,
  {Mack}~K.-H.,    {Vigotti}~M.,  1999, \aap, 345, 769

\bibitem[\protect\citename{{Muxlow} et~al.{\ }}{1988}]{mpr88}
{Muxlow}~T. W.~B.,  {Pelletier}~G.,    {Roland}~J.,  1988, \aap, 206, 237

\bibitem[\protect\citename{{Naundorf} et~al.{\ }}{1992}]{nar92}
{Naundorf}~C.~E.,  {Alexander}~P.,  {Riley}~J.~M.,    {Eales}~S.~A.,  1992,
  \mnras, 258, 647, (NAR92)

\bibitem[\protect\citename{{Nolan} et~al.{\ }}{2000}]{nolan2000}
{Nolan}~L.~A.,  {Dunlop}~J.~S.,  {Kukula}~M.~J.,  {Hughes}~D.~H.,
  {Boroson}~T.,    {Jimenez}~R.,  2000, MNRAS, to be published,
  astro-ph/0002020

\bibitem[\protect\citename{{Omont} et~al.{\ }}{1996}]{omc96}
{Omont}~A.,  {McMahon}~R.~G.,  {Cox}~P.,  {Kreysa}~E.,  {Bergeron}~J.,
  {Pajot}~F.,    {Storrie-Lombardi}~L.~J.,  1996, \aap, 315, 1

\bibitem[\protect\citename{{Oort} \& {van Langevelde}{\ }}{1987}]{ovl87}
{Oort}~M. J.~A.,  {van Langevelde}~H.~J.,  1987, \aaps, 71, 25, (OvL87)

\bibitem[\protect\citename{{Oort} et~al.{\ }}{1987}]{oks87}
{Oort}~M. J.~A.,  {Katgert}~P.,  {Steeman}~F. W.~M.,    {Windhorst}~R.~A.,
  1987, \aap, 179, 41, (OKS87)

\bibitem[\protect\citename{{Perlmutter} et~al.{\ }}{1999}]{perl99}
{Perlmutter}~S. et al., 1999, \apj, 517, 565

\bibitem[\protect\citename{{Pilkington} \& {Scott}{\ }}{1965}]{ps65}
{Pilkington}~J. D.~H.,  {Scott}~P.~F.,  1965, \memras, 69, 183, (PS65)

\bibitem[\protect\citename{{Rawlings} et~al.{\ }}{1990}]{rew90}
{Rawlings}~S.,  {Eales}~S.,    {Warren}~S.,  1990, \mnras, 243, L14, (REW90)

\bibitem[\protect\citename{{Rawlings} et~al.{\ }}{1996a}]{rlb96}
{Rawlings}~S.,  {Lacy}~M.,  {Blundell}~K.~M.,  {Eales}~S.~A.,  {Bunker}~A.~J.,
    {Garrington}~S.~T.,  1996a, \nat, 383, 502, (RLB96)

\bibitem[\protect\citename{{Rawlings} et~al.{\ }}{1996b}]{rll96}
{Rawlings}~S.,  {Lacy}~M.,  {Leahy}~J.~P.,  {Dunlop}~J.~S.,
  {Garrington}~S.~T.,    {L\"udke}~E.,  1996b, \mnras, 279, L13, (RLL96)

\bibitem[\protect\citename{{Rawlings} et~al.{\ }}{2000}]{rel00}
{Rawlings}~S.,  {Eales}~S.,    {Lacy}~M.,  2000, MNRAS, submitted, (REL00)

\bibitem[\protect\citename{{Rengelink} et~al.{\ }}{1997}]{wenss}
{Rengelink}~R.~B.,  {Tang}~Y.,  {de Bruyn}~A.~G.,  {Miley}~G.~K.,
  {Bremer}~M.~N.,  {R\"ottgering}~H. J.~A.,    {Bremer}~M. A.~R.,  1997,
  \aaps, 124, 259, (WENSS)

\bibitem[\protect\citename{{Ridgway} \& {Stockton}{\ }}{1997}]{rs97}
{Ridgway}~S.~E.,  {Stockton}~A.,  1997, \aj, 114, 511, (RS97)

\bibitem[\protect\citename{{Robson} et~al.{\ }}{1998}]{rlsh98}
{Robson}~E.~I.,  {Leeuw}~L.~L.,  {Stevens}~J.~A.,    {Holland}~W.~S.,  1998,
  \mnras, 301, 935

\bibitem[\protect\citename{{R\"ottgering} et~al.{\ }}{1995}]{rmc95}
{R\"ottgering}~H. J.~A.,  {Miley}~G.~K.,  {Chambers}~K.~C.,    {Macchetto}~F.,
  1995, \aaps, 114, 51, (RMC95)

\bibitem[\protect\citename{{R\"ottgering} et~al.{\ }}{1997}]{RvO97}
{R\"ottgering}~H. J.~A.,  {van Ojik}~R.,  {Miley}~G.~K.,  {Chambers}~K.~C.,
  {van Breugel}~W. J.~M.,    {de Koff}~S.,  1997, \aap, 326, 505, (RvO97)

\bibitem[\protect\citename{{Rudnick} et~al.{\ }}{1986}]{rjf86}
{Rudnick}~L.,  {Jones}~T.~W.,    {Fiedler}~R.,  1986, \aj, 91, 1011

\bibitem[\protect\citename{{Smail} et~al.{\ }}{2000a}]{sibk2000}
{Smail}~I.,  {Ivison}~R.,  {Blain}~A.,  {Kneib}~J.-P.,    {Owen}~F.,  2000a,
in van Breugel~W.,  Bland-Hawthorn~J.,  eds, ASP Conf. Ser. 195: Imaging the
  Universe in Three Dimensions. p.~248

\bibitem[\protect\citename{{Smail} et~al.{\ }}{2000b}]{smailzdistrib}
{Smail}~I.,  {Ivison}~R.~J.,  {Owen}~F.~N.,  {Blain}~A.~W.,    {Kneib}~J.~.,
  2000b, \apj, 528, 612

\bibitem[\protect\citename{{Smith} et~al.{\ }}{1979}]{sj79}
{Smith}~H.~E.,  {Junkkarinen}~V.~T.,  {Spinrad}~H.,  {Grueff}~G.,
  {Vigotti}~M.,  1979, \apj, 231, 307, (SJ79)

\bibitem[\protect\citename{{Spencer} et~al.{\ }}{1991}]{spencer91}
{Spencer}~R.~E. et al.,  1991, \mnras, 250, 225

\bibitem[\protect\citename{{Spinrad}{\ }}{a}]{spriv}
{Spinrad}~H., private communication, (Spriv)

\bibitem[\protect\citename{{Spinrad}{\ }}{b}]{sunpub}
{Spinrad}~H., unpublished, (S-unpub)

\bibitem[\protect\citename{{Spinrad}{\ }}{1982}]{S82}
{Spinrad}~H.,  1982, \pasp, 94, 397, (S82)

\bibitem[\protect\citename{{Spinrad} \& {Djorgovski}{\ }}{1984a}]{SD84a}
{Spinrad}~H.,  {Djorgovski}~S.,  1984a, \apjl, 280, L9, (SD84a)

\bibitem[\protect\citename{{Spinrad} \& {Djorgovski}{\ }}{1984b}]{SD84b}
{Spinrad}~H.,  {Djorgovski}~S.,  1984b, \apjl, 285, L49, (SD84b)

\bibitem[\protect\citename{{Spinrad} et~al.{\ }}{1985}]{SDM85}
{Spinrad}~H.,  {Djorgovski}~S.,  {Marr}~J.,    {Aguilar}~L.,  1985, \pasp, 97,
  932, (SDM85)

\bibitem[\protect\citename{{Spinrad} et~al.{\ }}{1995}]{SDG95}
{Spinrad}~H.,  {Dey}~A.,    {Graham}~J.~R.,  1995, \apjl, 438, L51, (SDG95)

\bibitem[\protect\citename{{Steidel} et~al.{\ }}{1999}]{sag99}
{Steidel}~C.~C.,  {Adelberger}~K.~L.,  {Giavalisco}~M.,  {Dickinson}~M.,
  {Pettini}~M.,  1999, \apj, 519, 1

\bibitem[\protect\citename{{Stern} et~al.{\ }}{1999}]{sds99}
{Stern}~D.,  {Dey}~A.,  {Spinrad}~H.,  {Maxfield}~L.,  {Dickinson}~M.,
  {Schlegel}~D.,    {Gonz{\'a}lez}~R.~A.,  1999, \aj, 117, 1122, (SDS99)

\bibitem[\protect\citename{{Strom} et~al.{\ }}{1990}]{SRS90}
{Strom}~R.~G.,  {Riley}~J.~M.,  {Spinrad}~H.,  {van Breugel}~W. J.~M.,
  {Djorgovski}~S.,  {Liebert}~J.,    {McCarthy}~P.~J.,  1990, \aap, 227, 19,
  (SRS90)

\bibitem[\protect\citename{{Tielens} et~al.{\ }}{1979}]{tmw79}
{Tielens}~A. G. G.~M.,  {Miley}~G.~K.,    {Willis}~A.~G.,  1979, \aaps, 35,
  153, (TMW79)

\bibitem[\protect\citename{{van Breugel} et~al.{\ }}{1992}]{vbf92}
{van Breugel}~W. J.~M.,  {Fanti}~C.,  {Fanti}~R.,  {Stanghellini}~C.,
  {Schilizzi}~R.~T.,    {Spencer}~R.~E.,  1992, \aap, 256, 56, (vBF92)

\bibitem[\protect\citename{{van Breugel} et~al.{\ }}{1998}]{vbs98}
{van Breugel}~W. J.~M.,  {Stanford}~S.~A.,  {Spinrad}~H.,  {Stern}~D.,
  {Graham}~J.~R.,  1998, \apj, 502, 614, (vBS98)

\bibitem[\protect\citename{{van Ojik} et~al.{\ }}{1996}]{vor96}
{van Ojik}~R.,  {R\"ottgering}~H. J.~A.,  {Carilli}~C.~L.,  {Miley}~G.~K.,
  {Bremer}~M.~N.,    {Macchetto}~F.,  1996, \aap, 313, 25, (vOR96)

\bibitem[\protect\citename{{Veron} et~al.{\ }}{1972}]{vvw72}
{Veron}~M.~P.,  {Veron}~P.,    {Witzel}~A.,  1972, \aap, 18, 82

\bibitem[\protect\citename{{Waddington}{\ }}{1999}]{wthesis}
{Waddington}~I.,  1999, The cosmological evolution of galaxies, Ph.D. thesis,
  University of Edinburgh, (Wthesis)

\bibitem[\protect\citename{{Waddington} et~al.{\ }}{2000}]{wadlbds}
{Waddington}~I.,  {Windhorst}~R.~A.,  {Dunlop}~J.~S.,  {Koo}~D.~C.,
  {Peacock}~J.~A.,  2000, MNRAS, accepted, astro-ph/0006169

\bibitem[\protect\citename{{Waldram} et~al.{\ }}{1996}]{wyr96}
{Waldram}~E.~M.,  {Yates}~J.~A.,  {Riley}~J.~M.,    {Warner}~P.~J.,  1996,
  \mnras, 282, 779, (WYR96)

\bibitem[\protect\citename{{White} \& {Becker}{\ }}{1992}]{wb92}
{White}~R.~L.,  {Becker}~R.~H.,  1992, \apjs, 79, 331, (WB92)

\bibitem[\protect\citename{{Willott} et~al.{\ }}{1999}]{wrbl99}
{Willott}~C.~J.,  {Rawlings}~S.,  {Blundell}~K.~M.,    {Lacy}~M.,  1999,
  \mnras, 309, 1017

\bibitem[\protect\citename{{Willott} \& et~al.{\ }}{2000}]{willott7crs}
{Willott}~C.~J.,  et~al. 2000, MNRAS, to be submitted, (W7CRS)

\bibitem[\protect\citename{{Windhorst} et~al.{\ }}{1984}]{wvhk84}
{Windhorst}~R.~A.,  {van Heerde}~G.~M.,    {Katgert}~P.,  1984, \aaps, 58, 1,
  (WvHK84)

\bibitem[\protect\citename{{Windhorst} et~al.{\ }}{1991}]{wbm91}
{Windhorst}~R.~A. et al.,  1991, \apj, 380, 362, (WBM91)

\bibitem[\protect\citename{{Wright} \& {Otrupcek}{\ }}{1990}]{p90}
{Wright}~A.,  {Otrupcek}~R.,  1990, Parkes Catalog.
Australia Telescope National Facility, (P90)

\end{thebibliography}

\appendix
\section{Radio-submillimetre spectral energy distributions}
\label{sedfig1}

\begin{table*}
\caption{Best-fit parameters (including the reduced $\chi^2$ of the
fit) of both straight-line and parabolic fits to the radio spectra
(the fits were made in log-space).  Note, the linear fits were applied
to $\nu > 1\; {\rm GHz}$ observations only.  In addition, if only two
data points were available (i.e. zero degrees of freedom), the slope
and y-intercept of the line joining the two data points were
calculated instead of using a fitting algorithm.  No value of
$\chi^2_{red}$ is given in this case.  Finally, all best-fit
parameters are quoted to two decimal places; therefore errors less
than 0.005 are quoted as being 0.00.}
\label{allradiofits}
\begin{tabular}{lrrrccrrrr}
\hline
&\multicolumn{3}{c}{Linear Fit: $y = mx + k$}&&&\multicolumn{4}{c}{Parabolic Fit: $y = ax^2+bx+c$}\\
Source&\multicolumn{1}{c}{$m$}&\multicolumn{1}{c}{$k$}&\multicolumn{1}{c}{$\chi^2_{red}$}&&&\multicolumn{1}{c}{$a$}&\multicolumn{1}{c}{$b$}&\multicolumn{1}{c}{$c$}&\multicolumn{1}{c}{$\chi^2_{red}$}\\
\hline
3C277.2      &$  -1.00\pm   0.03$   &$   0.41\pm   0.02$   &    2.8   &&&$  -0.03\pm   0.02$   &$  -0.96\pm   0.01$   &$   0.40\pm   0.01$   &    1.7\\
3C340        &$  -1.03\pm   0.02$   &$   0.56\pm   0.01$   &    2.9   &&&$  -0.12\pm   0.01$   &$  -0.86\pm   0.01$   &$   0.51\pm   0.01$   &    2.8\\
3C265        &$  -1.14\pm   0.03$   &$   0.63\pm   0.01$   &    1.3   &&&$  -0.07\pm   0.01$   &$  -1.05\pm   0.01$   &$   0.60\pm   0.01$   &    0.9\\
3C217        &$  -1.24\pm   0.03$   &$   0.52\pm   0.01$   &    2.3   &&&$  -0.16\pm   0.02$   &$  -1.02\pm   0.01$   &$   0.48\pm   0.01$   &    2.5\\
3C356        &$  -1.17\pm   0.06$   &$   0.33\pm   0.04$   &    1.3   &&&$  -0.09\pm   0.02$   &$  -1.10\pm   0.02$   &$   0.32\pm   0.01$   &    1.1\\
3C368        &$  -1.34\pm   0.10$   &$   0.23\pm   0.05$   &    0.2   &&&$  -0.13\pm   0.03$   &$  -1.34\pm   0.03$   &$   0.26\pm   0.02$   &    1.8\\
3C267        &$  -0.94\pm   0.03$   &$   0.49\pm   0.02$   &    2.9   &&&$  -0.04\pm   0.01$   &$  -0.92\pm   0.01$   &$   0.50\pm   0.01$   &    1.3\\
3C324        &$  -1.20\pm   0.02$   &$   0.59\pm   0.01$   &    5.6   &&&$  -0.15\pm   0.01$   &$  -1.01\pm   0.01$   &$   0.55\pm   0.01$   &    1.8\\
3C266        &$  -1.09\pm   0.04$   &$   0.30\pm   0.03$   &    4.0   &&&$  -0.06\pm   0.02$   &$  -1.04\pm   0.01$   &$   0.31\pm   0.01$   &    3.9\\
4C13.66      &$  -1.25\pm   0.08$   &$   0.42\pm   0.04$   &    1.2   &&&$  -0.25\pm   0.04$   &$  -1.04\pm   0.03$   &$   0.39\pm   0.02$   &    1.4\\
3C437        &$  -0.95\pm   0.02$   &$   0.58\pm   0.01$   &    0.5   &&&$  -0.11\pm   0.01$   &$  -0.83\pm   0.01$   &$   0.57\pm   0.01$   &    6.2\\
3C241        &$  -1.32\pm   0.05$   &$   0.43\pm   0.02$   &    0.2   &&&$  -0.20\pm   0.03$   &$  -1.09\pm   0.02$   &$   0.38\pm   0.01$   &    0.5\\
6C0919+38    &$  -1.16$             &$  -0.36$             &     --   &&&$  -0.09\pm   0.04$   &$  -1.06\pm   0.03$   &$  -0.37\pm   0.00$   &    0.1\\
3C470        &$  -1.15\pm   0.03$   &$   0.50\pm   0.01$   &    2.9   &&&$  -0.14\pm   0.01$   &$  -0.94\pm   0.01$   &$   0.45\pm   0.01$   &    6.0\\
3C322        &$  -1.18\pm   0.05$   &$   0.46\pm   0.03$   &    0.8   &&&$  -0.10\pm   0.02$   &$  -0.99\pm   0.02$   &$   0.39\pm   0.01$   &    2.3\\
6C1204+37    &$  -1.17$             &$  -0.04$             &     --   &&&$  -0.40\pm   0.03$   &$  -1.10\pm   0.02$   &$  -0.04\pm   0.00$   &    22\\
3C239        &$  -1.22\pm   0.05$   &$   0.34\pm   0.02$   &    0.1   &&&$  -0.09\pm   0.02$   &$  -1.13\pm   0.02$   &$   0.34\pm   0.01$   &    0.5\\
3C294        &$  -1.32\pm   0.05$   &$   0.32\pm   0.03$   &    0.3   &&&$  -0.07\pm   0.02$   &$  -1.16\pm   0.02$   &$   0.26\pm   0.01$   &    2.7\\
6C0820+36    &$  -1.14$             &$  -0.40$             &     --   &&&$  -0.17\pm   0.05$   &$  -1.09\pm   0.03$   &$  -0.40\pm   0.01$   &    2.2\\
6C0905+39    &$  -1.50$             &$  -0.37$             &     --   &&&$  -0.09\pm   0.04$   &$  -1.13\pm   0.03$   &$  -0.42\pm   0.01$   &    7.4\\
6C0901+35    &$  -1.17$             &$  -0.44$             &     --   &&&$  -0.23\pm   0.04$   &$  -1.12\pm   0.03$   &$  -0.44\pm   0.01$   &    7.7\\
5C7.269      &$  -0.98$             &$  -1.03$             &     --   &&&$  -0.01\pm   0.10$   &$  -0.98\pm   0.06$   &$  -1.03\pm   0.01$   &    0.1\\
4C40.36      &$  -1.62\pm   0.03$   &$  -0.06\pm   0.02$   &    0.7   &&&$  -0.23\pm   0.03$   &$  -1.38\pm   0.02$   &$  -0.10\pm   0.01$   &    0.2\\
MG1744+18    &$  -1.16\pm   0.00$   &$   0.30\pm   0.00$   &    335   &&&$  -0.14\pm   0.00$   &$  -0.94\pm   0.00$   &$   0.22\pm   0.00$   &    62\\
MG1248+11    &$  -1.27$             &$  -0.34$             &     --   &&&$  -0.41\pm   0.15$   &$  -0.96\pm   0.04$   &$  -0.36\pm   0.04$   &    0.9\\
4C48.48      &$  -1.19\pm   0.04$   &$  -0.28\pm   0.02$   &    0.8   &&&$  -0.10\pm   0.03$   &$  -1.10\pm   0.01$   &$  -0.29\pm   0.01$   &    0.8\\
53W002       &$  -1.26\pm   0.05$   &$  -1.11\pm   0.02$   &    0.5   &&&$  -0.09\pm   0.04$   &$  -1.15\pm   0.03$   &$  -1.13\pm   0.01$   &    0.5\\
6C0930+38    &$  -1.00$             &$  -0.40$             &     --   &&&$  -0.10\pm   0.04$   &$  -0.99\pm   0.02$   &$  -0.39\pm   0.01$   &    3.5\\
6C1113+34    &$  -1.14$             &$  -0.23$             &     --   &&&$  -0.29\pm   0.04$   &$  -0.98\pm   0.03$   &$  -0.24\pm   0.00$   &    3.9\\
MG2305+03    &$  -0.97\pm   0.09$   &$  -0.15\pm   0.05$   &    0.2   &&&$  -0.02\pm   0.14$   &$  -0.95\pm   0.05$   &$  -0.15\pm   0.03$   &    0.2\\
3C257        &$  -1.03\pm   0.06$   &$   0.39\pm   0.03$   &    0.4   &&&$  -0.12\pm   0.04$   &$  -0.92\pm   0.02$   &$   0.37\pm   0.02$   &    0.3\\
4C23.56      &$  -1.40\pm   0.03$   &$  -0.19\pm   0.02$   &     18   &&&$  -0.16\pm   0.02$   &$  -1.27\pm   0.02$   &$  -0.20\pm   0.01$   &    11\\
8C1039+68    &$  -1.12$             &$  -0.44$             &     --   &&&$  -0.08\pm   0.05$   &$  -1.09\pm   0.05$   &$  -0.43\pm   0.04$   &    0.2\\
MG1016+058   &$  -1.10\pm   0.02$   &$  -0.25\pm   0.01$   &    7.9   &&&$  -0.27\pm   0.02$   &$  -0.79\pm   0.02$   &$  -0.30\pm   0.01$   &    5.4\\
4C24.28      &$  -1.31\pm   0.02$   &$  -0.08\pm   0.01$   &    0.8   &&&$  -0.15\pm   0.02$   &$  -1.18\pm   0.01$   &$  -0.09\pm   0.00$   &    7.9\\
4C28.58      &$  -1.50\pm   0.04$   &$  -0.40\pm   0.02$   &    2.0   &&&$  -0.16\pm   0.05$   &$  -1.37\pm   0.03$   &$  -0.41\pm   0.02$   &    2.3\\
6C1232+39    &$  -1.70\pm   0.00$   &$  -0.23\pm   0.00$   &    659   &&&$  -0.30\pm   0.00$   &$  -1.29\pm   0.00$   &$  -0.37\pm   0.00$   &    25\\
6C1159+36    &$  -1.11$             &$  -0.27$             &     --   &&&$  -0.34\pm   0.04$   &$  -1.03\pm   0.03$   &$  -0.27\pm   0.01$   &    19\\
6C0902+34    &$  -0.96$             &$  -0.33$             &     --   &&&$  -0.14\pm   0.03$   &$  -0.93\pm   0.02$   &$  -0.34\pm   0.00$   &    2.5\\
TX1243+036   &$  -1.31\pm   0.06$   &$  -0.32\pm   0.04$   &    2.9   &&&$  -0.08\pm   0.06$   &$  -1.33\pm   0.03$   &$  -0.27\pm   0.02$   &    3.4\\
MG2141+192   &$  -1.57\pm   0.00$   &$  -0.12\pm   0.00$   &     25   &&&$  -0.27\pm   0.02$   &$  -1.14\pm   0.02$   &$  -0.29\pm   0.01$   &    11\\
6C0032+412   &$  -1.12\pm   0.03$   &$  -0.85\pm   0.02$   &    0.3   &&&$   0.03\pm   0.02$   &$  -1.14\pm   0.01$   &$  -0.85\pm   0.01$   &    0.7\\
4C60.07      &$  -1.48\pm   0.03$   &$  -0.60\pm   0.02$   &    0.4   &&&$  -0.05\pm   0.02$   &$  -1.44\pm   0.01$   &$  -0.60\pm   0.01$   &    0.7\\
4C41.17      &$  -1.49\pm   0.03$   &$  -0.41\pm   0.02$   &    1.6   &&&$  -0.15\pm   0.02$   &$  -1.33\pm   0.01$   &$  -0.44\pm   0.01$   &    1.1\\
8C1435+635   &$  -1.85\pm   0.00$   &$   0.09\pm   0.00$   &    4.3   &&&$  -0.35\pm   0.01$   &$  -1.30\pm   0.01$   &$  -0.12\pm   0.00$   &    3.3\\
6C0140+326   &$  -1.39\pm   0.02$   &$  -0.83\pm   0.01$   &    4.4   &&&$  -0.19\pm   0.02$   &$  -1.21\pm   0.01$   &$  -0.86\pm   0.01$   &    1.1\\
\hline
\end{tabular}
\end{table*}

We applied two fits to the radio spectra of our sample:

{\bf 1) Linear Fit -} In log-space, a straight line is fitted to radio
data with $\nu > 1\; {\rm GHz}$.  For all galaxies in the sample, at
least two data points satisfy this condition.  Except for the
noticeably curved spectra of 3C324, MG1744+18, 4C23.56, and
MG1016+058, the straight-line fit seems to mimic the existing
high-frequency data very well.  

{\bf 2) Parabolic Fit -} A physical model of synchrotron ageing could
be applied to the spectra in order to determine how they might
steepen.  However, the fit is only required to accurately predict the
shape of the expected steepening; the physics of the ageing
synchrotron population does not need to be extracted from the fit.
Blundell et al. \shortcite{brw99} have studied the
properties and evolution of over 300 FRII radio sources selected from
the 3C, 6C, and 7C surveys.  They applied a Bayesian polynomial
regression analysis to the fitting the flux densities.  The process
involved fitting polynomials of different degrees and associating a
likelihood or probability to each.  In some cases a first-order
polynomial, or straight-line was the preferred choice.  However, in
most cases a second-order polynomial, or parabola, accurately modelled
the curvature of the spectrum.  A parabolic fit seems to be the best
way of parameterising the curvature, as higher-order polynomials were
never chosen by the regression analysis.  Thus a polynomial of degree
$m=2$ was fitted, in log space, to the entire radio spectrum.

The best-fit parameters of the two models are shown in
\tab{allradiofits}.

For each galaxy in the sample, the radio-submillimetre spectral energy
distribution (SED) is presented in \fig{seds}.  The SEDs are arranged
in redshift order.  The solid circles denote integrated continuum flux
densities.  The open circles indicate radio core flux densities.  For
3C356, two radio core identifications exist.  The fainter core is
denoted by open stars.  If errors were unavailable, the error was
taken to be 10 per cent of the flux density.  Upper limits have been
taken at the 3-$\sigma$ level if S/N$<3$.  The two models fit to the
data are also displayed in \fig{seds}.  The solid line is the
parabolic fit to the spectrum, and the dash-dot line is the linear
fit.  The references for the data are:
{\bf 3C277.2} - BRR00\nocite{brr00}, LP\nocite{lp80};  
{\bf 3C340} - BRR00\nocite{brr00}, LP\nocite{lp80};  
{\bf 3C265} - BRR00\nocite{brr00}, LP\nocite{lp80};  
{\bf 3C217} - BRR00\nocite{brr00}, LP\nocite{lp80};  
{\bf 3C356} - BLR97\nocite{blr97}, F93\nocite{f93}, LP\nocite{lp80};
{\bf 3C368} - BLR97\nocite{blr97}, BRR00\nocite{brr00}, LP\nocite{lp80};
{\bf 3C267} - BLR97\nocite{blr97}, BRR00\nocite{brr00}, LP\nocite{lp80};
{\bf 3C324} - BCG98\nocite{bcg98}, BRB98\nocite{brb98}, LP\nocite{lp80};  
{\bf 3C266} - BLR97\nocite{blr97}, BRR00\nocite{brr00}, LP\nocite{lp80}, LPR92\nocite{lpr92};
{\bf 4C13.66} - BWE91\nocite{bwe91}, GC91\nocite{gc91}, BRR00\nocite{brr00}, LML81\nocite{lml81}, LRL, P90\nocite{p90}, WB92\nocite{wb92};  
{\bf 3C437} - BRR00\nocite{brr00}, LP\nocite{lp80};  
{\bf 3C241} - ASZ91\nocite{asz91}, LP\nocite{lp80}, vBF92\nocite{vbf92};
{\bf 6C0919+38} - 6C-II\nocite{6cII}, CC98\nocite{cc98}, FGT85\nocite{fgt85}, GC91\nocite{gc91}, GSDC96\nocite{gsdc96};
{\bf 3C470} - BLR97\nocite{blr97}, LP\nocite{lp80};  
{\bf 3C322} - BRR00\nocite{brr00}, LP\nocite{lp80};  
{\bf 6C1204+37} - 6C-II\nocite{6cII}, CC98\nocite{cc98}, DBB96\nocite{dbb96}, GC91\nocite{gc91}, GSDC96\nocite{gsdc96}, LLA95\nocite{lla95};  
{\bf 3C239} - BRR00\nocite{brr00}, LP\nocite{lp80};  
{\bf 3C294} - BRR00\nocite{brr00}, LP\nocite{lp80};  
{\bf 6C0820+36} - 6C-VI\nocite{6cVI}, CC98\nocite{cc98}, DBB96\nocite{dbb96}, GC91\nocite{gc91}, GSDC96\nocite{gsdc96}, LLA95\nocite{lla95};
{\bf 6C0905+39} - 6C-II\nocite{6cII}, CC98\nocite{cc98}, DBB96\nocite{dbb96}, FGT85\nocite{fgt85}, GC91\nocite{gc91}, GSDC96\nocite{gsdc96}, LGE95\nocite{lge95};  
{\bf 6C0901+35} - 6C-II\nocite{6cII}, CC98\nocite{cc98}, DBB96\nocite{dbb96}, GC91\nocite{gc91}, GSDC96\nocite{gsdc96}, NAR92\shortcite{nar92};  
{\bf 5C7.269} - BRR00\nocite{brr00}, CC98\nocite{cc98}, DBB96\nocite{dbb96}, GSDC96\nocite{gsdc96};
{\bf 4C40.36} - C\&C69\nocite{cc69}, CMvB96\nocite{cmvb96}, CRvO97\nocite{crvo97}, FGT85\nocite{fgt85}, TMW79\nocite{tmw79}, WENSS\nocite{wenss};  
{\bf MG1744+18} - BWE91\nocite{bwe91}, CRvO97\nocite{crvo97}, GC91\nocite{gc91}, GSW67\nocite{gsw67}, KB17\nocite{kb17}, LML81\nocite{lml81}, P90\nocite{p90}, WB92\nocite{wb92};  
{\bf MG1248+11} - DBB96\nocite{dbb96}, LML81\nocite{lml81}, LBH86\nocite{lbh86}, WB92\nocite{wb92};
{\bf 4C48.48} - 6C-V\nocite{6cV}, CMvB96\nocite{cmvb96}, CRvO97\nocite{crvo97}, GSW67\nocite{gsw67}, TMW79\nocite{tmw79}, WENSS\nocite{wenss};  
{\bf 53W002} - 6C-II\nocite{6cII}, HDR97\nocite{hdr97}, WBM91\nocite{wbm91}, WvHK84\nocite{wvhk84}, OKS87\nocite{oks87}, OvL87\nocite{ovl87};  
{\bf 6C0930+38} - 6C-II\nocite{6cII}, CC98\nocite{cc98}, DBB96\nocite{dbb96}, FGT85\nocite{fgt85}, GC91\nocite{gc91}, GSDC96\nocite{gsdc96}, NAR92\shortcite{nar92};  
{\bf 6C1113+34} - 6C-II\nocite{6cII}, CC98\nocite{cc98}, DBB96\nocite{dbb96}, GC91\nocite{gc91}, GSDC96\nocite{gsdc96}, LLA95\nocite{lla95}A;
{\bf MG2305+03} - BWE91\nocite{bwe91}, GC91\nocite{gc91}, GWBE95\nocite{gwbe95}, LML81\nocite{lml81}, P90\nocite{p90}, WB92\nocite{wb92};
{\bf 3C257} - BWE91\nocite{bwe91}, CC98\nocite{cc98}, DBB96\nocite{dbb96}, GC91\nocite{gc91}, GSW67\nocite{gsw67}, GWBE95\nocite{gwbe95}, HDR97\nocite{hdr97}, LML81\nocite{lml81}, P90\nocite{p90}, vBS98\nocite{vbs98}, WB92\nocite{wb92};  
{\bf 4C23.56} - CMvB96\nocite{cmvb96}, CRvO97\nocite{crvo97}, P90\nocite{p90}, PS65\nocite{ps65}, TMW79\nocite{tmw79};  
{\bf 8C1039+68} - HWR95\nocite{hwr95}, Lthesis\nocite{lthesis};  
{\bf MG1016+058} - BWE91\nocite{bwe91}, CC98\nocite{cc98}, CFR98\nocite{cfr98}, DBB96\nocite{dbb96}, DSD95\nocite{dsd95}, GC91\nocite{gc91}, GSW67\nocite{gsw67}, GWBE95\nocite{gwbe95}, LML81\nocite{lml81}, P90\nocite{p90}, WB92\nocite{wb92};  
{\bf 4C24.28} - CMvB96\nocite{cmvb96}, CRvO97\nocite{crvo97}, P90\nocite{p90}, PS65\nocite{ps65}, TMW79\nocite{tmw79}, WYR96\nocite{wyr96};  
{\bf 4C28.58} - CC98\nocite{cc98}, CMvB96\nocite{cmvb96}, DBB96\nocite{dbb96}, PS65\nocite{ps65}, TMW79\nocite{tmw79}; 
{\bf 6C1232+39} - 6C-II\nocite{6cII}, CC98\nocite{cc98}, C\&K94\nocite{ck94}, CRvO97\nocite{crvo97}, DBB96\nocite{dbb96}, FGT85\nocite{fgt85}, GC91\nocite{gc91}, GSDC96\nocite{gsdc96}, NAR92\shortcite{nar92}; 
{\bf 6C1159+36} - 6C-II\nocite{6cII}, CC98\nocite{cc98}, DBB96\nocite{dbb96}, GC91\nocite{gc91}, GSDC96\nocite{gsdc96}, LLA95\nocite{lla95}; 
{\bf 6C0902+34} - 6C-II\nocite{6cII}, C95\nocite{c95}, CC98\nocite{cc98}, C\&K94\nocite{ck94}, COH94\nocite{coh94}, DBB96\nocite{dbb96}, DSS96\nocite{dss96}, GC91\nocite{gc91}, GSDC96\nocite{gsdc96}, HDR97\nocite{hdr97}, LLA95\nocite{lla95}; 
{\bf TX1243+036} - CC98\nocite{cc98}, CFR98\nocite{cfr98}, DBB96\nocite{dbb96}, GSW67\nocite{gsw67}, LML81\nocite{lml81}, P90\nocite{p90}, vOR96\nocite{vor96}, WB92\nocite{wb92}; 
{\bf MG2141+192} - BWE91\nocite{bwe91}, CC98\nocite{cc98}, CRvO97\nocite{crvo97}, DBB96\nocite{dbb96}, GC91\nocite{gc91}, GSW67\nocite{gsw67}, HDR97\nocite{hdr97}, WB92\nocite{wb92}; 
{\bf 6C0032+412} - 6C-VI\nocite{6cVI}, BRE98\nocite{bre98}, DBB96\nocite{dbb96}, FGT85\nocite{fgt85}, GC91\nocite{gc91}, GSDC96\nocite{gsdc96}, HDR97\nocite{hdr97}; 
{\bf 4C60.07} - 6C-V\nocite{6cV}, CC98\nocite{cc98}, CMvB96\nocite{cmvb96}, CRvO97\nocite{crvo97}, DBB96\nocite{dbb96}, TMW79\nocite{tmw79}, WENSS\nocite{wenss}; 
{\bf 4C41.17} - 6C-VI\nocite{6cVI}, BCO99\nocite{bco99}, CC98\nocite{cc98}, C\&K94\nocite{ck94}, CMvB90\nocite{cmvb90}, COH94\nocite{coh94}, DBB96\nocite{dbb96}, DHR94\nocite{dhr94}, FGT85\nocite{fgt85}, GSW67\nocite{gsw67}, TMW79\nocite{tmw79}, WENSS\nocite{wenss}; 
{\bf 8C1435+635} - 6C-III\nocite{6cIII}, CC98\nocite{cc98}, CRvO97\nocite{crvo97}, DBB96\nocite{dbb96}, HDR97\nocite{hdr97}, HWR95\nocite{hwr95}, I95\nocite{i95}, IDHA98, LMR94\nocite{lmr94}; 
{\bf 6C0140+326} - 6C-VI\nocite{6cVI}, BRE98\nocite{bre98}, CC98\nocite{cc98}, DBB96\nocite{dbb96}, FGT85\nocite{fgt85}, RLB96\nocite{rlb96}.

\begin{figure*}
\epsfig{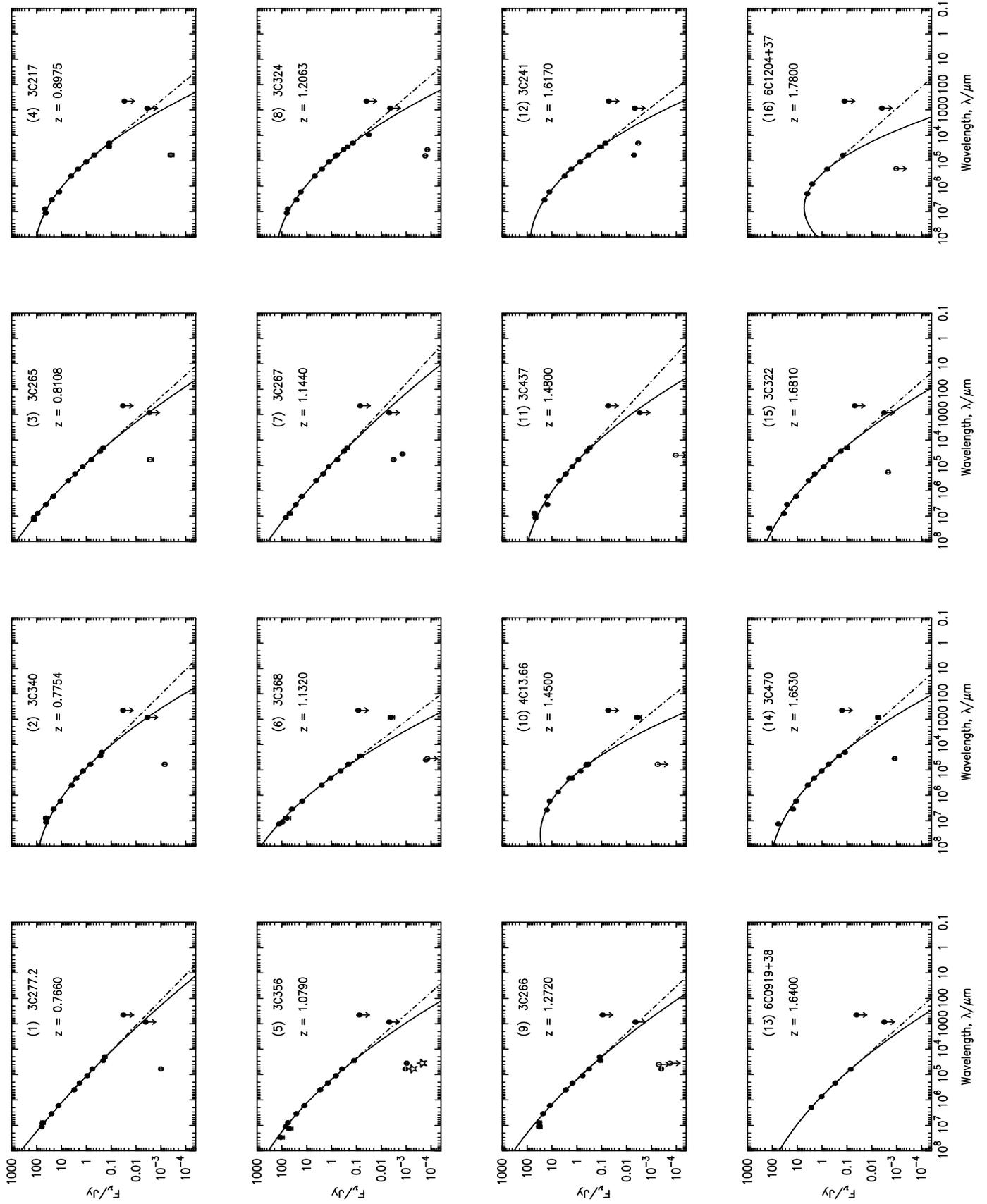}
\label{seds}
\caption{Spectral Energy Distributions}
\end{figure*}

\begin{figure*}
\epsfig{file=figA1_b.eps,height=23cm,width=18.5cm}
\contcaption{}
\end{figure*}

\begin{figure*}
\epsfig{file=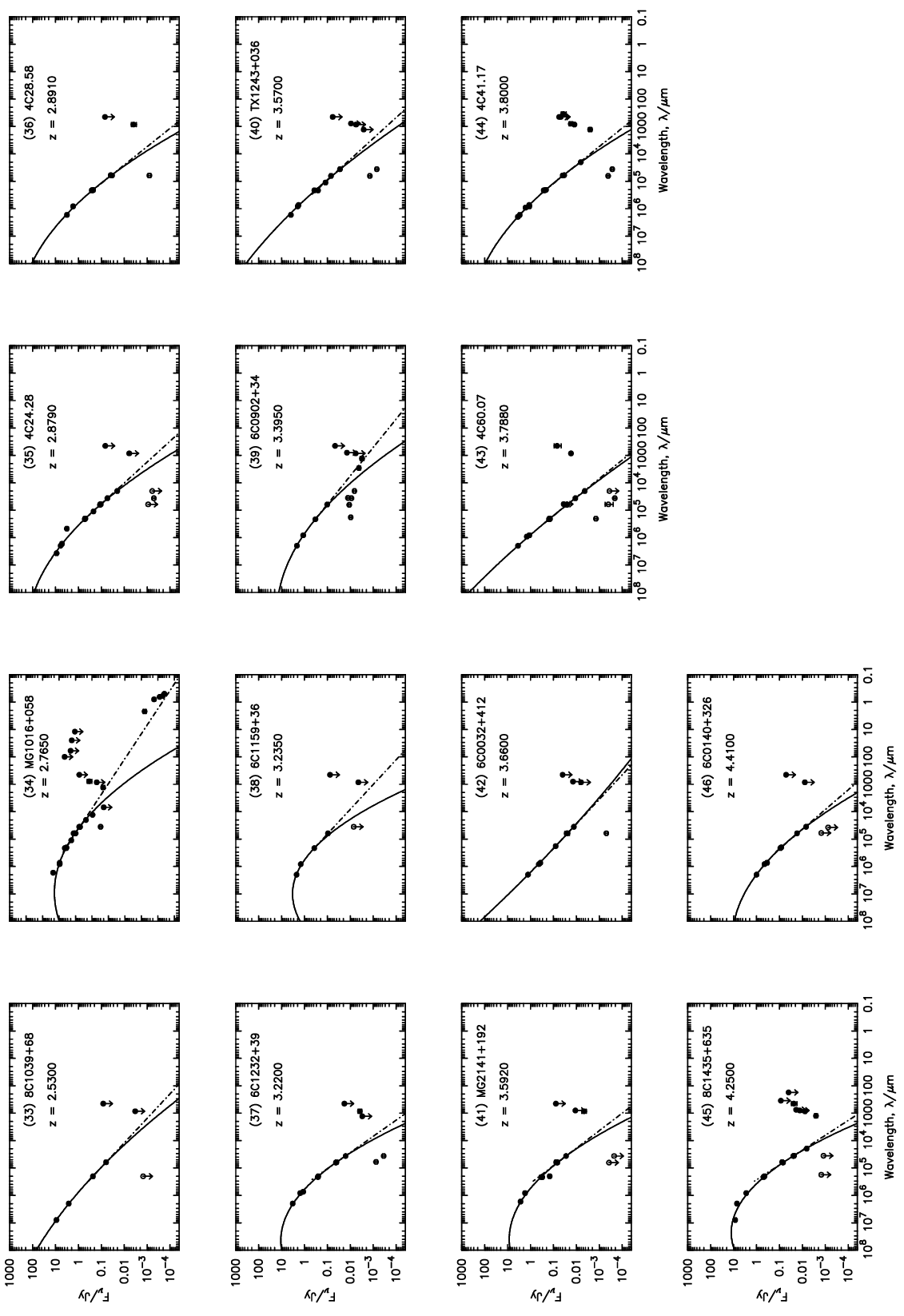,height=23cm,width=18.5cm}
\contcaption{}
\end{figure*}

\label{lastpage}

\end{document}